\documentclass[journal]{IEEEtran}
\usepackage{graphicx}
\usepackage{subfigure}
\usepackage{cite}
\usepackage{bm}
\usepackage{amsmath}   % From the American Mathematical Society
\usepackage{amsthm}
\usepackage{cases}
\usepackage{amsfonts,amssymb}
\usepackage{cuted}\stripsep-3pt
\usepackage{lipsum}

\usepackage{color}
\usepackage[normalem]{ulem}
\usepackage{float}

\usepackage[absolute,showboxes]{textpos}

\TPMargin{3pt}

\newcommand{\copyrightstatement}{
    \begin{textblock}{0.84}(0.08,0.95) % tweak here: {box width}(left position, bottom position)
         \noindent
         \footnotesize
         \copyright 2024 IEEE. Personal use of this material is permitted. Permission from IEEE must be obtained for all other uses, in any current or future media, including reprinting/republishing this material for advertising or promotional purposes, creating new collective works, for resale or redistribution to servers or lists, or reuse of any copyrighted component of this work in other works. DOI: 10.1109/JSAC.2024.3389118.
    \end{textblock}
}
\ifCLASSINFOpdf
\else
\fi
\begin{document}
\copyrightstatement
\title{Fractal OAM Generation and Detection Schemes}


\author{\IEEEauthorblockN{Runyu Lyu, \textit{Student Member, IEEE}, Wenchi Cheng, \textit{Senior Member, IEEE}, Muyao Wang, \textit{Student Member, IEEE}, and Wei Zhang, \textit{Fellow, IEEE}~}% <-this % stops an unwanted space

%\IEEEauthorblockA{State Key Laboratory of Integrated Services Networks, Xidian University, Xi'an, China}
%\vspace{-20pt}
\thanks{
Manuscript received March 31th, 2023; revised August 15th, 2023; accepted October 19th, 2023. This work was supported in part by National Key R\&D Program of China under Grant 2021YFC3002102, in part by the Key R\&D Plan of Shaanxi Province under Grant 2022ZDLGY05-09, and in part by Kry Area Research and Development Program of Guangdong Province under Grant 2020B0101110003. {\it(Corresponding author: Wenchi Cheng.)}

Runyu Lyu, Wenchi Cheng, and Muyao Wang are with Xidian University, Xi'an, 710071, China (e-mails: rylv@stu.xidian.edu.cn; wccheng@xidian.edu.cn).

Wei Zhang is with the School of Electrical Engineering and Telecommunications, the University of New South Wales, Sydney, Australia (e-mail: w.zhang@unsw.edu.au)
}
}

%{Shell \MakeLowercase{\textit{et al.}}: Bare Demo of IEEEtran.cls for IEEE Transactions on Magnetics Journals}



% for Transactions on Magnetics papers, we must declare the abstract and
% index terms PRIOR to the title within the \IEEEtitleabstractindextext
% IEEEtran command as these need to go into the title area created by
\IEEEtitleabstractindextext{%
%\vspace{-10pt}
\begin{abstract}
Orbital angular momentum (OAM) carried electromagnetic waves have the potential to improve spectrum efficiency in optical and radio-frequency communications due to the orthogonal wavefronts of different OAM modes. However, OAM beams are vortically hollow and divergent, which significantly decreases the capacity of OAM transmissions. In addition, unaligned transceivers in OAM transmissions can result in a high bit error rate (BER). The Talbot effect is a self-imaging phenomenon that can be used to generate optical or radio-frequency OAM beams with periodic repeating structures at multiples of a certain distance along the propagation direction. These periodic structures make it unnecessary for the transceiver antennas to be perfectly aligned and can also alleviate the hollow divergence of OAM beams. In this paper, we propose Talbot-effect-based fractal OAM generation and detection schemes using a uniform circular array (UCA) to significantly improve capacity and BER performance in unaligned OAM transmissions. We first provide a brief overview of fractal OAM. Then, we propose the fractal OAM beam generation and detection schemes. Numerical analysis and simulations verify the effectiveness of our proposed fractal OAM generation scheme and also demonstrate improved capacity and BER performance compared to normal OAM transmissions. We also analyze how the receive UCA radius and the distance between the UCAs impact the capacity and BER performances.
\end{abstract}

%\vspace{-5pt}
\begin{IEEEkeywords}
Orbital angular momentum (OAM), Talbot effect, uniform circular array (UCA).
\end{IEEEkeywords}}

\maketitle

\IEEEdisplaynontitleabstractindextext

% \fi
\IEEEpeerreviewmaketitle

\vspace{15pt}
\section{Introduction}
\IEEEPARstart{O}{rbital} angular momentum (OAM)\cite{oam_light} is the angular momentum of electromagnetic waves around the propagation axis. It can be used in optical and radio-frequency communications to improve spectrum efficiency\cite{oam_low_freq_radio,OAM_NFC_mine} and increase energy efficiency\cite{SWIPT_OAM_mine} by multiplexing and demultiplexing different OAM modes. However, OAM beams are vortically hollow and divergent\cite{oam_hollow}. Also, the divergence becomes more severe as the OAM mode index increases. This divergence significantly reduces the capacity for OAM transmissions. Moreover, OAM transmissions are originally designed for transceiver-aligned scenarios\cite{oam_detection}, where multiple OAM modes can be easily distinguished at the receiver. Unaligned transceiver-based OAM transmissions can result in a high bit error rate (BER)\cite{oam_AoAEs_UCA,oam_align_beamform}.

To address the divergence of OAM beams, many studies have focused on designing high-gain OAM generation antennas that can generate less divergent OAM beams in different frequency bands\cite{oam_metasurface2,oam_Higher_Order}. However, the designs of OAM antennas are complex and are not compatible with traditional antennas. The authors of \cite{oam_zc_scra} and \cite{oam_zc} propose schemes to generate non-hollow OAM beams using uniform circular arrays (UCAs) while maintaining the orthogonality among different OAM modes. However, these schemes require the transceiver antennas to be perfectly aligned. To decompose OAM beams with multiple OAM modes for transceiver-unaligned scenarios, the authors of \cite{oam_align_beamform} propose a joint beamforming and pre-detection scheme. This scheme can efficiently detect the signals of multiple OAM modes for unaligned transceivers, significantly increasing the spectrum efficiency. However, the beamforming and pre-detection rely on geometric parameters such as the radii of transceiver antennas and the distance between them, which are difficult to accurately obtain. The authors of \cite{oam_AoAEs_UCA} propose an OAM reception scheme that estimates the angle of arrival (AOA) and distance. This scheme can eliminate the effect of misalignment errors and approaches the performance of an ideally aligned OAM channel. However, the estimations of AOA and distance increase the complexity of OAM transmissions.

\begin{figure*}[htbp]
\centering
%\vspace{-10pt}
\includegraphics[scale=0.77]{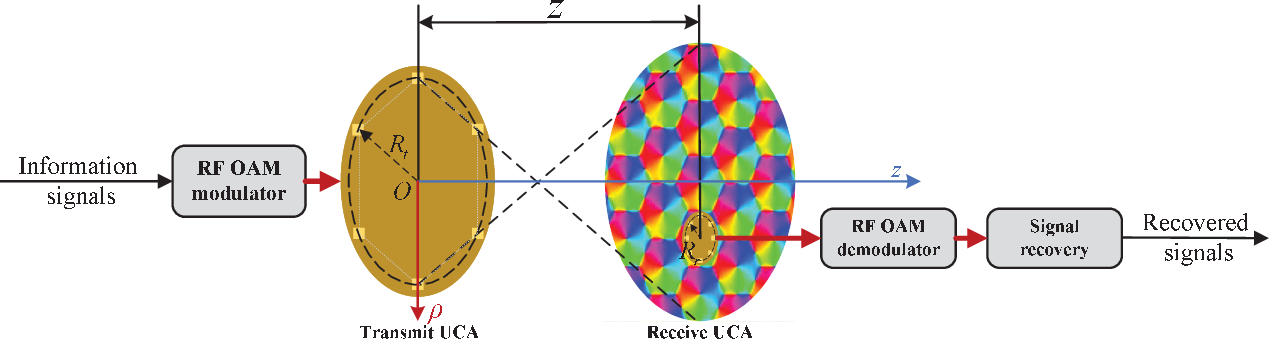}
%\vspace{-5pt}
\caption{The fractal OAM system model.} \label{fig:system_model}
%\vspace{-15pt}
\end{figure*}
The Talbot effect is an optical self-imaging effect originally discovered by Henry Fox Talbot in the 1830s\cite{Talbot}. When a light wave passes through a periodic or quasi-periodic object, the wave is scattered in many directions. Some of the scattered light waves interfere with each other constructively or destructively, producing a diffraction pattern. If the object is placed at a certain distance from the light source, known as the Talbot distance, the diffraction pattern will be a periodic replica of the object itself. Furthermore, at half or a quarter of the Talbot distance, there are double or quadruple self-imaging structures\cite{Talbot_quadruple}. Talbot effect can be used to generate optical fractal OAM beams as proposed in \cite{Quasi-Talbot_OAM}. However, since OAM beams are not periodic and the distance between the focal plane and the self-imaging plane is not fixed, the effect used in \cite{Quasi-Talbot_OAM} is referred to as the quasi-Talbot effect, rather than a strict Talbot effect. The Talbot effect can also be observed at radio frequency bands \cite{Talbot_mmWave,Talbot_antenna}, where it is known as the microwave Talbot effect. Like with visible light, microwaves can be diffracted by periodic or quasi-periodic objects, producing a diffraction pattern that is a periodic or quasi-periodic replica of the object. The authors of \cite{OAM_mobileTalbot} discuss using the quasi-Talbot effect to generate OAM beams in radio frequency based on the reconfigurable intelligent surface (RIS). While simulations show the feasibility of using the quasi-Talbot effect to generate OAM beams in radio frequency based on RIS, the complexity and energy utilization are worse compared to using UCA according to the authors. It should be noted that, the fractal OAM beam generations described in \cite{Quasi-Talbot_OAM} and \cite{OAM_mobileTalbot} are based on a UCA consisting of six OAM generators. Each of these OAM generators has the capability to independently generate OAM beams. While in this paper, we will propose a fractal OAM beam generation scheme based on conventional antenna array, reducing the complexity of antenna array compared with schemes in \cite{Quasi-Talbot_OAM} and \cite{OAM_mobileTalbot}.

In this paper, we propose Talbot-effect-based fractal OAM beam generation and detection schemes based on UCA. These schemes can greatly improve the capacity and BER performances for unaligned OAM transmissions as well as reduce the divergence of OAM beams. We find that periodic OAM phase and amplitude structures can be generated in planes parallel to the transmit UCA by setting suitable radii for the transmit UCA. The radius and phase structures of fractal OAM beams are nearly identical, allowing the receive UCA to be unaligned with the transmit UCA and reducing the hollow divergence of OAM beams. We analyze the relationship of the transmit UCA radius, the transmission distance, the wavelength, the fractal OAM center coordinates, and the fractal OAM radius. Based on the analysis, we derive the coordinates and radii for fractal OAM beams and propose the fractal OAM beam generation and detection schemes. Then, we derive the channel capacity and BER of our proposed fractal OAM generation and detection scheme. Numerical results are given to demonstrate the improved capacity and BER performances compared to normal OAM transmissions. We also analyze how the receive UCA radius and the distance between the UCAs impact the capacity and BER performances. Simulations based on ANSYS high-frequency structure simulator (HFSS) verify our proposed fractal OAM generation scheme and validate the above analysis.

The rest of the paper is organized as follows. Section~\ref{sec:systemModel} gives the fractal OAM system model. Section~\ref{sec:FractalOAM_GenerationDetection} introduces the UCA-based fractal OAM phenomenon and gives the fractal OAM generation and detection schemes. Section~\ref{sec:performanceAnalysis} derives and gives numerical analysis for the capacity and BER of our proposed fractal OAM generation and detection scheme. In Section~\ref{sec:simulations}, simulations are given to verify our proposed fractal OAM generation scheme and the analysis given in Section~\ref{sec:performanceAnalysis}. The conclusion is given in Section~\ref{sec:Conclusion}.

\section{Fractal OAM System Model}\label{sec:systemModel}
Figure~\ref{fig:system_model} shows the fractal OAM system model, where the transmit antenna is a UCA with six elements and the receive antenna is a UCA with $N_r$ elements. The radius of the transmit UCA is denoted by $R_t$ and the radius of the receive UCA is denoted by $R_r$. The wavelength of the electromagnetic waves is denoted by $\lambda$. The multi-stream information signals are first modulated into excitations corresponding to the six elements of the transmit UCA using unit inverse discrete Fourier transform (IDFT). This transform is equivalent to simultaneously feeding the equidistant transmit elements around the transmit ring with equal-amplitude excitations with linearly increasing phases. The phase of adjacent elements increases linearly by $\frac{\pi}{3}l$, where $l$ is the index of the OAM mode with $0\le l\le5$. The excitations of the multiple OAM modes are then emitted via the transmit UCA and form the OAM beams\cite{oam_low_freq_radio}. After passing through the wireless channel, the received RF signals are sent to the signal recovery unit after unit discrete Fourier transform (DFT) in the OAM demodulator.

To make it easier to illustrate, we also establish a cylindrical coordinate system in Fig.~\ref{fig:system_model}. We set the coordinate origin at the transmit UCA center. The $z$-axis is perpendicular to the transmit UCA plane and points to the receive UCA. The polar axis is defined as the direction from the origin to the first transmit UCA element and is perpendicular to the $z$-axis. Therefore, the cylindrical coordinate of the $n_t$th transmit UCA element can be given as $\left(R_t,\frac{\pi}{3}n_t,0\right)$ with $n_t=0,1,\cdots,5$. Also, the distance from the transmit UCA to the receive UCA can be denoted by $z$ as shown in Fig.~\ref{fig:system_model}. %In addition, we set the cylindrical coordinate of receive UCA center as $(\rho,\phi,z)$.

\section{Fractal OAM Generation and Detection Scheme}\label{sec:FractalOAM_GenerationDetection}
In this section, we first briefly introduce the UCA-based fractal OAM phenomenon. Then, we propose the fractal OAM beam generation scheme by analyzing the minimum transmit UCA radius as well as the relationship of the transmit UCA radius, the transmission distance, the wavelength, the fractal OAM center coordinates, and the fractal OAM radius. Finally, we propose the fractal OAM beams detection scheme.

\subsection{UCA-based Fractal OAM Phenomenon}
\begin{figure}[htbp]
\centering
\vspace{-10pt}
\subfigure[OAM-mode 1 Power.]{
\begin{minipage}{0.45\linewidth}
\centering
\includegraphics[scale=0.3105]{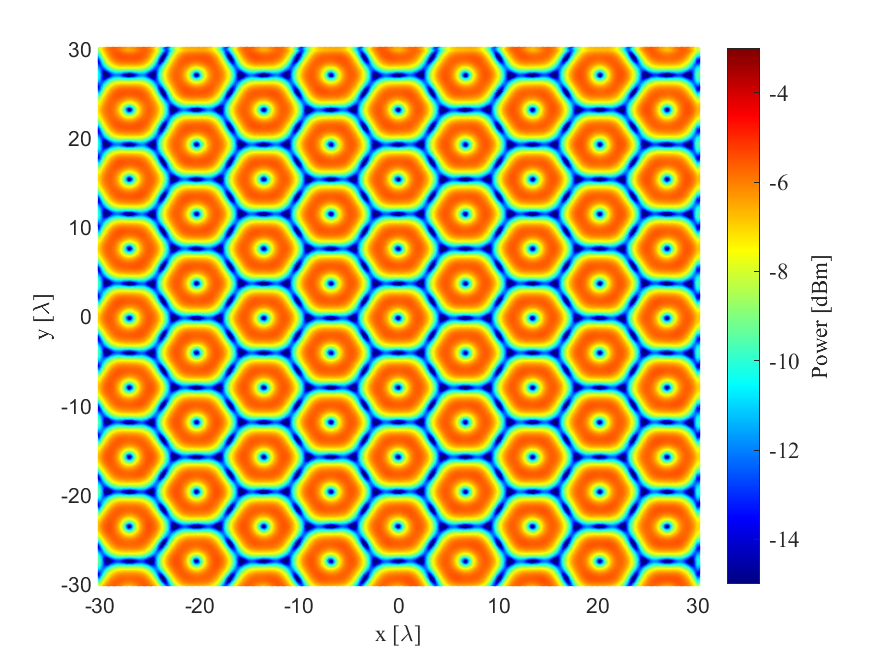}\label{fig:mode1Power}
%\vspace{-20pt}
\end{minipage}
}
\subfigure[OAM-mode 1 Phase.]{
\begin{minipage}{0.45\linewidth}
\centering
\includegraphics[scale=0.3105]{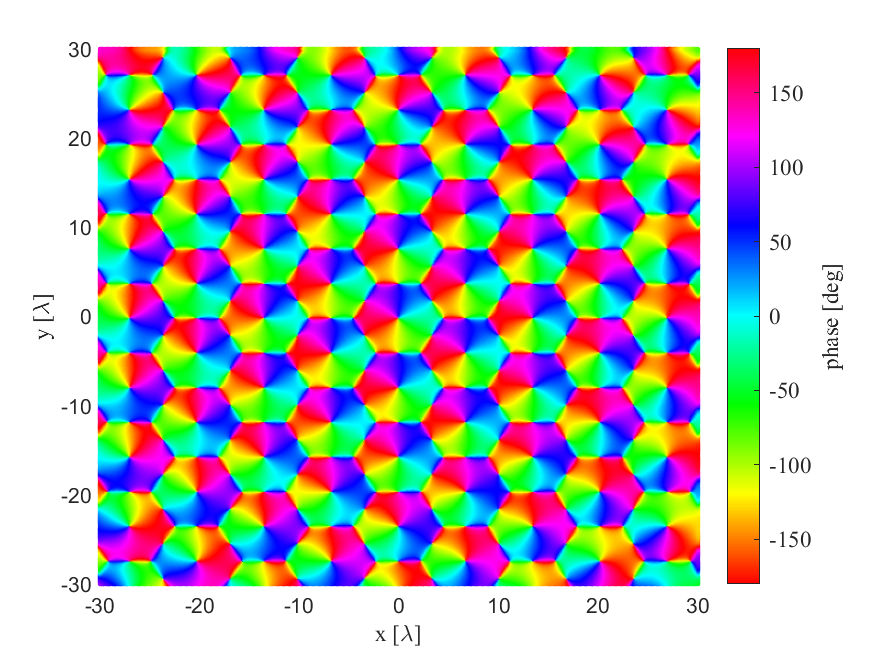}\label{fig:mode1Phase}
%\vspace{-20pt}
\end{minipage}
}\\
\vspace{-10pt}
\subfigure[OAM-mode 2 Power.]{
\begin{minipage}{0.45\linewidth}
\centering
\includegraphics[scale=0.3105]{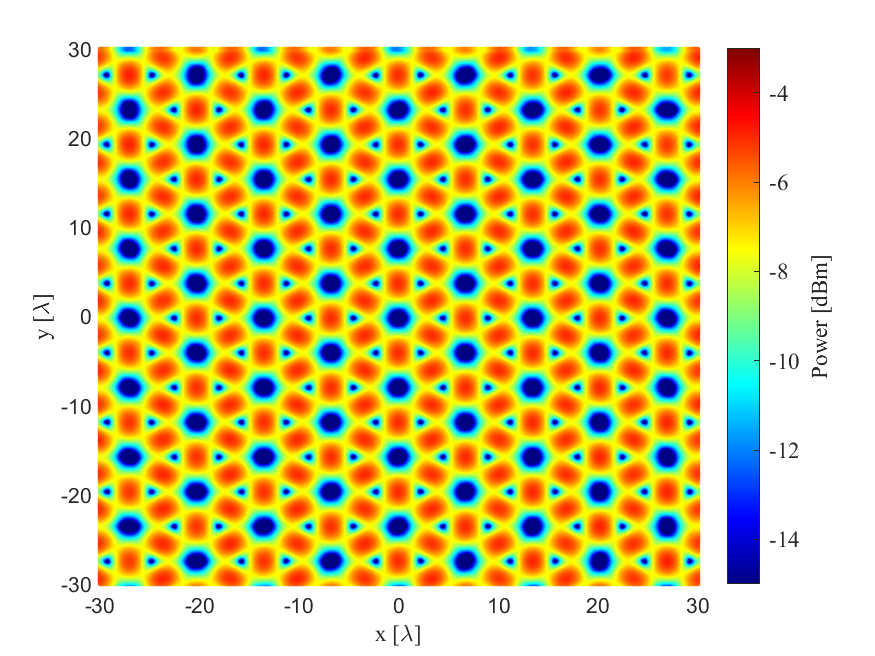}\label{fig:mode2Power}
%\vspace{-20pt}
\end{minipage}
}
\subfigure[OAM-mode 2 Phase.]{
\begin{minipage}{0.45\linewidth}
\centering
\includegraphics[scale=0.3105]{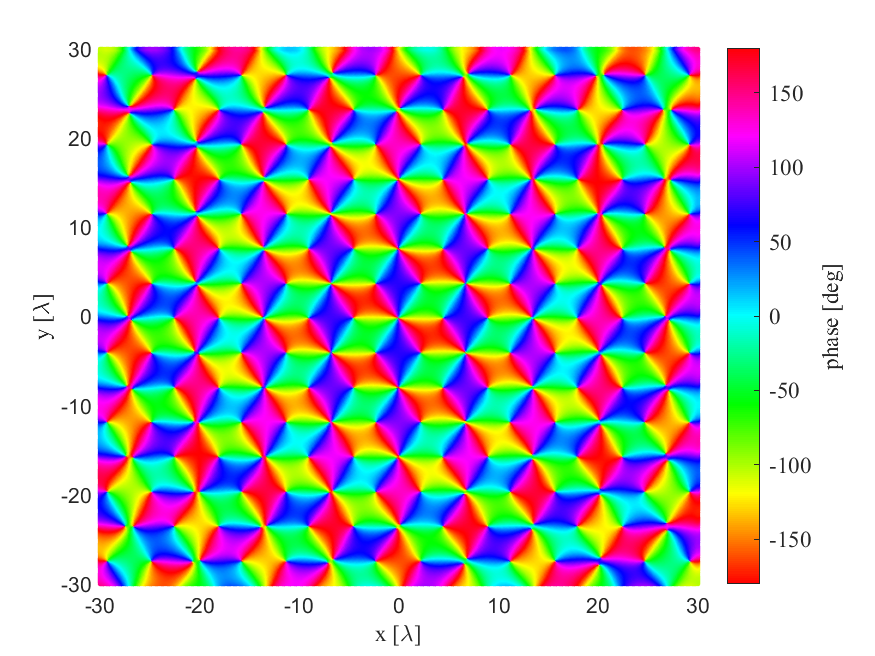}\label{fig:mode2Phase}
%\vspace{-20pt}
\end{minipage}
}
\centering
%\vspace{-5pt}
\caption{Simulated power and phase distribution generated by UCAs with $6$ elements.}\label{fig:TalbotEffect_OAM}
\vspace{-10pt}
\end{figure}
We discover that periodic OAM phase and amplitude structures can be generated in planes parallel to the transmit UCA by setting suitable radii for the transmit UCA. This is because, the electromagnetic waves generated by the UCA elements are self-imaged after free-space propagation and inference. It should be noted that the generation of fractal OAM beams by UCA are different from that based on quis-Talbot effect in \cite{Quasi-Talbot_OAM} and \cite{OAM_mobileTalbot}. The aforementioned UCA-based fractal OAM beam generation relies on a UCA composed of six conventional antennas. In contrast, the fractal OAM beam generations described in \cite{Quasi-Talbot_OAM} and \cite{OAM_mobileTalbot} are based on a UCA consisting of six OAM generators. Each of these OAM generators has the capability to independently generate OAM beams. Figure~\ref{fig:TalbotEffect_OAM} depicts the periodic OAM structures generated by a UCA with six elements on a plane that is $1000\lambda$ far away from the transmit UCA. The transmit UCA radius is set as $150\lambda$ and the OAM-modes are set as $1$ and $2$. The fractal OAM structures are in a hexagonal close-packing. The radius and phase structure of fractal OAM beams are nearly identical. Therefore, the receive UCA no longer needs to be perfectly aligned with the transmit UCA, but only needs to be aligned with a fractal OAM. Also, because if the receive UCA radius exceeds the fractal OAM radius, it may receive adjacent fractal OAM phase information, which can lead to potential errors in information extraction, the receive UCA radius should be smaller than the fractal OAM radius. In addition, since each fractal OAM beam has a much smaller radius compared to the normal OAM beams, enabling more energy to be received by a smaller receive antenna, the hollow divergence of OAM beams can be alleviated.

\begin{figure}[!h]
\centering
\setcounter {subfigure} {0}
\vspace{-10pt}
\subfigure[OAM-mode 1 Power of $8$-element UCA.]{
\begin{minipage}{0.45\linewidth}
\centering
\includegraphics[scale=0.3105]{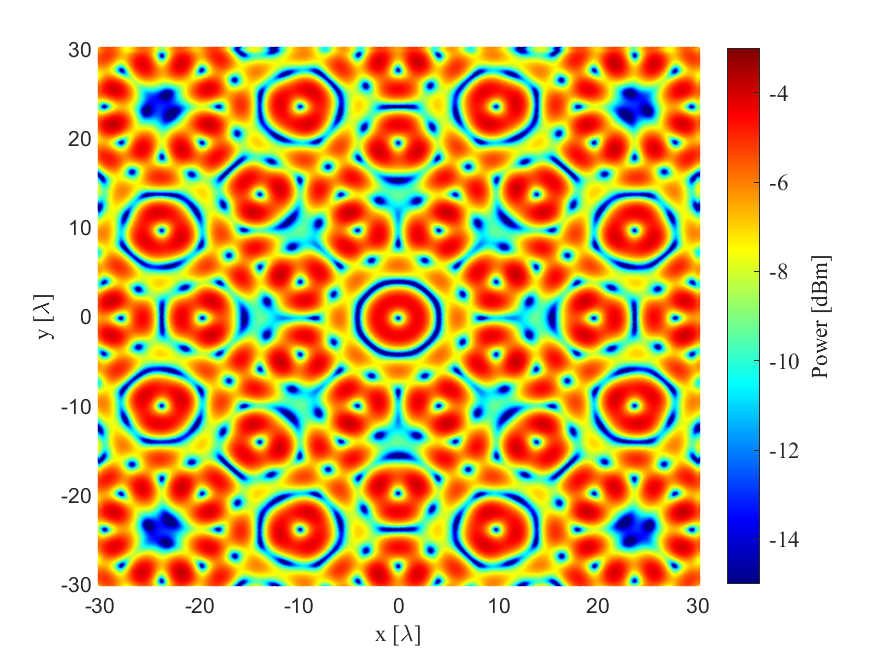}\label{fig:mode1Power_8}
%\vspace{-20pt}
\end{minipage}
}
\subfigure[OAM-mode 1 Phase of $8$-element UCA.]{
\begin{minipage}{0.45\linewidth}
\centering
\includegraphics[scale=0.3105]{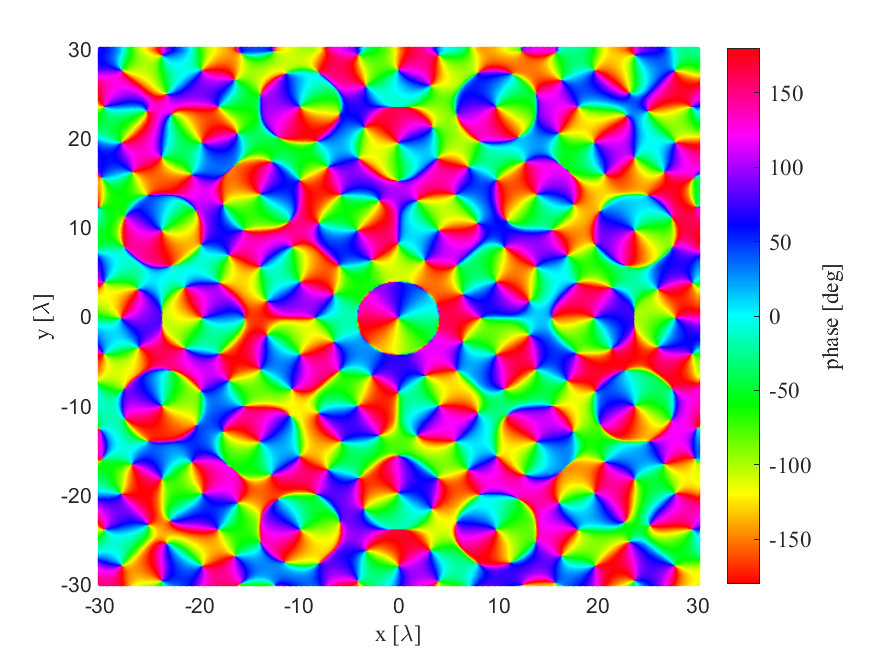}\label{fig:mode1Phase_8}
%\vspace{-20pt}
\end{minipage}
}\\
\vspace{-10pt}
\subfigure[OAM-mode 1 Power of $4$-element UCA.]{
\begin{minipage}{0.45\linewidth}
\centering
\includegraphics[scale=0.3105]{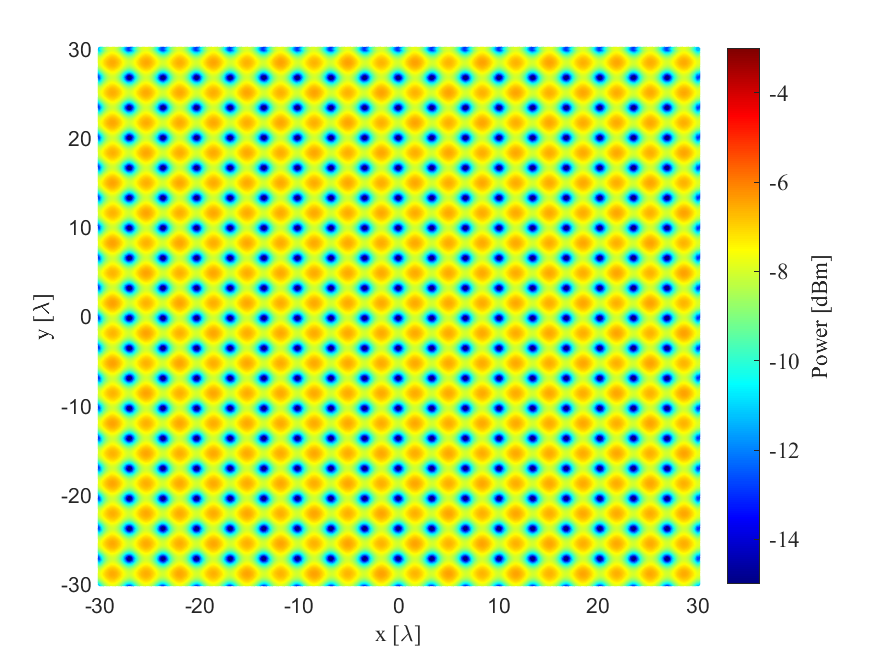}\label{fig:mode1Power_4}
%\vspace{-20pt}
\end{minipage}
}
\subfigure[OAM-mode 1 Phase of $4$-element UCA.]{
\begin{minipage}{0.45\linewidth}
\centering
\includegraphics[scale=0.3105]{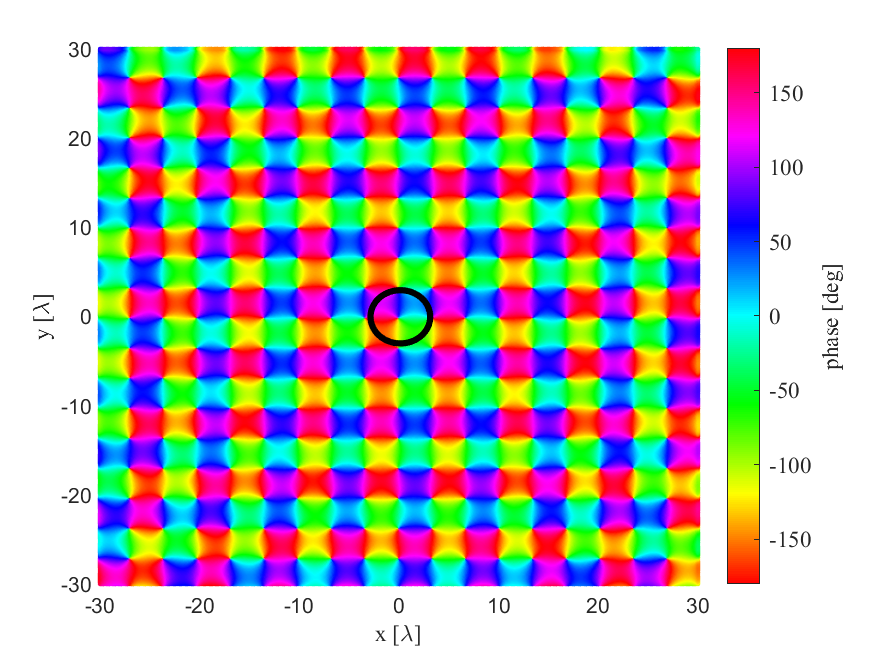}\label{fig:mode1Phase_4}
%\vspace{-20pt}
\end{minipage}
}
\centering
%\vspace{-5pt}
\caption{Simulated power and phase distribution beams generated by UCAs with $8$ and $4$ elements.}\label{fig:TalbotEffect_OAM_2}
\vspace{-10pt}
\end{figure}
However, if we change the number of UCA elements to $8$, closely packed OAM phase and amplitude structures cannot be generated. As shown in Figs.~\ref{fig:mode1Power_8} and \ref{fig:mode1Phase_8}, where we set the number of transmit UCA elements to $8$ and keep other variables the same as in Figs.~\ref{fig:mode1Power} to \ref{fig:mode2Phase}, an $8$-element UCA cannot generate closely packed fractal OAM beams. Instead, other types of phase structures will be present between different fractal OAM patterns. We also provide an example where the number of transmit UCA elements is set to $4$, and the power and phase results are shown in Figs.~\ref{fig:mode1Power_4} and \ref{fig:mode1Phase_4}, respectively. Due to the small number of antennas in the UCA, the OAM structures are not clearly visible in Fig.~\ref{fig:mode1Power_4}. However, within the black circle in Fig.~\ref{fig:mode1Phase_4}, a distinct phase structure of OAM-mode $1$ can be observed and different fractal OAM phase structures are closely packed. This phenomenon can be attributed to the problem of equilateral polygon tessellation. Only antennas arranged in equilateral triangles, equilateral quadrilaterals (squares), or equilateral hexagons can generate closely packed fractal OAM beams. Therefore, in the rest of this paper, we set the umber of UCA elements as $6$ to generate relatively complete OAM phase structures.

\subsection{Fractal OAM Beams Generation Scheme}\label{sec:fractalOAMGen}
To generate fractal OAM beams, we need to know the minimum transmit UCA radius. Also, to generate needed radii and density of fractal OAM beams at suitable distances, it is necessary to derive the relationship between the transmit UCA radius, the transmission distance, the wavelength, the fractal OAM center coordinates, and the fractal OAM radius. We first give the complex electric field strength generated by the transmit UCA at a receive point with its cylindrical coordinate given as $(\rho,\phi,z)$. The complex electric field strength is denoted by $E(\rho,\phi,z)$ and is given as follows:
\begin{align}
E(\rho,\phi,z) = \sum_{n_t = 0}^{5}\frac{\lambda}{4\pi d_{n_t}}\exp{\left(-j\frac{2\pi}{\lambda}d_{n_t}\right)},
\label{eq:E}
\end{align}
where $j$ denotes the imaginary symbol and $d_{n_t}$ denotes the distance from the $n_t$th transmit UCA element to the receive point. $d_{n_t}$ is given as follows:
\begin{align}
d_{n_t} \hspace{-0.2cm}=\hspace{-0.1cm} \sqrt{\left[\rho\cos\phi\hspace{-0.1cm}-\hspace{-0.1cm}R_t\cos\left(\frac{\pi}{3}n_t\right)\right]^2
                \hspace{-0.3cm}+\hspace{-0.1cm} \left[\rho\sin\phi\hspace{-0.1cm}-\hspace{-0.1cm}R_t\sin\left(\frac{\pi}{3}n_t\right)\right]^2 \hspace{-0.3cm}+\hspace{-0.1cm} z^2}.
\end{align}
$d_{n_t}$ can be approximated and simplified by binomial expansion as follows:
\begin{footnotesize}
\begin{align}
d_{n_t}
&\hspace{-0.1cm}=\hspace{-0.1cm}z\hspace{-0.1cm}\sum_{k=0}^{\infty}\hspace{-0.1cm}\tbinom{1/2}{k}\hspace{-0.1cm}\left[\frac{\left[\rho\cos\phi\hspace{-0.1cm}-\hspace{-0.1cm}R_t\cos\left(\frac{\pi}{3}n_t\right)\right]^2
                \hspace{-0.18cm}+\hspace{-0.1cm} \left[\rho\sin\phi\hspace{-0.1cm}-\hspace{-0.1cm}R_t\sin\left(\frac{\pi}{3}n_t\right)\right]^2}{z^2}\hspace{-0.1cm}\right]^k\nonumber\\
&\quad\hspace{-0.1cm}=\hspace{-0.1cm}z\bigg\{1 \hspace{-0.1cm}+\hspace{-0.1cm} \frac{1}{2}\frac{\left[\rho\cos\phi\hspace{-0.1cm}-\hspace{-0.1cm}R_t\cos\left(\frac{\pi}{3}n_t\right)\right]^2
                \hspace{-0.18cm}+\hspace{-0.1cm} \left[\rho\sin\phi\hspace{-0.1cm}-\hspace{-0.1cm}R_t\sin\left(\frac{\pi}{3}n_t\right)\right]^2}{z^2}
                \nonumber\\&\quad\quad\hspace{-0.1cm}-\hspace{-0.1cm} \frac{1}{8}\left[\frac{\left[\rho\cos\phi\hspace{-0.1cm}-\hspace{-0.1cm}R_t\cos\left(\frac{\pi}{3}n_t\right)\right]^2 \hspace{-0.18cm}+\hspace{-0.1cm} \left[\rho\sin\phi\hspace{-0.1cm}-\hspace{-0.1cm}R_t\sin\left(\frac{\pi}{3}n_t\right)\right]^2}{z^2}\hspace{-0.1cm}\right]^{2}
                \hspace{-0.28cm}+\hspace{-0.1cm} \cdots\hspace{-0.1cm}\bigg\}\nonumber\\
&\quad\quad\quad\hspace{-0.1cm}\approx\hspace{-0.1cm} z \hspace{-0.1cm}+\hspace{-0.1cm}
\frac{\left[\rho\cos\phi\hspace{-0.1cm}-\hspace{-0.1cm}R_t\cos\left(\frac{\pi}{3}n_t\right)\right]^2
                \hspace{-0.18cm}+\hspace{-0.1cm} \left[\rho\sin\phi\hspace{-0.1cm}-\hspace{-0.1cm}R_t\sin\left(\frac{\pi}{3}n_t\right)\right]^2}{2z}\nonumber\\
&\quad\quad\quad\quad\hspace{-0.1cm}=\hspace{-0.1cm} z \hspace{-0.1cm}+\hspace{-0.1cm} \frac{\rho^2 \hspace{-0.1cm}+\hspace{-0.1cm} R_t^2 \hspace{-0.1cm}-\hspace{-0.1cm} 2\rho R_t\cos\left(\phi\hspace{-0.1cm}-\hspace{-0.1cm}\frac{\pi}{3}n_t\right)}{2z},
\label{eq:d_nt_simp}
\end{align}
\end{footnotesize}
where $\tbinom{1/2}{k}$ denotes a combinatorics symbol given as $\tbinom{1/2}{k} = \frac{\frac{1}{2}(\frac{1}{2}-1)\cdots(\frac{1}{2}-k+1)}{k!}$. Therefore, for $z\gg\rho$ and $R_t$, Eq.~\eqref{eq:E} can be rewritten as follows:
\begin{align}
&E(\rho,\phi,z)\nonumber\\ &\quad\approx\hspace{-0.1cm} \frac{\lambda}{4\pi z}\hspace{-0.1cm}\sum_{n_t = 0}^{5}\hspace{-0.1cm}\exp{\left[-j\frac{2\pi}{\lambda}\hspace{-0.1cm}\left(\hspace{-0.1cm}z \hspace{-0.1cm}+\hspace{-0.1cm} \frac{\rho^2 \hspace{-0.1cm}+\hspace{-0.1cm} R_t^2 \hspace{-0.1cm}-\hspace{-0.1cm} 2\rho R_t\cos\left(\phi\hspace{-0.1cm}-\hspace{-0.1cm}\frac{\pi}{3}n_t\right)}{2z}\hspace{-0.1cm}\right)\right]}.
\label{eq:E2}
\end{align}

\begin{figure}[!h]
\centering
%\vspace{-10pt}
\subfigure[$R_t=0.5\lambda$.]{
\begin{minipage}{0.45\linewidth}
\centering
\includegraphics[scale=0.302]{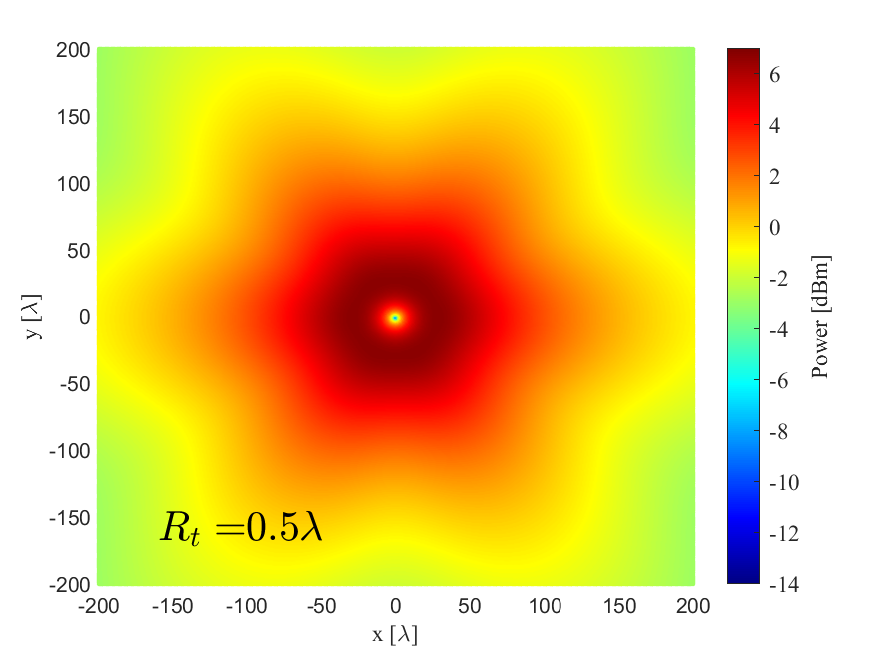}\label{fig:gratingLobes_halfLambda}
%\vspace{-20pt}
\end{minipage}
}
%\vspace{-10pt}
\subfigure[$R_t=\lambda$.]{
\begin{minipage}{0.45\linewidth}
\centering
\includegraphics[scale=0.302]{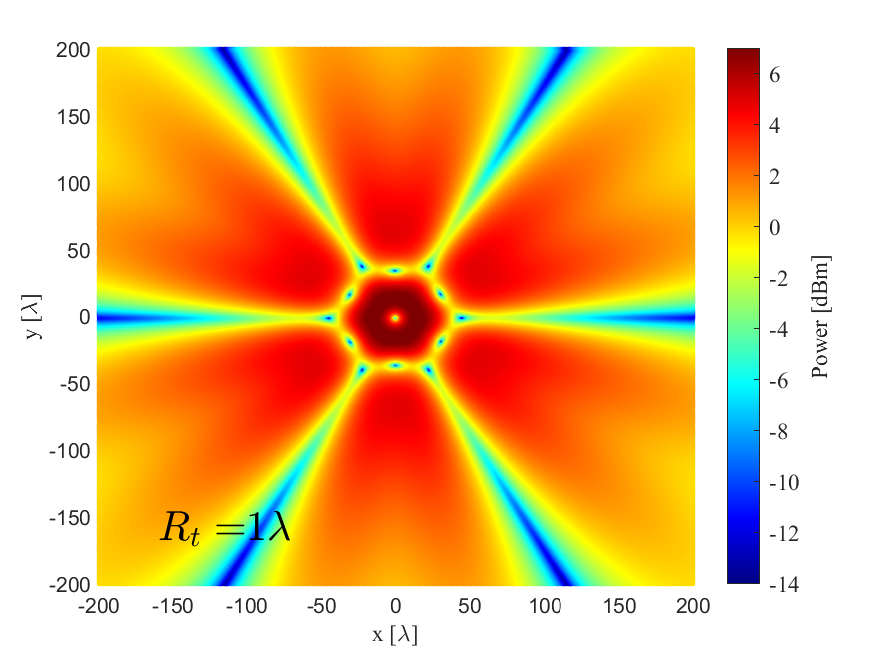}\label{fig:gratingLobes_1Lambda}
%\vspace{-20pt}
\end{minipage}
}\\
\subfigure[$R_t=1.5\lambda$.]{
\begin{minipage}{0.45\linewidth}
\centering
\includegraphics[scale=0.302]{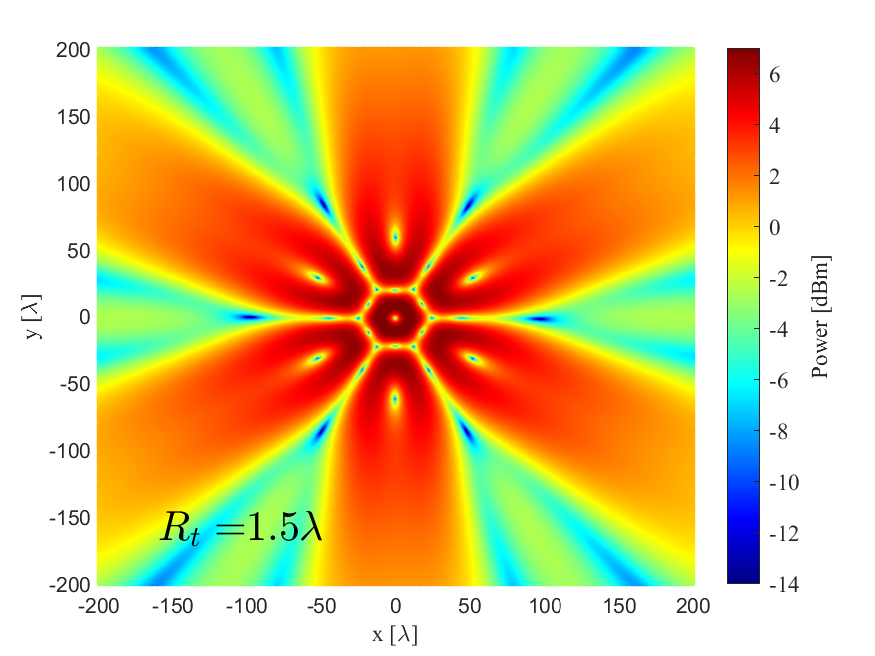}\label{fig:gratingLobes_1andHalfLambda}
%\vspace{-20pt}
\end{minipage}
}
%\vspace{-10pt}
\subfigure[$R_t=2\lambda$.]{
\begin{minipage}{0.45\linewidth}
\centering
\includegraphics[scale=0.302]{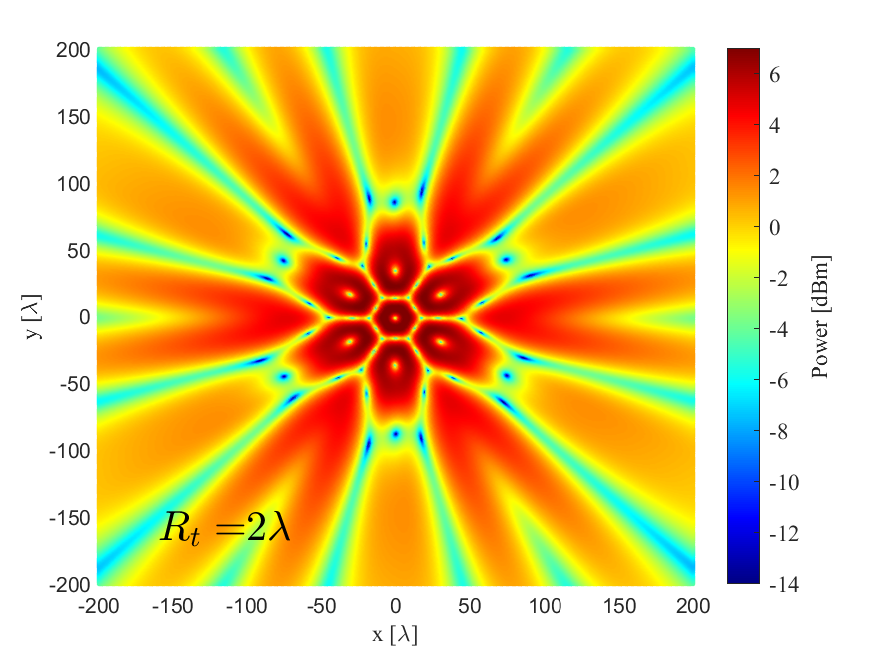}\label{fig:gratingLobes_2Lambda}
%\vspace{-20pt}
\end{minipage}
}\\
\subfigure[$R_t=2.5\lambda$.]{
\begin{minipage}{0.45\linewidth}
\centering
\includegraphics[scale=0.302]{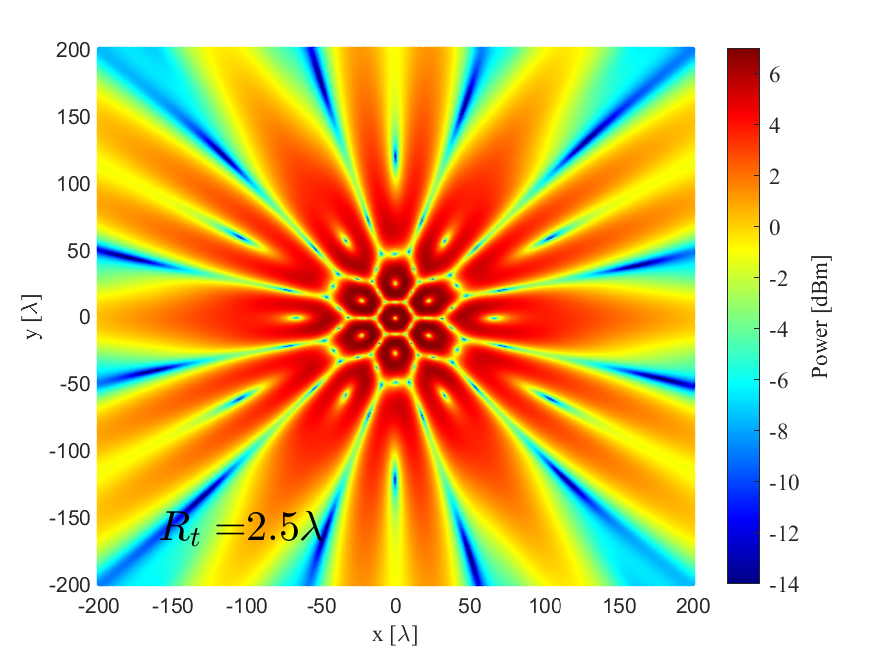}\label{fig:gratingLobes_2andHalfLambda}
%\vspace{-20pt}
\end{minipage}
}
%\vspace{-10pt}
\subfigure[$R_t=3\lambda$.]{
\begin{minipage}{0.45\linewidth}
\centering
\includegraphics[scale=0.302]{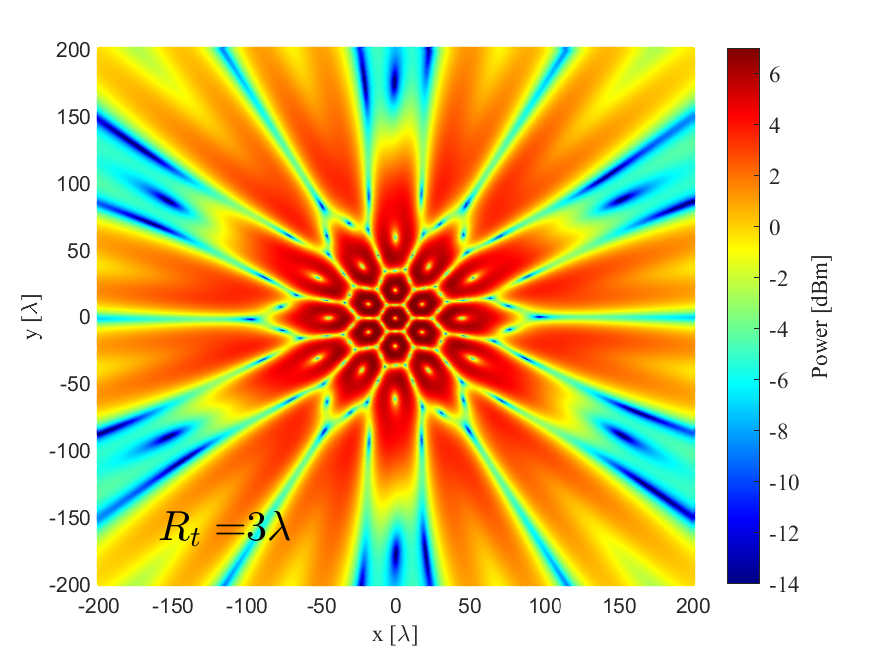}\label{fig:gratingLobes_3Lambda}
%\vspace{-20pt}
\end{minipage}
}
\centering
%\vspace{-10pt}
\caption{Received power of OAM beams with $R_t = 0.5,1,1.5,2,2.5,$ and $3\lambda$.}\label{fig:gratingLobes}
%\vspace{-10pt}
\end{figure}

Based on the complex electric field strength given in Eq.~\eqref{eq:E2}, we analyze the minimum transmit UCA radius to generate fractal OAM beams. Based on Eq.~\eqref{eq:E2}, we plot the received power for OAM beams of OAM-mode 1 with $R_t = 0.5,1,1.5,2,2.5,$ and $3\lambda$ in Fig.~\ref{fig:gratingLobes}. The receive plane is $50\lambda$ away from the transmit UCA. Figure~\ref{fig:gratingLobes_halfLambda} shows that there are no non-central fractal OAM beams for $R_t = 0.5\lambda$. With $R_t$ increases from $1$ to $3\lambda$, the number of fractal OAM beams increases and the shapes of the non-central OAM beams become more regular. By comparing Fig.~\ref{fig:gratingLobes_halfLambda} with Fig.~\ref{fig:gratingLobes_3Lambda}, a noticeable reduction in the central hollow was observed. The reduction of central hollow in the fractal OAM beams indicated an improvement in the concentration of energy, thereby mitigating the impact of hollow divergence of OAM beams. It should be noted that, we chose to set $R_t$ as integer multiples of half wavelength for the sake of using a minimal number of figures to present the increasing number of fractal OAM beams as $R_t$ increases. However, it should be noted that the same conclusion can also be derived even if $R_t$ is not an integer multiple of half wavelength. Actually, we can think of a UCA as a spatial sampling of a signal and the generation of fractal OAM beams is similar to aliasing that occurs in the time domain. As we know, the sampling rate must be at least twice the highest frequency of the signal to avoid spectral aliasing. Therefore, by analogy, we can conclude that to generate fractal OAM beams, the sampling rate should be larger than twice the highest frequency of the signal. Hence, to generate fractal OAM beams, the distance between each transmit UCA element should be larger than $\lambda/2$. Since the transmit UCA elements are located at the six vertices of a hexagon, the distance between adjacent elements is equal to $R_t$. Therefore $R_t$ should satisfy $R_t > \lambda/2$. However, in order to ensure the shape of the fractal OAM beams, $R_t$ should be set much larger than $\lambda/2$. In this paper, we set $R_t\ge3\lambda$ for all fractal OAM transmit UCA.

\begin{figure}[!h]
\centering
%\vspace{-15pt}
\subfigure[The fractal OAM hexagonal grid schematic diagram.]{
\begin{minipage}{1\linewidth}
\centering
\includegraphics[scale=0.82]{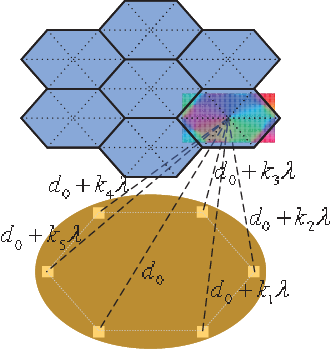}\label{fig:OAM_Grid}
%\vspace{-20pt}
\end{minipage}
}\\
%\vspace{-10pt}
\subfigure[Superposition of fractal OAM hexagonal grid and simulated power.]{
\begin{minipage}{1\linewidth}
\centering
\includegraphics[scale=0.45]{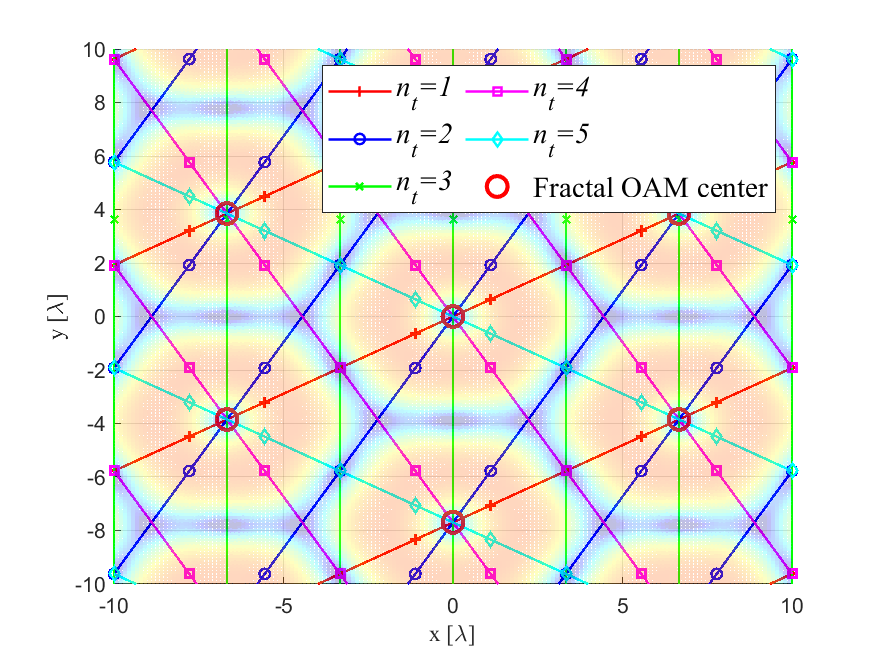}\label{fig:OAM_Grid_Power_mode1}
%\vspace{-20pt}
\end{minipage}
}
\centering
%\vspace{-10pt}
\caption{The fractal OAM hexagonal grid.}
%\vspace{-15pt}
\end{figure}

Then, we analyze the relationship between the transmit UCA radius, the transmission distance, the wavelength, the fractal OAM center coordinates, and the fractal OAM radius. As shown in Fig.~\ref{fig:TalbotEffect_OAM}, each fractal OAM has a center point where the phase superposition is $0$, which means the differences between the distances from the fractal OAM center to six transmit UCA elements are integer multiples of the wavelength. Without loss of generality, the distance between the fractal OAM center and the second to the sixth transmit UCA element should be different from the distance between the fractal OAM center and the first transmit UCA element by an integer multiple of the wavelength. This makes the center points appear in a hexagonal arrangement. Figure~\ref{fig:OAM_Grid} shows the hexagonal arrangement of fractal OAM centers and the distances between center points and transmit UCA elements. The adjacent six center points also limit the radius of the fractal OAM structure in the center. Therefore, the radius of each fractal OAM is $\sqrt{3}/3$ times that of the distance between two fractal OAM centers. This relationship between the radius and the distance between the center points of fractal OAM structures is important for accurately generating and detecting the fractal OAM beams.

Next, we give the formula derivations for the above analysis. Base on Eq.~\eqref{eq:d_nt_simp} and Fig.~\ref{fig:OAM_Grid}, the difference between $d_0$ and $d_{n_t}$ with $n_t = 1,2,\cdots,5$ can be given as follows:
\begin{align}
d_0 - d_{n_t} &= \frac{\rho R_t}{z}\left[\cos\left(\phi-\frac{\pi}{3}n_t\right) - \cos\phi\right]\nonumber\\
&\quad\quad= \frac{2\rho R_t}{z}\sin\left(\phi-\frac{\pi}{6}n_t\right)\sin\left(\frac{\pi}{6}n_t\right).
\label{eq:d_n_difference}
\end{align}
Let $d_0 - d_{n_t} = k_{n_t}\lambda$, where $k_{n_t}$ is an integer, and let
$y=\rho\sin\left(\phi\right)$, $x=\rho\cos\left(\phi\right)$, and $z=z$, Eq.~\eqref{eq:d_n_difference} can be rewritten as follows:
\begin{align}
%d_0 - d_{n_t} &= k_{n_t}\lambda\nonumber\\
%\sin\left(\frac{\pi}{6}n_t\right)\rho\sin\left(\phi-\frac{\pi}{6}n_t\right)&=\frac{k_{n_t}\lambda z}{2R_t}\nonumber\\
%\left[\sin\left(\frac{\pi}{6}n_t\right)\cos\left(\frac{\pi}{6}n_t\right)\right]y-\left[\sin^2\left(\frac{\pi}{6}n_t\right)\right]x&=\frac{k_{n_t}\lambda z}{2R_t},
\left[\sin^2\left(\frac{\pi}{6}n_t\right)\right]x-\left[\sin\left(\frac{\pi}{6}n_t\right)\cos\left(\frac{\pi}{6}n_t\right)\right]y+\frac{k_{n_t}\lambda z}{2R_t}&=0,
\label{eq:OAM_grid_lines}
\end{align}
which describes five sets of parallel lines in a Cartesian coordinate system. The Cartesian coordinate system has its $x$ axis overlapped with the polar axis of the cylindrical coordinate system given in Section~\ref{sec:systemModel} and shares the same $z$-axis with the cylindrical coordinate system. Therefore, the coordinates for fractal OAM centers can be obtained by calculating the intersection points of the five sets of parallel lines. Solve Eq.~\eqref{eq:OAM_grid_lines} and we get the following solutions:
\begin{align}
(x,y)=
\begin{cases}
\left(m\frac{2\lambda z}{R_t},n\frac{\sqrt{3}}{3}\frac{\lambda z}{R_t}\right),\ &{\rm for}\ n\ {\rm even};\\ \\
\left(\left(\frac{1}{2}+m\right)\frac{2\lambda z}{R_t},n\frac{\sqrt{3}}{3}\frac{\lambda z}{R_t}\right),\ &{\rm for}\ n\ {\rm odd},
\end{cases}
\label{eq:OAM_grid_CarCorrdinates}
\end{align}
where $m$ is an integer denoting the index of rows and $n$ is also an integer denoting the index of columns for the fractal OAM beams in a hexagonal arrangement. Based on Eq.~\eqref{eq:OAM_grid_CarCorrdinates}, we further derive the radius of each fractal OAM. %\textcolor{blue}{In Eq.~(8), let $x=0$ to calculate the coordinates for the fractal OAM centers on the $y$ axis and results are given as follows:}
To determine the coordinates for the fractal OAM centers on the y-axis, we set $x=0$. Hence, we obtain
\begin{align}
y=n\frac{\sqrt{3}}{3}\frac{\lambda z}{R_t},
\label{eq:centerOny}
\end{align}
where $n$ is an even number. Thus, without loss of generality, the radius of each fractal OAM, which is also the side length of each hexagon, denoted by $r$, can be given as $\sqrt{3}/3$ the length of the distance between two adjacent fractal OAM centers on the $y$ axis as follows:
\begin{align}
r=\frac{\sqrt{3}}{3}\left[(n+2)\frac{\sqrt{3}}{3}\frac{\lambda z}{R_t} - n\frac{\sqrt{3}}{3}\frac{\lambda z}{R_t}\right]=\frac{2}{3}\frac{\lambda z}{R_t}.
\label{eq:OAM_grid_radius}
\end{align}

Based on Eqs.~\eqref{eq:E2}, \eqref{eq:OAM_grid_lines}, and \eqref{eq:OAM_grid_CarCorrdinates}, the overlapping image of the five sets of parallel lines and the fractal OAM amplitudes are drawn in Fig.~\ref{fig:OAM_Grid_Power_mode1}. It can be found in Fig.~\ref{fig:OAM_Grid_Power_mode1} that the intersection points of the five sets of parallel lines are also the centers of fractal OAM beams, validating the above analysis and derivations.

Finally, we give the fractal OAM beam generation scheme based on the above analysis and derivations. The fractal OAM beam generation scheme can be divided into three steps. In the first step, according to the wavelength, the transmission distance, and the required radius of each fractal OAM, the radius of the transmit UCA can be given based on Eq.~\eqref{eq:OAM_grid_radius} as follows:
\begin{align}
R_t=\frac{2}{3}\frac{\lambda z}{r}.
\end{align}
In the second step, the multi-stream information signals, denoted by $\boldsymbol{\mathrm x}\in \mathbb{C}^{6\times 1}$, are modulated into the excitations corresponding to six transmit UCA elements, denoted by $\boldsymbol{\mathrm s}\in \mathbb{C}^{6\times 1}$, using unit IDFT as follows:
\begin{align}
\boldsymbol{\mathrm s} = \boldsymbol{\mathrm W}\boldsymbol{\mathrm P}_t\boldsymbol{\mathrm x},
\end{align}
where $\boldsymbol{\mathrm P}_t\hspace{-0.1cm}=\hspace{-0.1cm}{\rm diag}\left\{\left[\sqrt{P_0}, \sqrt{P_1}, \cdots, \sqrt{P_l}, \cdots,\sqrt{P_{N_t-1}}\right]\right\}$ is the power-allocating matrix, $P_l$ denotes the power allocated to the $l$th OAM-mode, and $\boldsymbol{\mathrm W}\in \mathbb{C}^{6\times 6}$ is the unit IDFT matrix with the $n_1$th row and the $n_2$th column element given by ${W}_{n_1,n_2}\hspace{-0.1cm}=\hspace{-0.1cm}\frac{1}{\sqrt 6}{\rm exp}[j{\pi n_1n_2}/{3}]$. Here, we have $0\le n_1,n_2\le6$. In the third step, the excitations are sent to the transmit UCA and generate the fractal OAM beams in the free space.

\subsection{Fractal OAM Beams Detection Scheme}
Based on analysis and derivations in Section~\ref{sec:fractalOAMGen}, we propose the fractal OAM beams detection scheme in this part. The detection scheme can be divided into four steps. The first step is to determine the center coordinate and the radius of the receive UCA. To detect the fractal OAM signals, the receive UCA needs to be aligned with a fractal OAM and have a radius smaller than the fractal OAM radius. Based on Eqs.~\eqref{eq:OAM_grid_CarCorrdinates} and \eqref{eq:OAM_grid_radius}, we can obtain the center coordinates and the radius for the fractal OAM beams, respectively. Therefore, after determining the transmission distance, the Cartesian coordinates of receive UCA centers, denoted by $(x',y',z)$, can be given as a particular solution of following general solutions:
\begin{align}
(x',y',z) =
\begin{cases}
\left(m\frac{2\lambda z}{R_t},n\frac{\sqrt{3}}{3}\frac{\lambda z}{R_t},z\right),\ &{\rm for}\ n\ {\rm even};\\ \\
\left(\left(\frac{1}{2}+m\right)\frac{2\lambda z}{R_t},n\frac{\sqrt{3}}{3}\frac{\lambda z}{R_t},z\right),\ &{\rm for}\ n\ {\rm odd}.
\end{cases}
\label{eq:RrCenterCoordinate}
\end{align}
The receive UCA radius is given as follows:
\begin{align}
0<R_r\le\frac{2\sqrt{3}}{9}\frac{\lambda z}{R_t},
\label{eq:RrRadius}
\end{align}
which can be further given as follows:
\begin{align}
0<R_r\le\frac{\sqrt{3}}{3}r,
\end{align}

Therefore, the Cartesian coordinates for the receive UCA elements, denoted by $(x_{n_r},y_{n_r},z_{n_r})$ with $n_r = 0,1,\cdots,N_r-1$, can be given as follows:
\begin{subequations}\label{eq:R_r_coordinate}
\begin{numcases}{}
x_{n_r}=x'+R_r\cos\left(\frac{2\pi n_r}{N_r}\right);\\
y_{n_r}=y'+R_r\sin\left(\frac{2\pi n_r}{N_r}\right);\\
z_{n_r}=z.
\end{numcases}
\end{subequations}
The corresponding cylindrical coordinates for the receive UCA elements, denoted by $(\rho'_{n_r},\phi'_{n_r},z_{n_r})$ , can be given as follows:
\begin{subequations}\label{eq:R_r_coordinate_polar}
\begin{numcases}{}
\rho'_{n_r}\hspace{-0.1cm}=\hspace{-0.1cm}\sqrt{x_{n_r}^2+y_{n_r}^2}\nonumber\\
\hspace{0.5cm}\quad\hspace{-0.1cm}=\hspace{-0.1cm}\sqrt{x'^2\hspace{-0.15cm}+\hspace{-0.1cm}y'^2\hspace{-0.15cm}+\hspace{-0.1cm}R_r^2\hspace{-0.15cm}+\hspace{-0.1cm}2R_r\hspace{-0.1cm}\left[x\cos\hspace{-0.1cm}\left(\frac{2\pi n_r}{N_r}\right)\hspace{-0.1cm}+\hspace{-0.1cm}y\sin\hspace{-0.1cm}\left(\hspace{-0.1cm}\frac{2\pi n_r}{N_r}\hspace{-0.1cm}\right)\hspace{-0.1cm}\right]};\nonumber\\ \\
\phi'_{n_r}\hspace{-0.1cm}=\arcsin\hspace{-0.1cm}\left(\frac{y_{n_r}}{\rho'_{n_r}}\right)\hspace{-0.1cm}=\arcsin\left(\hspace{-0.1cm}\frac{y'\hspace{-0.1cm}+\hspace{-0.1cm}R_r\sin\left(\hspace{-0.1cm}\frac{2\pi n_r}{N_r}\hspace{-0.1cm}\right)}{\rho'_{n_r}}\hspace{-0.1cm}\right);\\
z_{n_r}\hspace{-0.1cm}=z.
\end{numcases}
\end{subequations}

In the second step, we calculate the free-space channel. Substitute Eq.~\eqref{eq:R_r_coordinate_polar} into Eq.~\eqref{eq:E2}, the free-space channel from the $n_t$th transmit UCA element to the $n_r$th receive UCA element, denoted by $H_{n_r,n_t}$, can be given as follows:
\begin{align}
&H_{n_r,n_t}\nonumber\\ &\quad=\hspace{-0.1cm} \frac{\lambda}{4\pi z}\hspace{-0.08cm}\exp{\hspace{-0.08cm}\left[\hspace{-0.1cm}-j\frac{2\pi}{\lambda}\hspace{-0.1cm}\left(\hspace{-0.1cm}z \hspace{-0.1cm}+\hspace{-0.1cm} \frac{(\rho'_{n_r})^2 \hspace{-0.2cm}+\hspace{-0.1cm} R_t^2 \hspace{-0.1cm}-\hspace{-0.1cm} 2\rho'_{n_r} R_t\cos\left(\phi'_{n_r}\hspace{-0.2cm}-\hspace{-0.1cm}\frac{\pi}{3}n_t\right)}{2z}\hspace{-0.1cm}\right)\hspace{-0.1cm}\right]}.
\label{eq:H_free}
\end{align}
\hspace{-0.1cm}For $n_t = 0$ to $5$ and $n_r = 0$ to $N_r$, the free-space channel from the $n_t$th transmit UCA element to the $n_r$th receive UCA element can be given in matrix form, denoted by $\boldsymbol{\mathrm H}\in \mathbb{C}^{N_r\times 6}$, as follows:
\begin{align}
\boldsymbol{\mathrm H}=
\left[ {\begin{array}{*{20}{c}}
H_{0,0}&\cdots&H_{0,5}\\
\vdots &\ddots &\vdots \\
H_{N_r,0}&\cdots&H_{N_r,5}\\
\end{array}}\right].
\end{align}
The receive OAM signals, denoted by $\boldsymbol{\mathrm r}\in \mathbb{C}^{N_r\times 1}$, can be then given as follows:
\begin{align}
\boldsymbol{\mathrm r} &= \boldsymbol{\mathrm H}\boldsymbol{\mathrm s} + \boldsymbol{\mathrm n}
= \boldsymbol{\mathrm H}\boldsymbol{\mathrm W}\boldsymbol{\mathrm P}_t\boldsymbol{\mathrm x} + \boldsymbol{\mathrm n},
\end{align}
where $\boldsymbol{\mathrm n}\in \mathbb{C}^{N_r\times 1}$ denotes the narrow-band additive white Gaussian noise (AWGN) with independent and identically distributed elements following $\mathcal{CN}(0,\sigma_n^2)$.

In the thrid step, the receive OAM signals are demodulated into the signals corresponding to different OAM-modes, denoted by $\boldsymbol{\mathrm y}$, using unit DFT as follows:
\begin{align}
\boldsymbol{\mathrm y} &= \boldsymbol{\mathrm W}'\boldsymbol{\mathrm r}
= \boldsymbol{\mathrm W}'\boldsymbol{\mathrm H}\boldsymbol{\mathrm W}\boldsymbol{\mathrm P}_t\boldsymbol{\mathrm x} + \boldsymbol{\mathrm W}'\boldsymbol{\mathrm n},
 \label{eq:y0}
\end{align}
where $\boldsymbol{\mathrm W}'\in \mathbb{C}^{N_r\times N_r}$ is the unit DFT matrix with the $m_1$th row and the $m_2$th column element given by ${W}'_{m_1,m_2}\hspace{-0.1cm}=\hspace{-0.1cm}\frac{1}{\sqrt N_r}{\rm exp}[-j{2\pi m_1m_2}/{N_r}]$. Here, we have $0\le m_1,m_2\le N_t$.

In the forth step, the signals corresponding to different OAM-modes are sent
to the signal recovery. For simplifying the following description, let $\boldsymbol{\mathrm H}'=\boldsymbol{\mathrm W}'\boldsymbol{\mathrm H}\boldsymbol{\mathrm W}$ denotes the OAM channel matrix with its $l_1$th row and the $l_2$th column element given by $H'_{l_1,l_2}$ denoting the complex channel from the $l_2$ transmitted OAM-mode to the $l_1$ received OAM-mode. Therefore, the recovered input signal, denoted by $\tilde{\boldsymbol{\mathrm x}}$, can be given as follows:
\begin{align}
&\tilde{\boldsymbol{\mathrm x}} = \mathop{\arg\min}\limits_{\boldsymbol{\mathrm x}\in \mathbb{C}_{6}}\left\Arrowvert{{\rm diag}\left(\boldsymbol{\mathrm h}'\right)^{\dag}\boldsymbol{\mathrm y}[0:5] - \boldsymbol{\mathrm P}_t\boldsymbol{\mathrm x}}\right\Arrowvert^2,
\label{eq:x_re}
\end{align}
where $\boldsymbol{\mathrm h}'$ denotes the vector formed by the diagonal elements of $\boldsymbol{\mathrm H}'$ and $\boldsymbol{\mathrm y}[0:5]$ denotes the first $6$ elements of $\boldsymbol{\mathrm y}$.

\subsection{Realization Challenges and Potential Solutions}
The above theoretical derivation gives our proposed fractal OAM beams generation and detection schemes. However, there are still some challenges in the practical implementation of our schemes. Here, we provide some further insights into the challenges and potential solutions for realizing and demonstrating our proposed fractal OAM generation and detection schemes.

\textbf{Realization Challenges:} One challenge with our proposed UCA-based fractal OAM communication is that the size of the transmit UCA is larger compared to a traditional antenna array with element spacing of half wavelength. This can present difficulties in implementing the hardware, especially in certain frequency bands. Another issue is that the transmit UCA and receive UCA are not reciprocal, which can create challenges in achieving duplex communication. The third challenge is that, while our proposed UCA-based fractal OAM communication offers the advantage of improved flexibility in the receive UCA arrangement, as it no longer requires perfect alignment with the transmit UCA, there is still a need to align the receive UCA with a specific fractal OAM pattern.

\textbf{Potential Solutions:} To address the first challenge of the larger size of the transmitting UCA, one approach is to utilize higher frequency band signals. Higher frequency signals allow for smaller antenna elements and can help reduce the overall size of the transmit UCA. Additionally, the use of metasurfaces can be explored to reproduce a desired aperture field, enabling radiation through the leaky wave effect\cite{Metasurface_Antennas_New,Modulated-Metasurface_Antennas}. Regarding the second challenge of non-reciprocity, one potential solution is to employ the fractal OAM communication as a downlink transmission scheme. The transmit end, equipped with the fractal OAM generation capability, can function as the base station. Different users can be served by utilizing different orthogonal OAM-modes, ensuring efficient and reliable communication from the base station to the users. By focusing on the downlink direction, the challenges associated with non-reciprocity can be mitigated. For the third challenge related to alignment requirements of the receive UCA, a transceiver beam steering scheme can be employed\cite{oam_beam_steering}. This technique allows for dynamic adjustment of the receive UCA's beam direction to align it with a specific fractal OAM pattern while still taking advantage of the improved flexibility in receive UCA arrangement.

\subsection{Comparison of Our Proposed Fractal OAM with Other Talbot-effect-based Fractal OAM}
\begin{figure}[htbp]
\centering
\setcounter {subfigure} {0}
%\vspace{-10pt}
\includegraphics[scale=0.6]{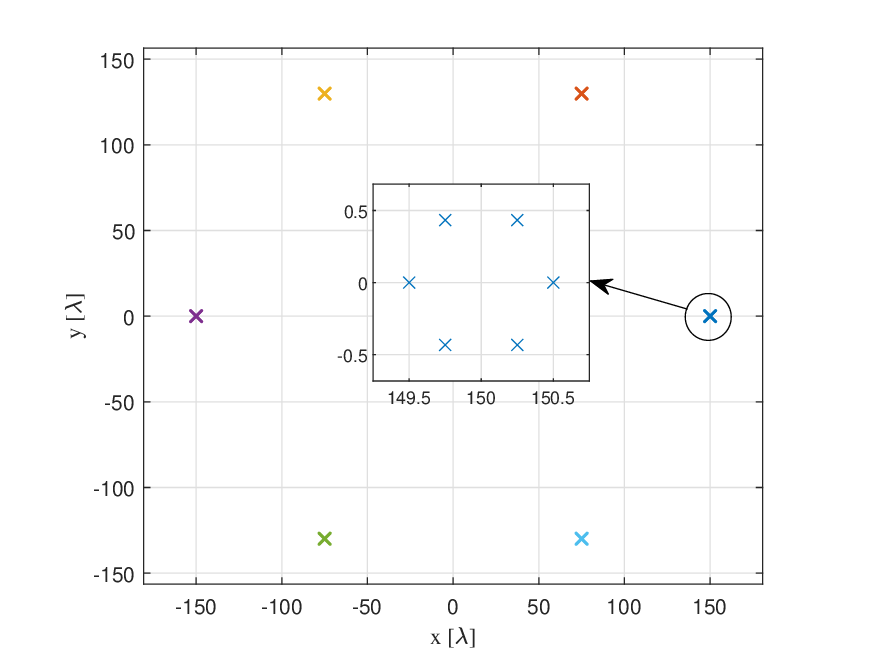}
\vspace{-10pt}
\caption{The two-layer transmit UCA of the baseline scheme.}\label{fig:DUCA}
%\vspace{-15pt}
\end{figure}
It should be noted again that, the fractal OAM beam generations described in \cite{Quasi-Talbot_OAM} and \cite{OAM_mobileTalbot} are based on a UCA consisting of six OAM generators. Each of these OAM generators has the capability to independently generate OAM beams. While in this paper, our proposed fractal OAM beam generation scheme is based on conventional antenna array. To provide a more comprehensive comparison of our proposed fractal OAM beam generation scheme with the fractal OAM beam generations described in \cite{Quasi-Talbot_OAM} and \cite{OAM_mobileTalbot}, we establish a fractal OAM generation scheme based on similar schemes described in references \cite{Quasi-Talbot_OAM} and \cite{OAM_mobileTalbot} as a baseline for comparison. In Fig.~\ref{fig:DUCA}, we design a two-layer transmit UCA, where the outer UCA consists of six $6$-element inner UCAs. The radius of the outer UCA is set to $150\lambda$, and the radius of the inner UCA is set to $0.5\lambda$. We maintained the transmission distance at $1000\lambda$ to ensure that the parameters match those in Fig.~\ref{fig:TalbotEffect_OAM}. Furthermore, we adjusted the transmit power of each element to be one-sixth of the transmit power of each element in Fig.~\ref{fig:TalbotEffect_OAM}, thus controlling the total transmit power to be the same.

\begin{figure}[htbp]
\centering
%\setcounter {subfigure} {0}
%%\vspace{-15pt}
\subfigure[OAM-mode 1 Power of the baseline.]{
\begin{minipage}{0.45\linewidth}
\centering
\includegraphics[scale=0.3105]{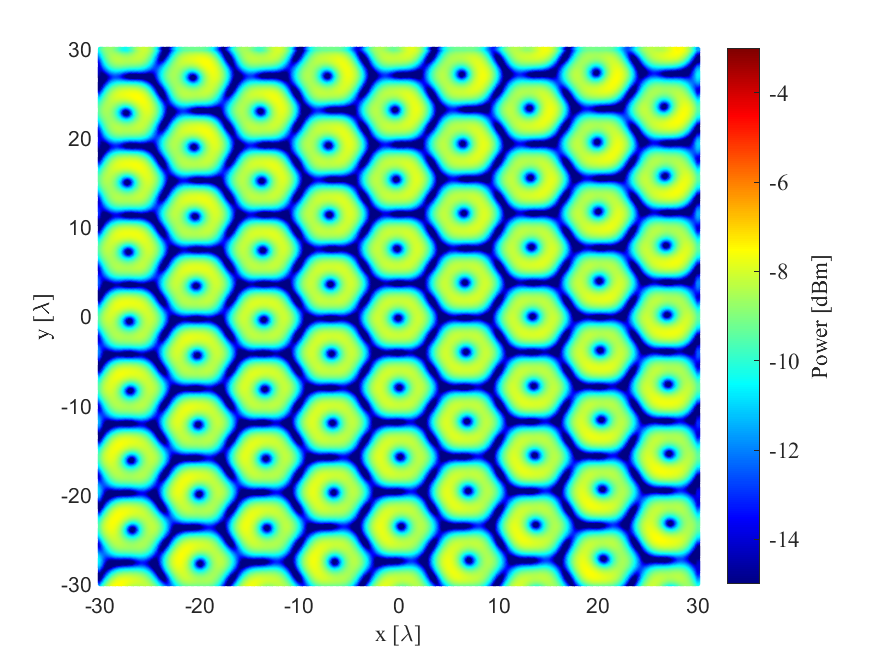}\label{fig:mode1Power_DUCA}
%\vspace{-20pt}
\end{minipage}
}
\subfigure[OAM-mode 1 Phase of the baseline.]{
\begin{minipage}{0.45\linewidth}
\centering
\includegraphics[scale=0.3105]{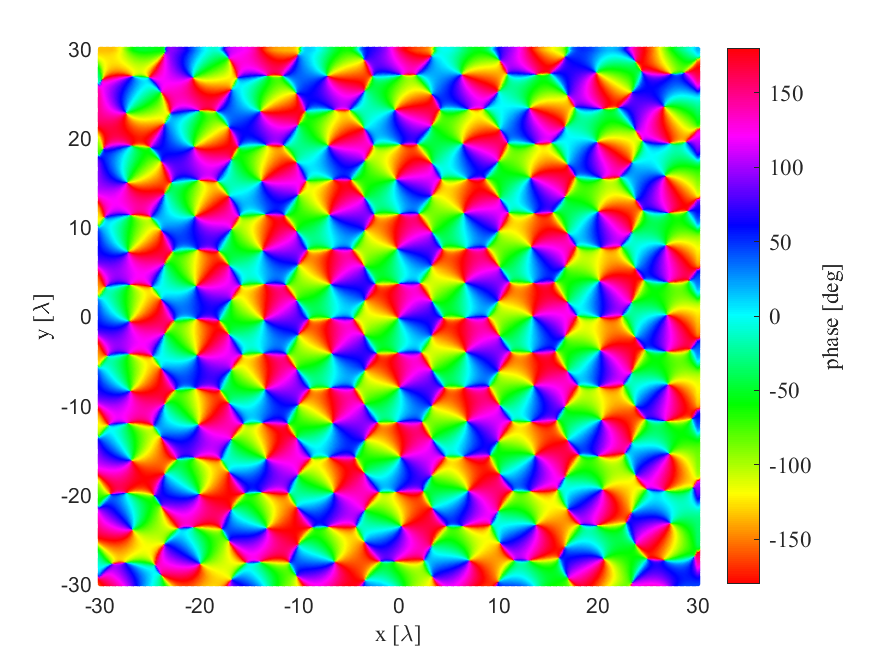}\label{fig:mode1Phase_DUCA}
%\vspace{-20pt}
\end{minipage}
}
\centering
%\vspace{-10pt}
\caption{Simulated power and phase distribution of the fractal OAM beams generated by the two-layer transmit UCA.}\label{fig:UCA_DUCA_PowerPhase}
%\vspace{-10pt}
\end{figure}
In Figs.~\ref{fig:mode1Power_DUCA} and \ref{fig:mode1Phase_DUCA}, we displayed the periodic OAM structures generated by the two-layer transmit UCA as shown in Fig.~\ref{fig:DUCA}. Comparing Figs.~\ref{fig:mode1Power_DUCA} and \ref{fig:mode1Phase_DUCA} with Figs.~\ref{fig:mode1Power} and \ref{fig:mode1Phase}, we observe that although the two-layer transmit UCA can also generate periodic OAM structures, the received power of our proposed fractal transmission is significantly higher than that of the baseline scheme. These results demonstrate that our proposed fractal OAM transmission scheme exhibits superior performance in terms of received power compared to the baseline scheme inspired by references \cite{Quasi-Talbot_OAM} and \cite{OAM_mobileTalbot}.

\section{Capacity and BER Analysis}\label{sec:performanceAnalysis}
%In this section, we first derive the channel capacity and BER of our proposed fractal OAM generation and detection scheme. Then, {\color{red}we simplify the capacity and BER} and compare the capacity and BER of our proposed fractal OAM transmission with normal OAM transmission. Also, we analyze how the receive UCA radius and the distance between the UCAs impact the capacity and BER performances. Numerical results are given for for better analysis.
In this section, we first derive the channel capacity and BER of our proposed fractal OAM generation and detection scheme. Then, we compare the capacity and BER of our proposed fractal OAM transmission with normal OAM transmission. Also, we analyze how the receive UCA radius and the distance between the UCAs impact the capacity and BER performances. Numerical results are given for better analysis.

\subsection{Capacity and BER Performances}
In this part, we give the capacity and BER for our proposed fractal OAM generation and detection scheme. We let $\boldsymbol{\mathrm x}$ be a complex Gaussian distributed vector with zero mean and $\mathbb{E}\left\{\boldsymbol{\mathrm x}\boldsymbol{\mathrm x}^H\right\}\hspace{-0.1cm}=\hspace{-0.1cm}\boldsymbol{\mathrm I}_{6}$, where $\boldsymbol{\mathrm I}_{6}$ denotes an identity matrix of size $6$. Based on Eq.~\eqref{eq:y0}, the signal-to-interference-plus-noise ratio (SINR) of the $l$th OAM mode, denoted by $\gamma_l$, can be given as follows:
\begin{align}
\gamma_l=\frac{P_l\left|\boldsymbol{\mathrm h}'(l)\right|^2}{\sigma^2_n+\sum_{k=0\atop k\ne l}^{5}P_k\left|\boldsymbol{\mathrm H}'(l,k)\right|^2},
\label{eq:fractal_OAM_SINR}
\end{align}
where $\boldsymbol{\mathrm h}'(l)$ represents the $(l)$th element of $\boldsymbol{\mathrm h}'$ and $\boldsymbol{\mathrm H}'(l,k)$ denotes the $l$th row and the $k$th column element in $\boldsymbol{\mathrm H}'$. $\boldsymbol{\mathrm H}'(l,k)$ also denotes the interference from the $k$th transmit OAM mode to the $l$th receive OAM mode. Furthermore, the capacity for our proposed fractal OAM generation and detection scheme in the free-space transmission is given as follows:
\begin{align}
C\hspace{-0.1cm}=\hspace{-0.1cm}\sum_{l=0}^{5}{\rm log}\left(1\hspace{-0.1cm}+\hspace{-0.1cm}\gamma_l\right)\hspace{-0.1cm}=\hspace{-0.1cm}\sum_{l=0}^{5}{\rm log}\left(\hspace{-0.1cm}1\hspace{-0.1cm}+\hspace{-0.1cm}\frac{P_l\left|\boldsymbol{\mathrm h}'(l)\right|^2}{\sigma^2_n\hspace{-0.18cm}+\hspace{-0.1cm}\sum_{k=0\atop k\ne l}^{5}\hspace{-0.1cm}P_k\left|\boldsymbol{\mathrm H}'(l,k)\right|^2}\hspace{-0.1cm}\right).
\label{eq:fractal_OAM_capacity}
\end{align}

Using binary phase shift keying (BPSK) for the input and output signals, the BER of our proposed fractal OAM generation and detection scheme, denoted by ${P_e}$, can be given as follows:
%\vspace{-5pt}
\begin{align}
{P_e}&=\frac{1}{6}\sum_{l=0}^{5}\frac{1}{2}{\rm{erfc}}\sqrt{\gamma_l}\nonumber\\
&\quad\quad=\frac{1}{12}\sum_{l=0}^{5}{\rm{erfc}}\sqrt{\frac{P_l\left|\boldsymbol{\mathrm h}'(l)\right|^2}{\sigma_n^2+\boldsymbol \sum_{k=0\atop k\ne l}^{5}P_k\left|\boldsymbol{\mathrm H}'(l,k)\right|^2}},
\label{eq:fractal_OAM_BER}
\end{align}
where erfc$(\alpha)=\frac{2}{\sqrt{\pi}}$$\int_{\alpha}^{\infty} e^{-t^2}dt$.

%{\color{red}SIMPLIFICATIONS AND ANALYSES FOR CAPACITY AND BER}
%
%Then, we simplify the capacity and BER and analyze the impact of each parameter on the capacity and BER. For the receive UCA center coordinate and radius satisfying Eqs.~\eqref{eq:RrCenterCoordinate} and \eqref{eq:RrRadius}, the receive UCA can receive UCA can receive a fractal OAM structure in alignment. Therefore, the interference from other OAM modes can be ignored and the simplified SINR of the $l$th OAM mode, denoted by $\widehat{\gamma}_l$, can be given by substituting Eq.~\eqref{eq:H_free} into Eq.~\eqref{eq:fractal_OAM_SINR} as follows:
%\begin{align}
%\widehat{\gamma}_l&=\frac{P_l\left|\boldsymbol{\mathrm h}'(l)\right|^2}{\sigma^2_n}
%=\frac{P_l}{\sigma^2_n}\left|\sum_{n_r=0}^{N_r-1}e^{-j\frac{2\pi}{N_r}n_rl}\sum_{n_t=0}^{5}e^{j\frac{\pi}{3}n_tl}H_{n_r,n_t}\right|^2\nonumber\\
%&\quad\quad=\frac{P_l}{\sigma^2_n}\left|\frac{\lambda}{4\pi z}e^{-j\frac{2\pi}{\lambda}\left(z+\frac{R_t^2}{2z}\right)}\sum_{n_r=0}^{N_r-1}e^{-j2\pi\left[\frac{(\rho'_{n_r})^2}{2z\lambda}+\frac{n_rl}{N_r}\right]}\sum_{n_t=0}^{5}e^{j2\pi\left[\frac{n_tl}{6}-\frac{R_t}{z\lambda}\rho'_{n_r}\cos\left(\phi'_{n_r}-\frac{\pi}{3}n_t\right)\right]}\right|^2\nonumber\\
%&\quad\quad\quad\quad=\frac{P_l\lambda^2}{16\pi^2\sigma^2_n z^2}\left|\sum_{n_r=0}^{N_r-1}e^{-j2\pi\left[\frac{(\rho'_{n_r})^2}{2z\lambda}+\frac{n_rl}{N_r}\right]}\sum_{n_t=0}^{5}e^{j2\pi\left[\frac{n_tl}{6}-\frac{R_t}{z\lambda}\rho'_{n_r}\cos\left(\phi'_{n_r}-\frac{\pi}{3}n_t\right)\right]}\right|^2,
%\label{eq:fractal_OAM_SINR2}
%\end{align}

\subsection{Numerical Results and Analyses}
In this part, we first compare the capacity and BER performances of fractal OAM with normal OAM. Then, we analyze how the receive UCA radius and the distance between the UCAs impact the capacity and BER performances.
\begin{figure}[htbp]
\centering
%\vspace{-15pt}
\subfigure[Capacity performance.]{
\begin{minipage}{1\linewidth}
\centering
\includegraphics[scale=0.54]{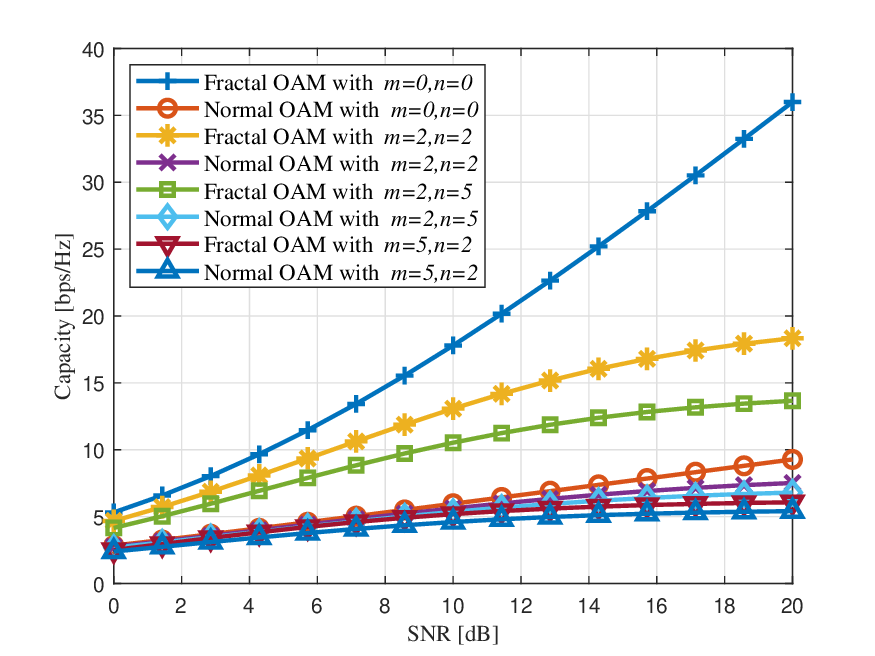}\label{fig:Fractal_OAM_capacity}
%\vspace{-20pt}
\end{minipage}
}\\
%\vspace{-10pt}
\subfigure[BER performance.]{
\begin{minipage}{1\linewidth}
\centering
\includegraphics[scale=0.54]{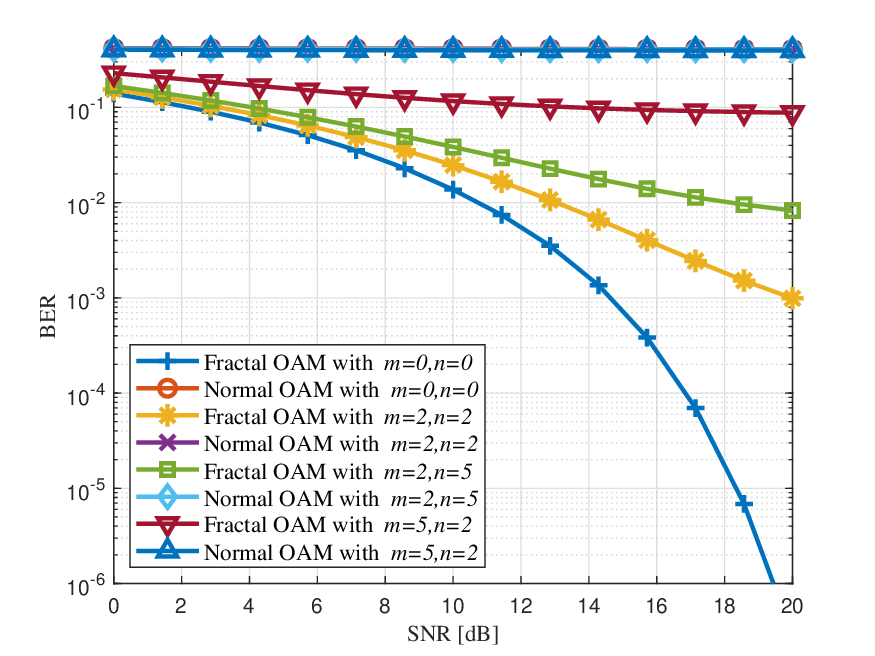}\label{fig:Fractal_OAM_BER}
%\vspace{-20pt}
\end{minipage}
}
\centering
%\vspace{-10pt}
\caption{Capacity and BER performances for normal OAM and fractal OAM transmissions.}\label{fig:Fractal_OAM_capacityBER}
%\vspace{-5pt}
\end{figure}

\textbf{Capacity and BER performances for normal OAM and fractal OAM transmissions:} Figure~\ref{fig:Fractal_OAM_capacityBER} shows the capacity and BER performances for normal OAM and fractal OAM transmissions. The wavelength is set to $\lambda=10$ mm, with $R_t=\lambda/2$ for normal OAM and $R_t=150\lambda$ for fractal OAM. The transmission distance is set to $z=1000\lambda$. The receive UCA radius is given as $R_r = 1.67\lambda$ according to $R_r = \frac{1}{4}\frac{\lambda z}{R_t}$, which satisfies the requirement of Eq.~\eqref{eq:RrRadius}. The transmit and receive UCAs are both with $6$ elements. The transmit power for each OAM mode is set to $1$ W. Then, we plot the capacity and BER performances for the normal OAM and fractal OAM with different receive UCA center coordinates by setting $m = \{0,2,2,5\}$ and $n=\{0,2,5,2\}$. The receive UCA center coordinates are given according to Eqs.~\eqref{eq:RrCenterCoordinate} and \eqref{eq:R_r_coordinate_polar}. The channels are calculated based on Eq.~\eqref{eq:H_free}. As shown in Fig.~\ref{fig:Fractal_OAM_capacityBER}, fractal OAM transmissions have better capacity and BER performances than normal OAM transmissions with the same $m$ and $n$. This indicates that fractal OAM not only alleviates the hollow divergence of OAM beams in the transceiver-aligned scenario, but it also improves the capacity and BER performance in transceiver-unaligned scenarios. We also observe that the BER for normal OAM is much higher than that for fractal OAM. This is because normal OAM beams are divergent, and the receive UCA does not cover the main lobe of the normal OAM beams. In addition, Fig.~\ref{fig:Fractal_OAM_capacityBER} shows that both fractal and normal OAM transmissions have worse capacity and BER performances with larger values of $m$ and $n$. This is because larger $m$ and $n$ lead to a farther distance between the receive UCA and the transmission axis, resulting in lower received power. Furthermore, both fractal and normal OAM transmissions have better capacity and BER performance with $m = 2$ and $n = 5$ compared to $m = 5$ and $n = 2$. This is because the receive UCA is farther from the transmission axis with $m = 5$ and $n = 2$ than with $m = 2$ and $n = 5$, leading to lower received power. Therefore, our proposed fractal OAM transmission has better capacity and BER performance than normal OAM transmission with the same receive UCA radius at the same transceiver distance.

\begin{figure}[htbp]
\centering
\vspace{-10pt}
\subfigure[Capacity performance for $0\le R_r\le 2.57\lambda$.]{
\begin{minipage}{1\linewidth}
\centering
\includegraphics[scale=0.5]{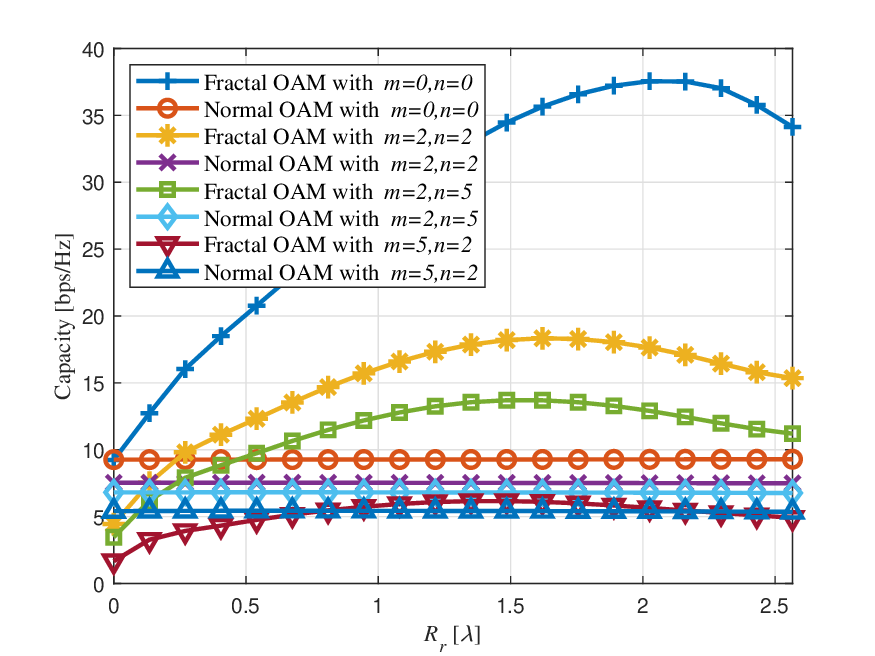}\label{fig:Fractal_OAM_capacity_RrLen}
%\vspace{-20pt}
\end{minipage}
}
\vspace{-10pt}
\subfigure[BER performance for $0\le R_r\le 2.57\lambda$.]{
\begin{minipage}{1\linewidth}
\centering
\includegraphics[scale=0.5]{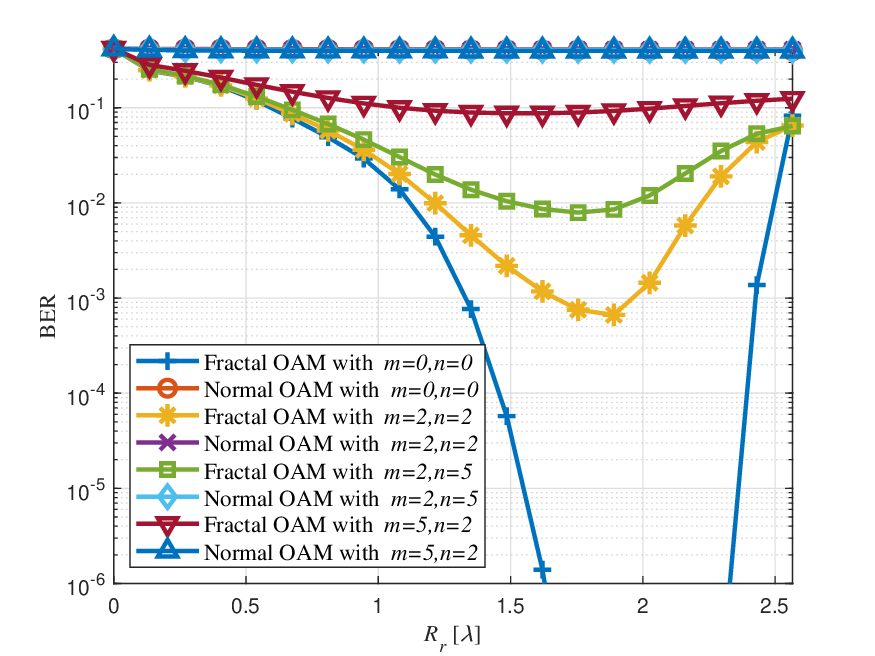}\label{fig:Fractal_OAM_BER_RrLen}
%\vspace{-20pt}
\end{minipage}
}
\vspace{-10pt}
\subfigure[Capacity performance for $0\le R_r\le 200\lambda$.]{
\begin{minipage}{1\linewidth}
\centering
\includegraphics[scale=0.5]{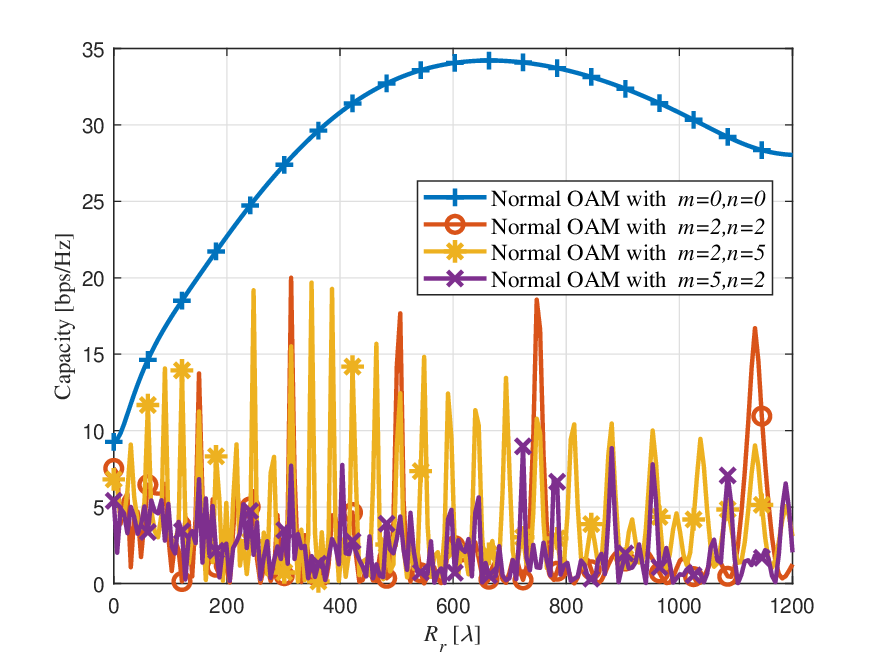}\label{fig:Fractal_OAM_capacity_RrLen2}
%\vspace{-20pt}
\end{minipage}
}
\vspace{-10pt}
\subfigure[BER performance for $0\le R_r\le 1200\lambda$.]{
\begin{minipage}{1\linewidth}
\centering
\includegraphics[scale=0.5]{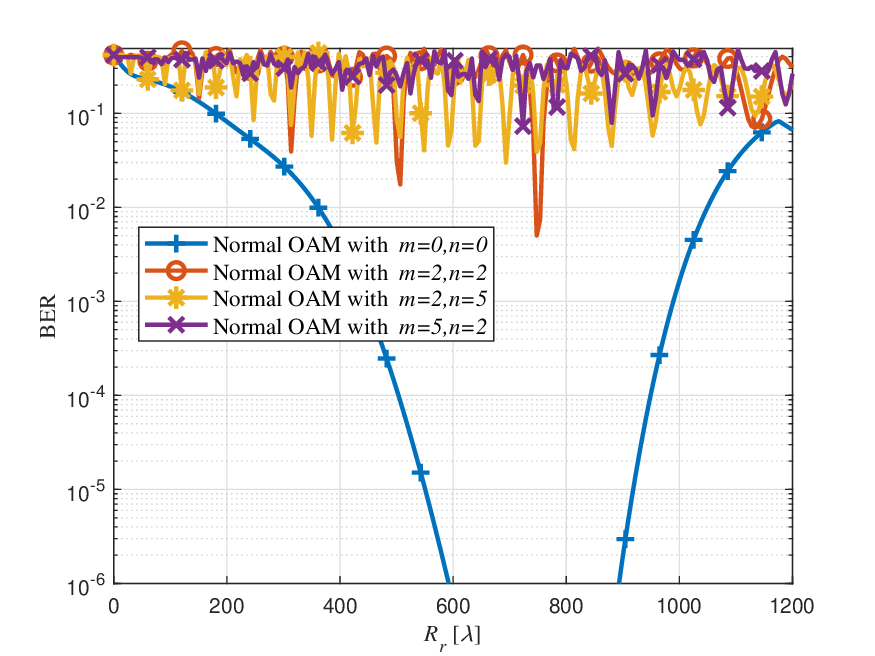}\label{fig:Fractal_OAM_BER_RrLen2}
%\vspace{-20pt}
\end{minipage}
}
\centering
\vspace{-5pt}
\caption{Impact of the receive UCA radius on capacity and BER performances for normal OAM and fractal OAM transmissions.}\label{fig:Fractal_OAM_capacityBER_RrLen}
%\vspace{-5pt}
\end{figure}
\textbf{Impact of the Receive UCA Radius:} Figure~\ref{fig:Fractal_OAM_capacityBER_RrLen} shows the impact of the receive UCA radius on the capacity and BER performance for normal OAM and fractal OAM transmissions. In Figs.~\ref{fig:Fractal_OAM_capacity_RrLen} and \ref{fig:Fractal_OAM_BER_RrLen}, we increase $R_r$ from $0$ to $2.57\lambda$ according to Eq.~\eqref{eq:RrRadius} to analyze the effect of the receive UCA radius on fractal OAM. In Figs.~\ref{fig:Fractal_OAM_capacity_RrLen2} and \ref{fig:Fractal_OAM_BER_RrLen2}, we increase $R_r$ from $0$ to $1200\lambda$ to further analyze the effect of the receive UCA radius on normal OAM. The SNR is set to $20$ dB, and all other variables remain the same as in Fig.~\ref{fig:Fractal_OAM_capacityBER}. Figure~\ref{fig:Fractal_OAM_capacity_RrLen} shows that as $R_r$ increases, the capacity of fractal OAM first increases to a maximum value and then decreases. Similarly, in Fig.~\ref{fig:Fractal_OAM_BER_RrLen}, the BER of fractal OAM first decreases to a minimum value and then increases as $R_r$ increases. This is because OAM beams are hollow, and there is an angle between the main lobe and the transmission axis. Therefore, as the receive UCA radius increases, fractal OAM experiences a maximum received power. However, for maximum power reception, the receive UCA radius should not be as small as possible. For normal OAM transmissions, the capacity and BER remain constant as $R_r$ increases in Figs.~\ref{fig:Fractal_OAM_capacity_RrLen} and \ref{fig:Fractal_OAM_BER_RrLen}. This is because the variation range of the receive UCA radius is too small to reflect the impact of $R_r$ on normal OAM transmissions. In Figs.~\ref{fig:Fractal_OAM_capacity_RrLen2} and \ref{fig:Fractal_OAM_BER_RrLen2}, we further analyze the effect of the receive UCA radius on normal OAM. It can be seen in Fig.~\ref{fig:Fractal_OAM_capacity_RrLen2} that as $R_r$ increases from $0$ to $1200\lambda$, the capacity of transceiver-aligned normal OAM first increases to a maximum value and then decreases. Similarly, in Fig.~\ref{fig:Fractal_OAM_BER_RrLen2}, the BER of transceiver-aligned normal OAM first decreases to a minimum value and then increases as $R_r$ increases. The variation of transceiver-aligned normal OAM capacity and BER in Figures Figs.~\ref{fig:Fractal_OAM_capacity_RrLen2} and \ref{fig:Fractal_OAM_BER_RrLen2} is similar to that of fractal OAM transmissions in Figs.~\ref{fig:Fractal_OAM_capacity_RrLen} and \ref{fig:Fractal_OAM_BER_RrLen}. This is because there is also an angle between the main lobe and the transmission axis for normal OAM beams, but with a much larger radius compared to fractal OAM beams. However, for transceiver-unaligned normal OAM transmissions, the received beam structures are incomplete and the orthogonality among different OAM modes cannot be achieved, resulting in poor capacity and BER performance.

\begin{figure}[htbp]
\centering
%\vspace{3pt}
\subfigure[Capacity performance with $R_r = 1.283\lambda$.]{
\begin{minipage}{1\linewidth}
\centering
\includegraphics[scale=0.5]{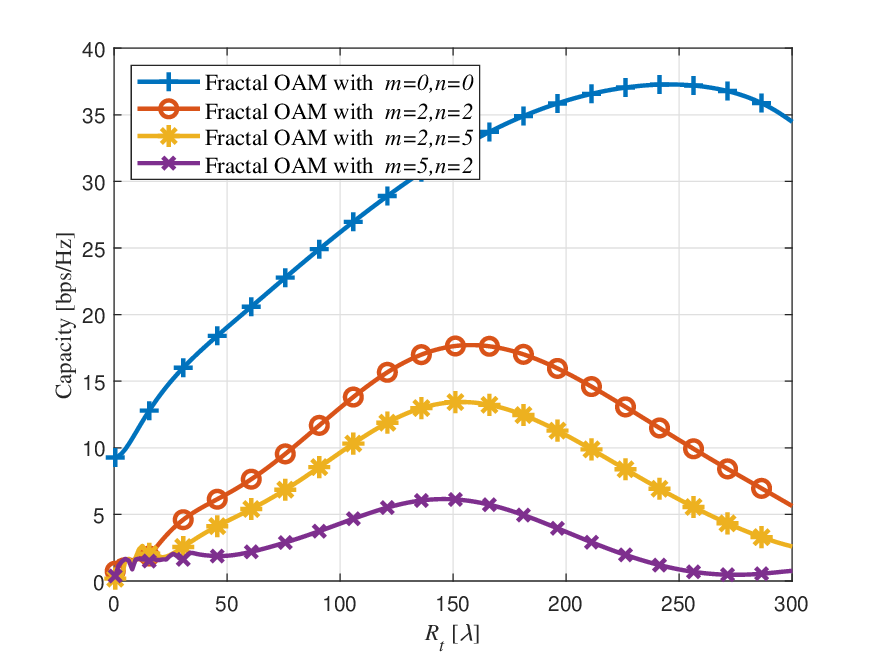}
\label{fig:Fractal_OAM_capacity_RtWithRrMin}
%\vspace{3pt}
\end{minipage}
}
\subfigure[BER performance with $R_r = 1.283\lambda$.]{
\begin{minipage}{1\linewidth}
\centering
\includegraphics[scale=0.5]{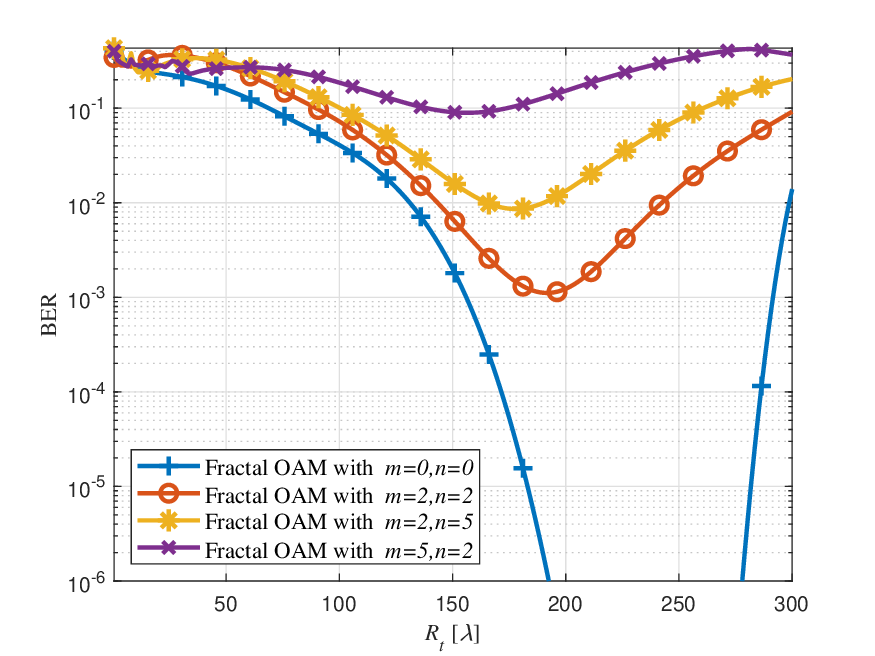}
\label{fig:Fractal_OAM_BER_RtWithRrMin}
%\vspace{3pt}
\end{minipage}
}
%\vspace{-5pt}
\centering
\caption{Impact of the transmit UCA radius on capacity and BER performances for normal OAM and fractal OAM transmissions.}\label{fig:Fractal_OAM_capacityBER_Rt}
\end{figure}
\textbf{Impact of the Transmit UCA Radius:} Figure~\ref{fig:Fractal_OAM_capacityBER_Rt} shows the impact of the transmit UCA radius on the capacity and BER performance of the fractal OAM transmission. In Fig.~\ref{fig:Fractal_OAM_capacityBER_Rt}, we increase $R_t$ from $0.5\lambda$ to $300\lambda$ to analyze the effect of the transmit UCA radius on the fractal OAM. According to Eq.~\eqref{eq:RrRadius}, we set $R_r = 1.283\lambda$ to make sure that the receive UCA radius is smaller than the smallest radius of the fractal OAM corresponding to the largest $R_t$. The SNR is set to $20$ dB and all other variables remain the same as in Fig.~\ref{fig:Fractal_OAM_capacityBER}. Figure~\ref{fig:Fractal_OAM_capacity_RtWithRrMin} shows that as $R_t$ increases, the capacity of fractal OAM first increases to a maximum value and then decreases. In Fig.~\ref{fig:Fractal_OAM_BER_RtWithRrMin}, the BER of fractal OAM first decreases to a minimum value and then increases as $R_t$ increases. These results are similar to the results given in Figs.~\ref{fig:Fractal_OAM_capacity_RrLen} and \ref{fig:Fractal_OAM_BER_RrLen}, which is also due to the hollow and divergent of OAM beams. To be specific, as $R_t$ increases, the radii of the received fractal OAM beams decreases, resulting in a higher concentration of energy and increased capacity, accompanied by a decreased BER. However, as $R_t$ continues to increase, the hollow region within each fractal OAM beam expands. This, in turn, reduces the energy collected by the receiving UCA since the radius of the receiving UCA remains constant. Consequently, the capacity and BER performance decrease.

\begin{figure}[htbp]
\centering
%\vspace{-15pt}
\subfigure[Capacity performance.]{
\begin{minipage}{1\linewidth}
\centering
\includegraphics[scale=0.5]{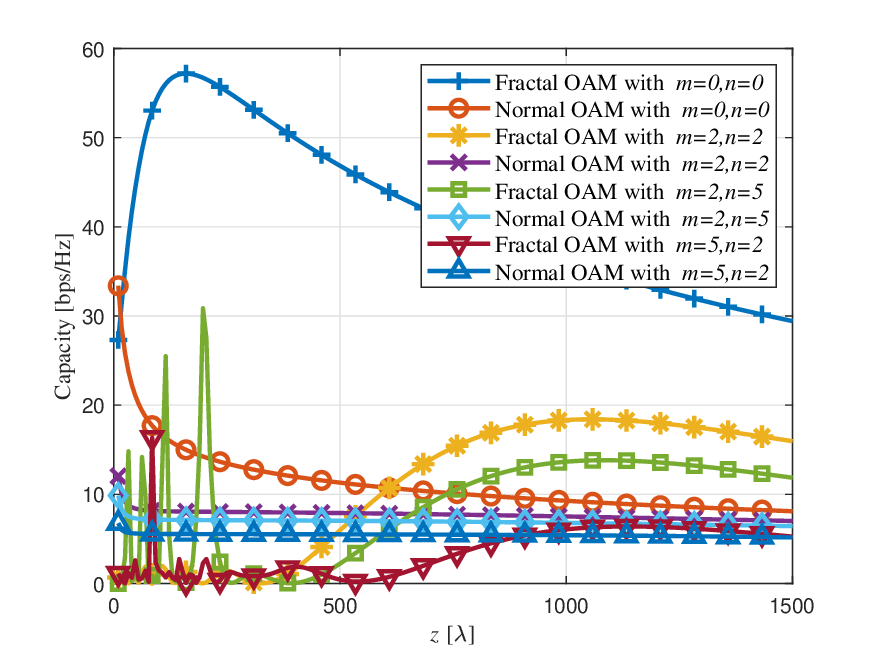}\label{fig:Fractal_OAM_capacity_z}
%\vspace{-20pt}
\end{minipage}
}
%\vspace{-10pt}
\subfigure[BER performance.]{
\begin{minipage}{1\linewidth}
\centering
\includegraphics[scale=0.5]{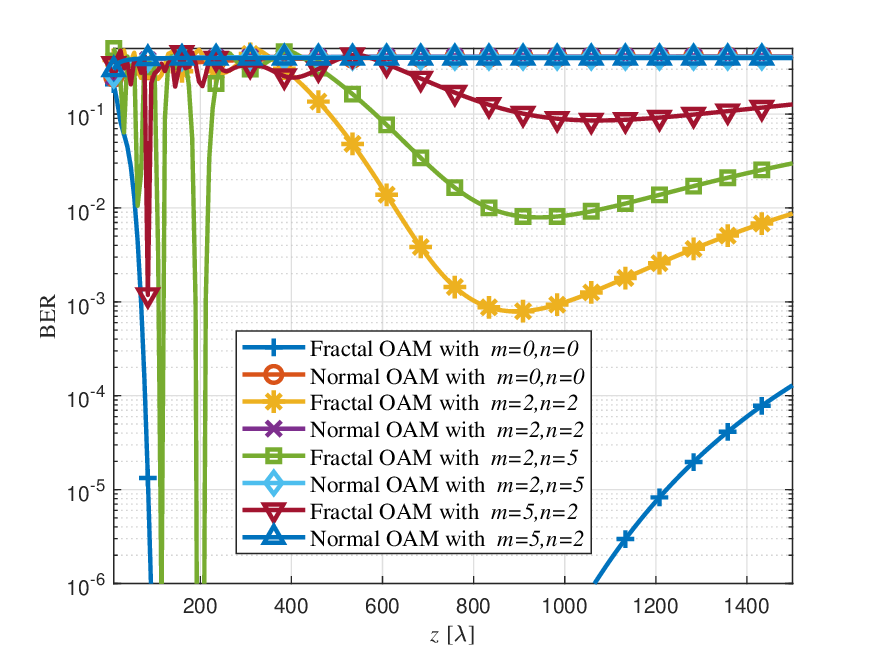}\label{fig:Fractal_OAM_BER_z}
%\vspace{-20pt}
\end{minipage}
}
\centering
%\vspace{-10pt}
\caption{Impact of the distance between the transceiver UCAs on capacity and BER performances for normal OAM and fractal OAM transmissions.}\label{fig:Fractal_OAM_capacityBER_z}
%\vspace{-5pt}
\end{figure}
\textbf{Impact of the Distance Between the Transceiver UCAs:} Figure~\ref{fig:Fractal_OAM_capacityBER_z} shows the impact of the distance between the transceiver UCAs on the capacity and BER performances for normal OAM and fractal OAM transmissions. We increase $z$ from $10\lambda$ to $1500\lambda$. The receive UCA radius $R_r$ is changed linearly with the distance between transceiver UCAs, ranging from $0.017\lambda$ to $2.5\lambda$ according to $R_r = \frac{1}{4}\frac{\lambda z}{R_t}$. The SNR is set to $20$ dB, and all other variables remain the same as in Fig.~\ref{fig:Fractal_OAM_capacityBER}. In Fig.~\ref{fig:Fractal_OAM_capacity_z}, the capacity of normal OAM transmissions decreases as the transceiver distance increases.
%\textcolor{blue}{This is because, as the transceiver distance increases, the attenuation and divergence of normal OAM beams become more severe\cite{OAM_Divergence}}.
This is because, the OAM beams are hollow and divergent, and the divergence angle increases as the transmit distance increases until it asymptotically reaches a limiting value when the beam is sufficiently far from the transmit UCA\cite{OAM_Divergence}. As the transmit distance grows, the OAM beams experience increased divergence, leading to a widening of the beam and a decrease in the power concentration at the receiver. Therefore, as the transceiver distance increases, the attenuation and divergence of normal OAM beams become more severe, decreasing the capacity of normal OAM transmissions\cite{oam_for_wireless_communication}. For transceiver-aligned fractal OAM, the capacity first increases to a maximum value and then decreases as the transceiver distance increases. For transceiver-unaligned fractal OAM, the capacity fluctuates severely for $z < 500\lambda$. This is because, for $z < 500\lambda$, the distance between the transceiver UCAs is not much greater than the transmit UCA radius, which is $150\lambda$. Therefore the magnitude part of the channel gain is no longer simply inversely proportional to $z$ when the transmit and receive UCAs are not aligned, leading to the severe fluctuation of the capacity. For $z \ge 500\lambda$, the capacity of transceiver-unaligned fractal OAM first increases to a maximum value and then decreases as the transceiver distance increases. The fractal OAM transmission with aligned transceivers has the highest capacity for $z \ge 40\lambda$. For $z \ge 900\lambda$, fractal OAM transmissions with unaligned transceivers also have higher capacities than normal OAM transmissions with the same $m$ and $n$.

\section{Simulations}\label{sec:simulations}
In this section, we first give our proposed fractal OAM transmit UCA simulation model in ANSYS HFSS. Then, we validate the feasibility of our proposed fractal OAM generation scheme by analyzing the amplitudes, phases, and antenna gains of the OAM beams generated by the modeled UCA. After that, we compare the capacity and BER performances of our proposed fractal OAM with the normal OAM transmissions. We also analyze how the receive UCA radius and the distance between the transceiver UCAs impact the capacity and BER performances of the fractal OAM transmission via simulations.

\subsection{HFSS Model}\label{sec:hfssModel}
\begin{figure}[htbp]
\centering
%\vspace{-15pt}
\subfigure[Patch model.]{
\begin{minipage}{0.45\linewidth}
\centering
\includegraphics[scale=0.2]{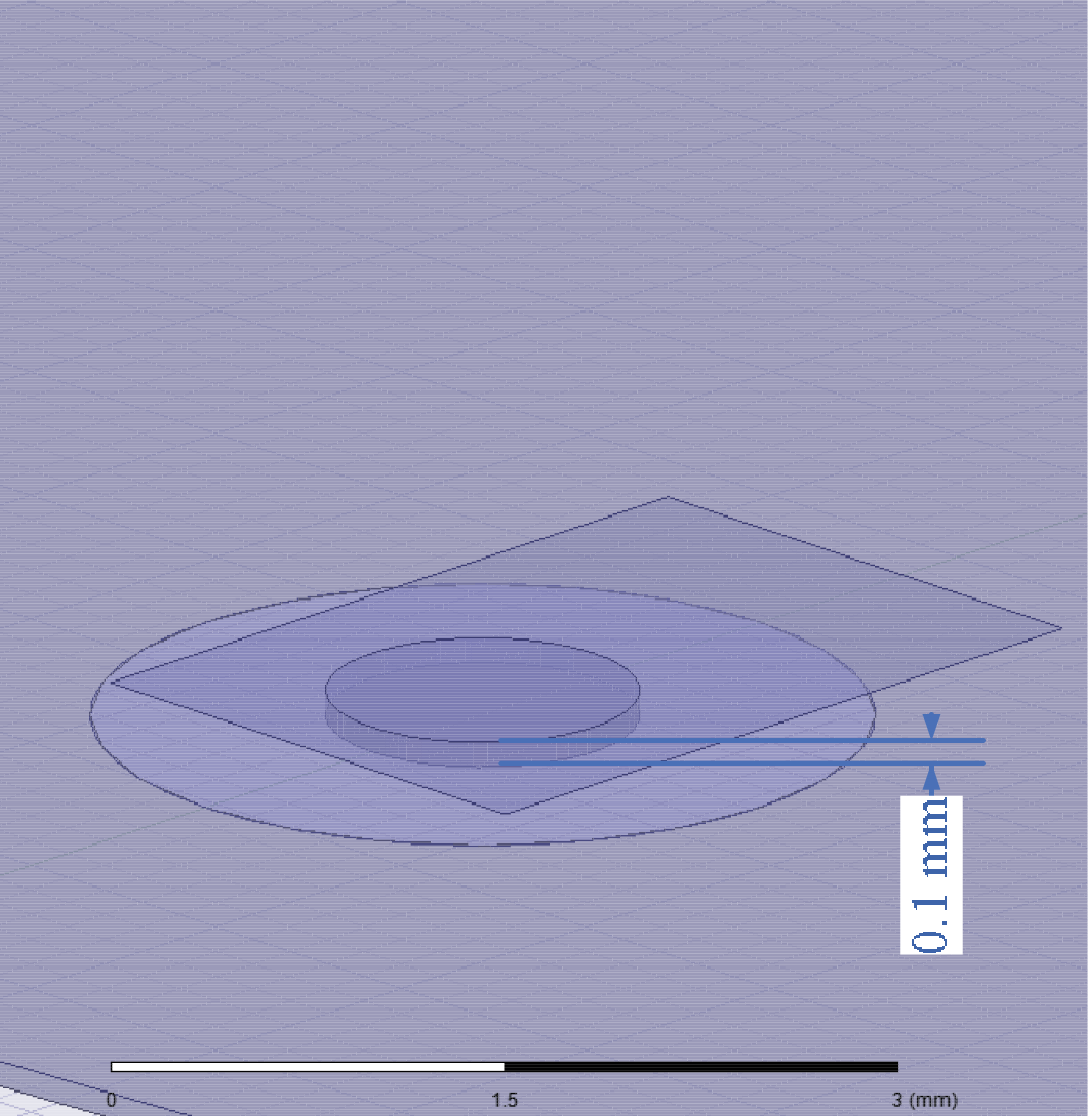}\label{fig:hfssPatchModel}
\vspace{3pt}
\end{minipage}
}
\subfigure[Top view of patch.]{
\begin{minipage}{0.45\linewidth}
\centering
\includegraphics[scale=0.2]{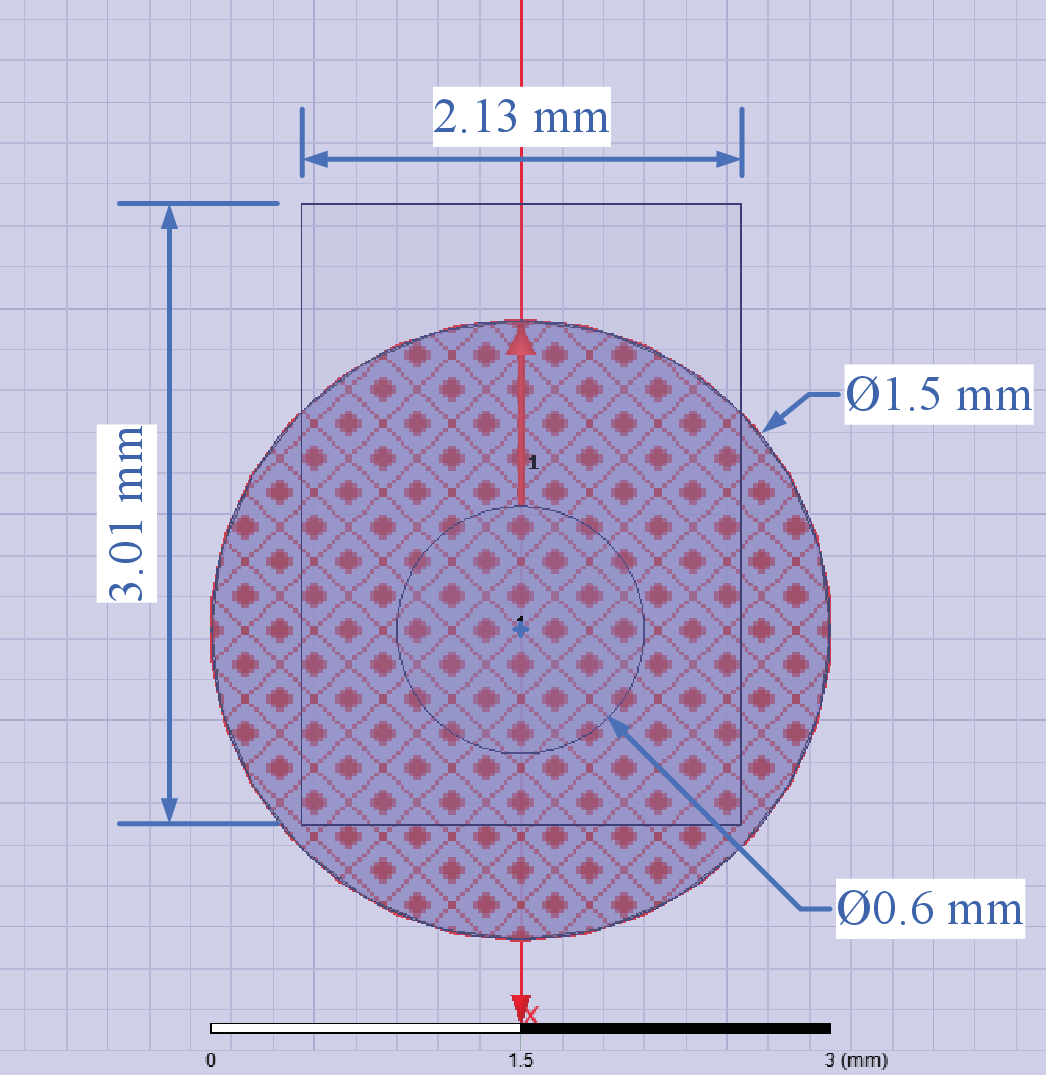}\label{fig:hfssPatchModelBird}
\vspace{3pt}
\end{minipage}
}\\
\subfigure[UCA model with $R_t=30$mm.]{
\begin{minipage}{0.45\linewidth}
\centering
\includegraphics[scale=0.2]{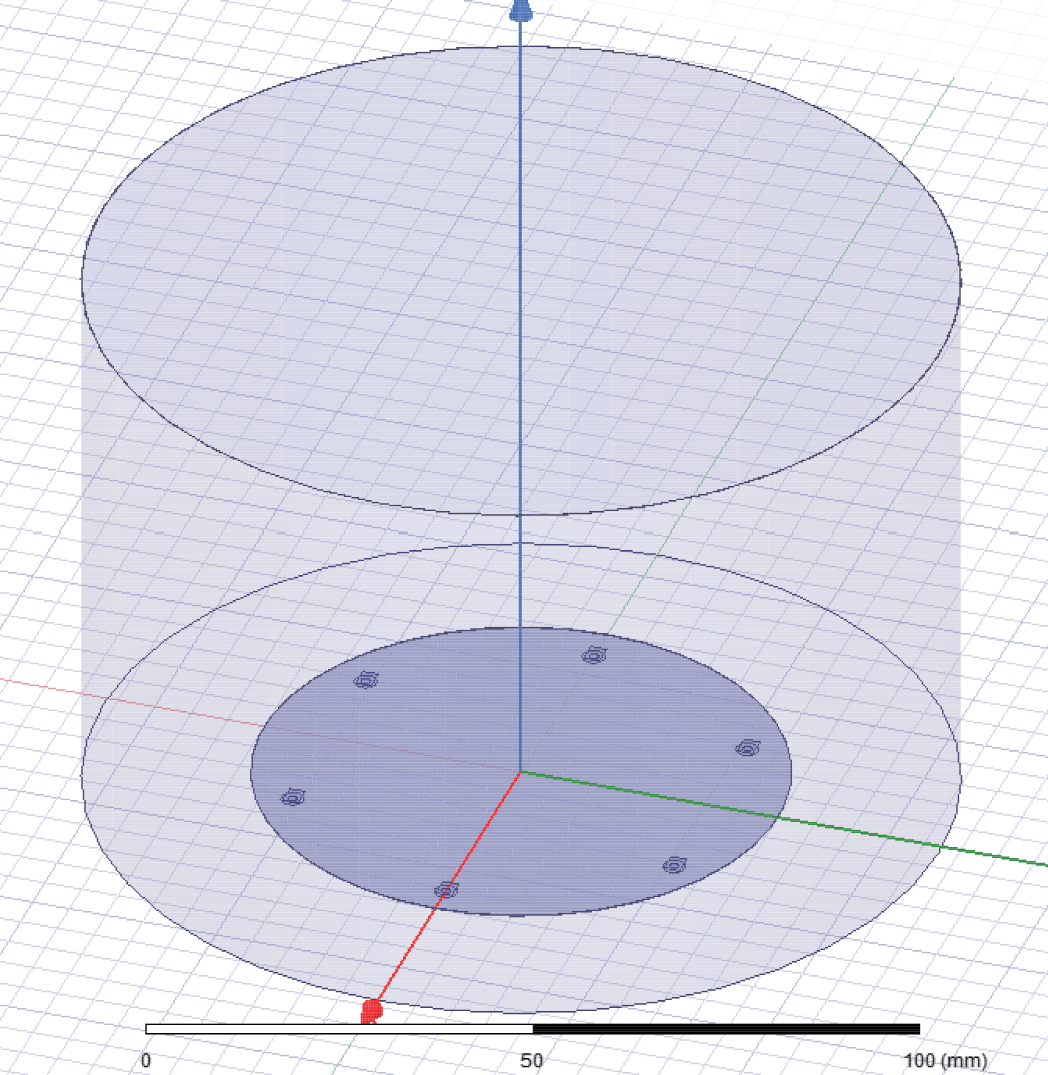}\label{fig:hfssUCAModel}
\vspace{3pt}
\end{minipage}
}
\subfigure[Top view of the UCA with $R_t=30$mm.]{
\begin{minipage}{0.45\linewidth}
\centering
\includegraphics[scale=0.2]{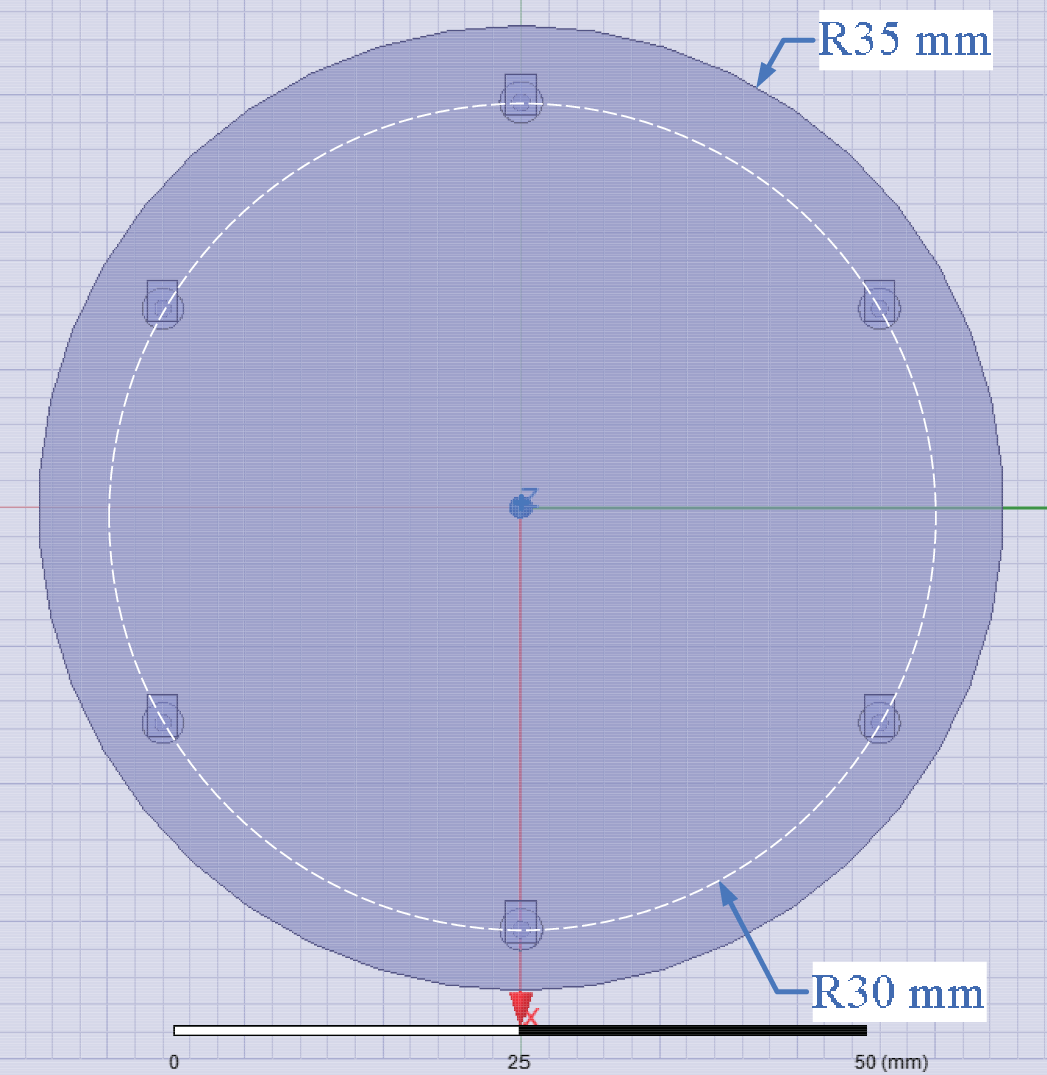}\label{fig:hfssUCAModelBird}
\vspace{3pt}
\end{minipage}
}\\
\subfigure[UCA model with $R_t=80$mm.]{
\begin{minipage}{0.45\linewidth}
\centering
\includegraphics[scale=0.2]{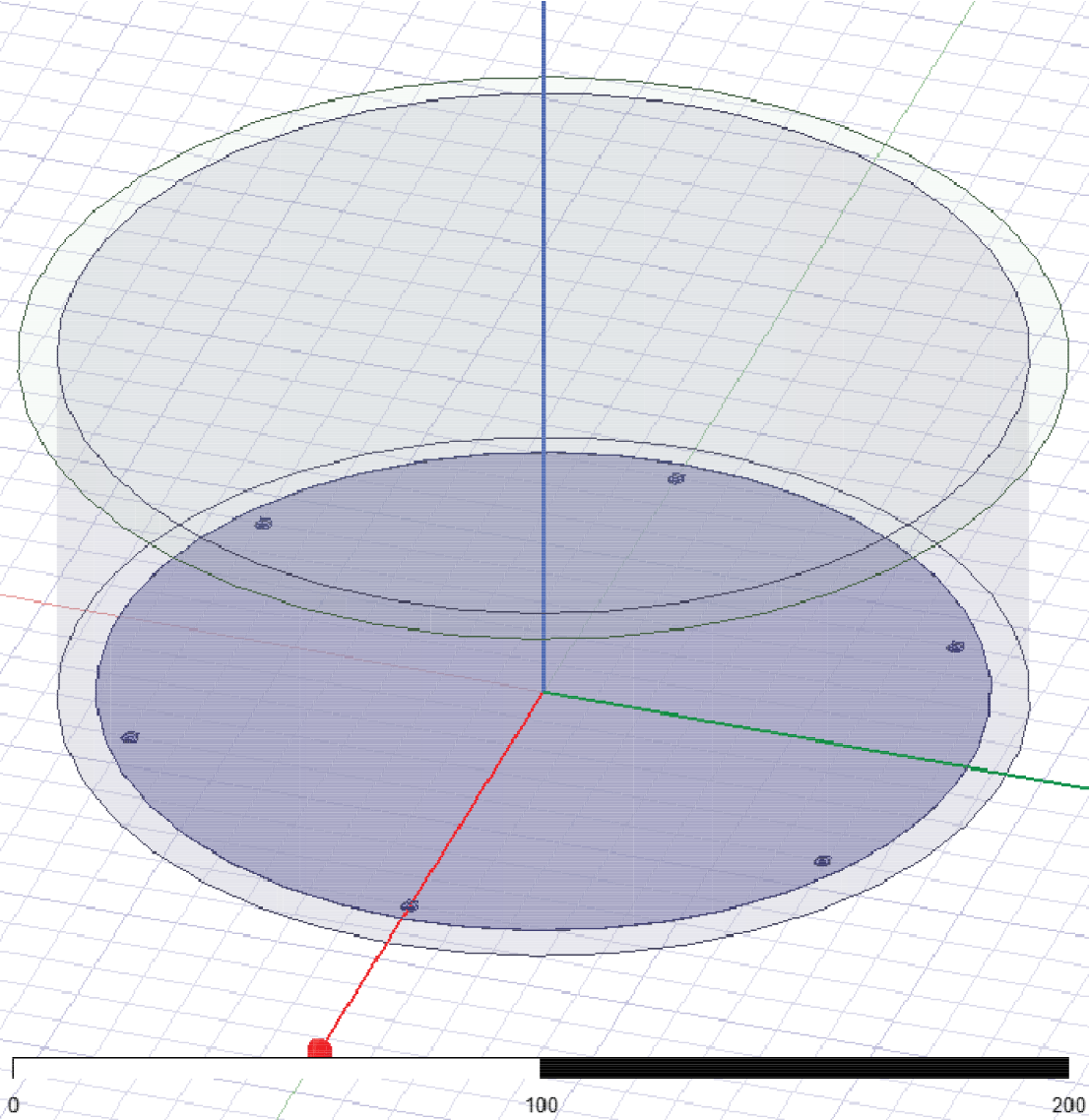}\label{fig:hfssUCAModel_Rt80mm}
\vspace{3pt}
\end{minipage}
}
\subfigure[Top view of the UCA with $R_t=80$mm.]{
\begin{minipage}{0.45\linewidth}
\centering
\includegraphics[scale=0.21]{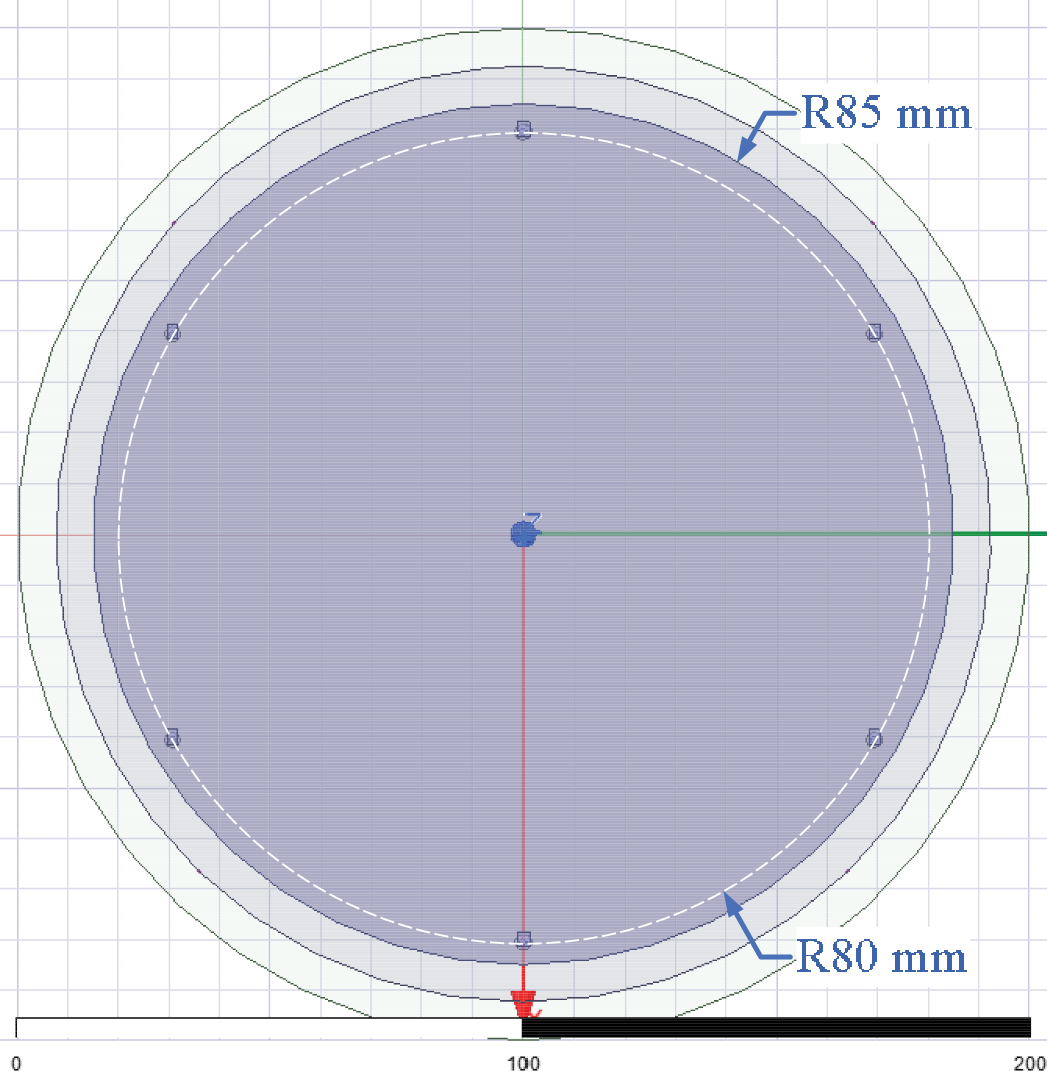}\label{fig:hfssUCAModelBird_Rt80mm}
\vspace{3pt}
\end{minipage}
}
\centering
\vspace{-5pt}
\caption{HFSS simulation model of the transmit UCA for the fractal OAM.}\label{fig:hfssModel}
%\vspace{-10pt}
\end{figure}
Figure~\ref{fig:hfssModel} shows the simulation models of the patches, the feeds, the cylindrical dielectric plate, and the transmit UCA. We set the operating frequency to $30$ GHz. The dielectric plate material is set as FR4 epoxy. We set the patches, feeds, and ground as perfect electrical conductors (PECs). As shown in Figs.~\ref{fig:hfssPatchModel} and \ref{fig:hfssPatchModelBird}, the length and wide of each patch is set as $3.01$ mm and $2.13$ mm, respectively. The radii and heights of cylindrical feeds are set as $0.6$ mm and $0.1$ mm, respectively. The ground is set as the lower surface of the dielectric plate. We also set $R_t=30$ mm for the UCA in Figs.~\ref{fig:hfssUCAModel} and \ref{fig:hfssUCAModelBird}. The radius and height of the cylindrical dielectric plate is set as $35$ mm and $0.1$ mm, respectively. In order to observe the impact of different transmit UCA radii on the generation of fractal OAM beams, we also give another simulation model with $R_t = 80$ mm as shown in Figs.~\ref{fig:hfssUCAModel_Rt80mm} and \ref{fig:hfssUCAModelBird_Rt80mm}. The radius of the cylindrical dielectric plate is set as $85$ mm. Other variables remain the same as in Figs.~\ref{fig:hfssUCAModel} and \ref{fig:hfssUCAModelBird}.

\subsection{Amplitude and Phase Structures for Fractal OAM Beams}
\begin{figure}[H]
\centering
\vspace{-10pt}
\subfigure[Mode 1 Power.]{
\begin{minipage}{0.4\linewidth}
\centering
\includegraphics[scale=0.157]{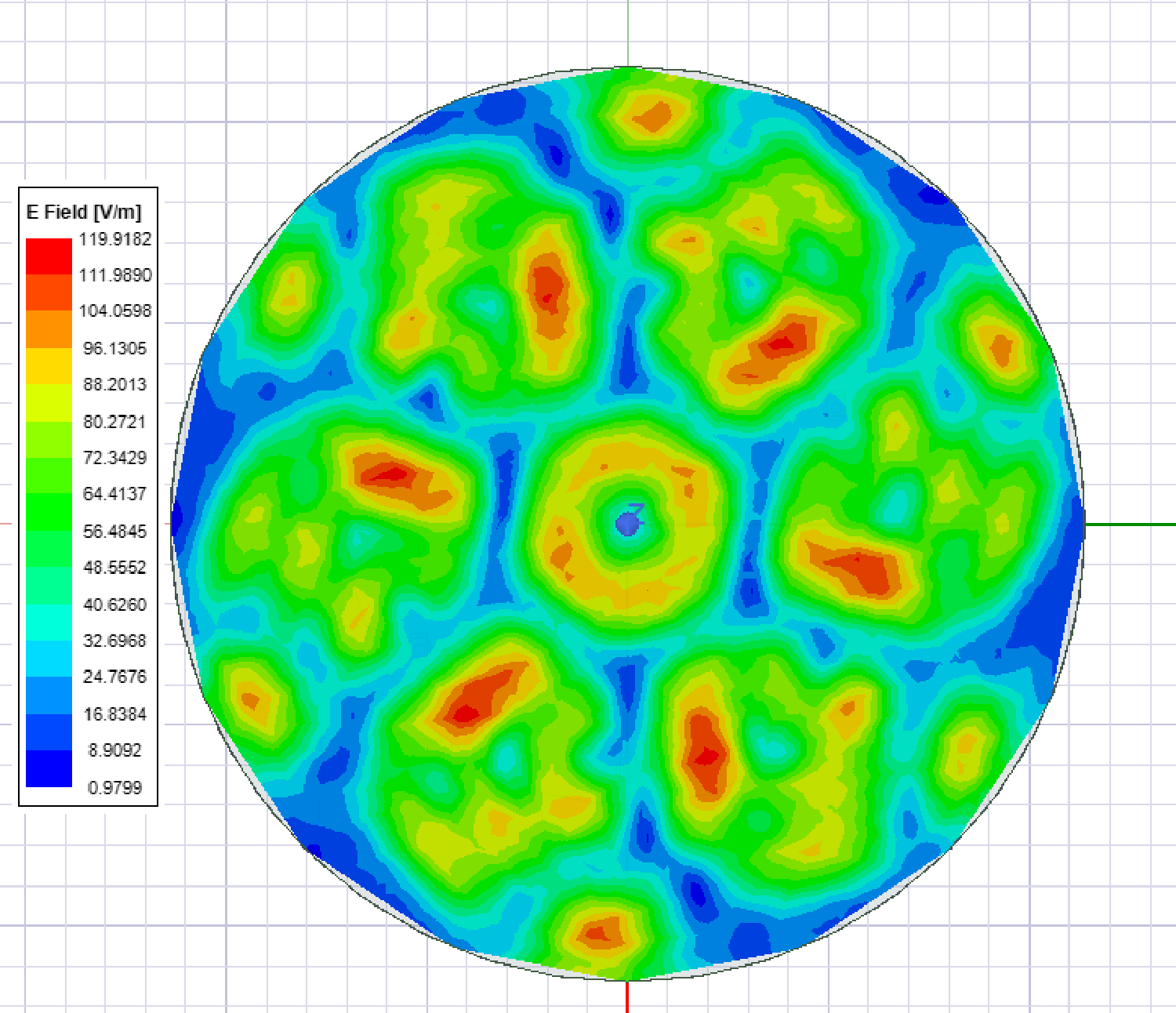}\label{fig:hfssMode1Power}
%\vspace{1pt}
\end{minipage}
}
\vspace{-5pt}
\subfigure[Mode 1 Phase.]{
\begin{minipage}{0.4\linewidth}
\centering
\includegraphics[scale=0.157]{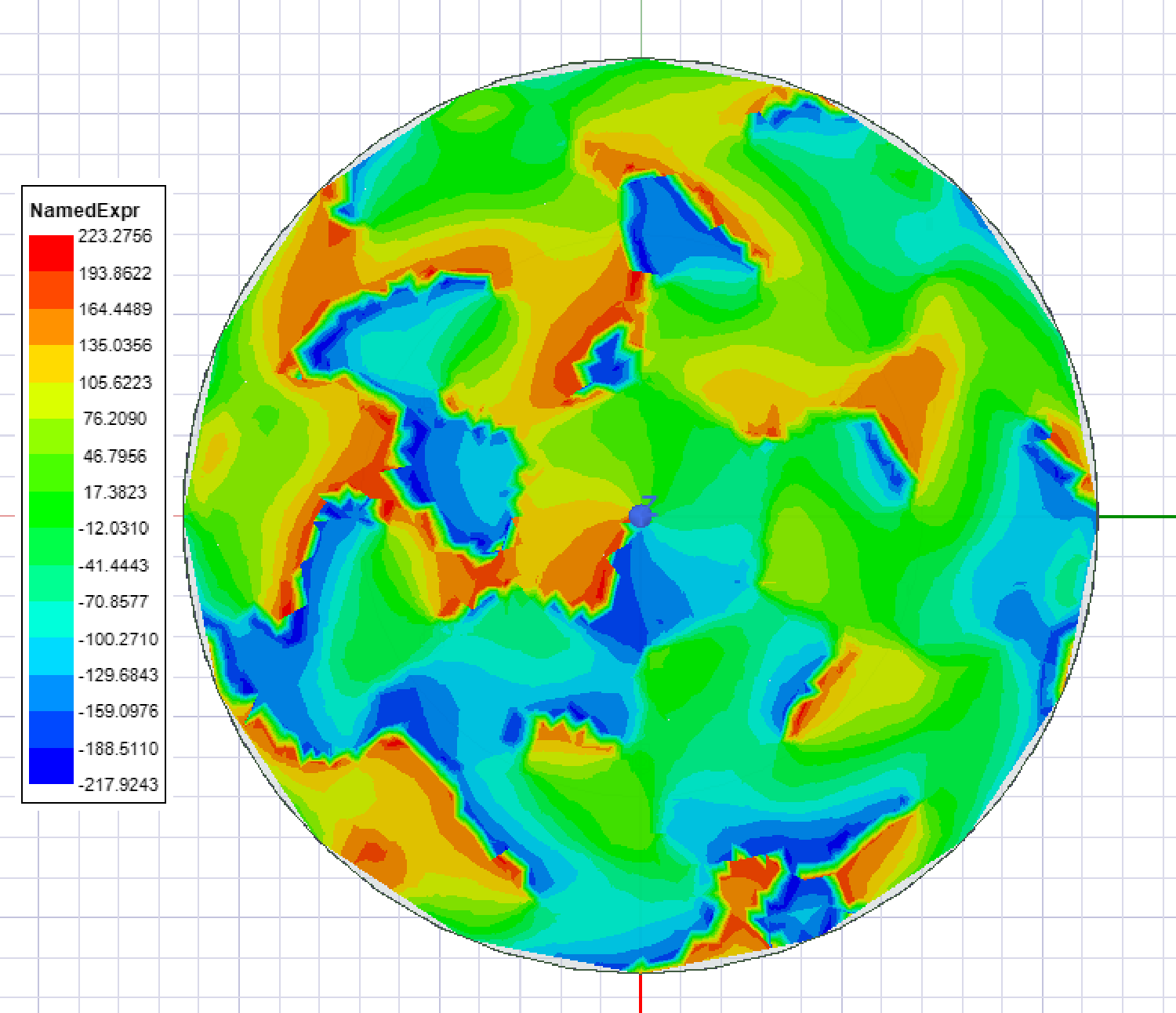}\label{fig:hfssMode1Phase}
%\vspace{1pt}
\end{minipage}
}\\
\vspace{-5pt}
\subfigure[Mode 1 gain.]{
\begin{minipage}{0.4\linewidth}
\centering
\includegraphics[scale=0.157]{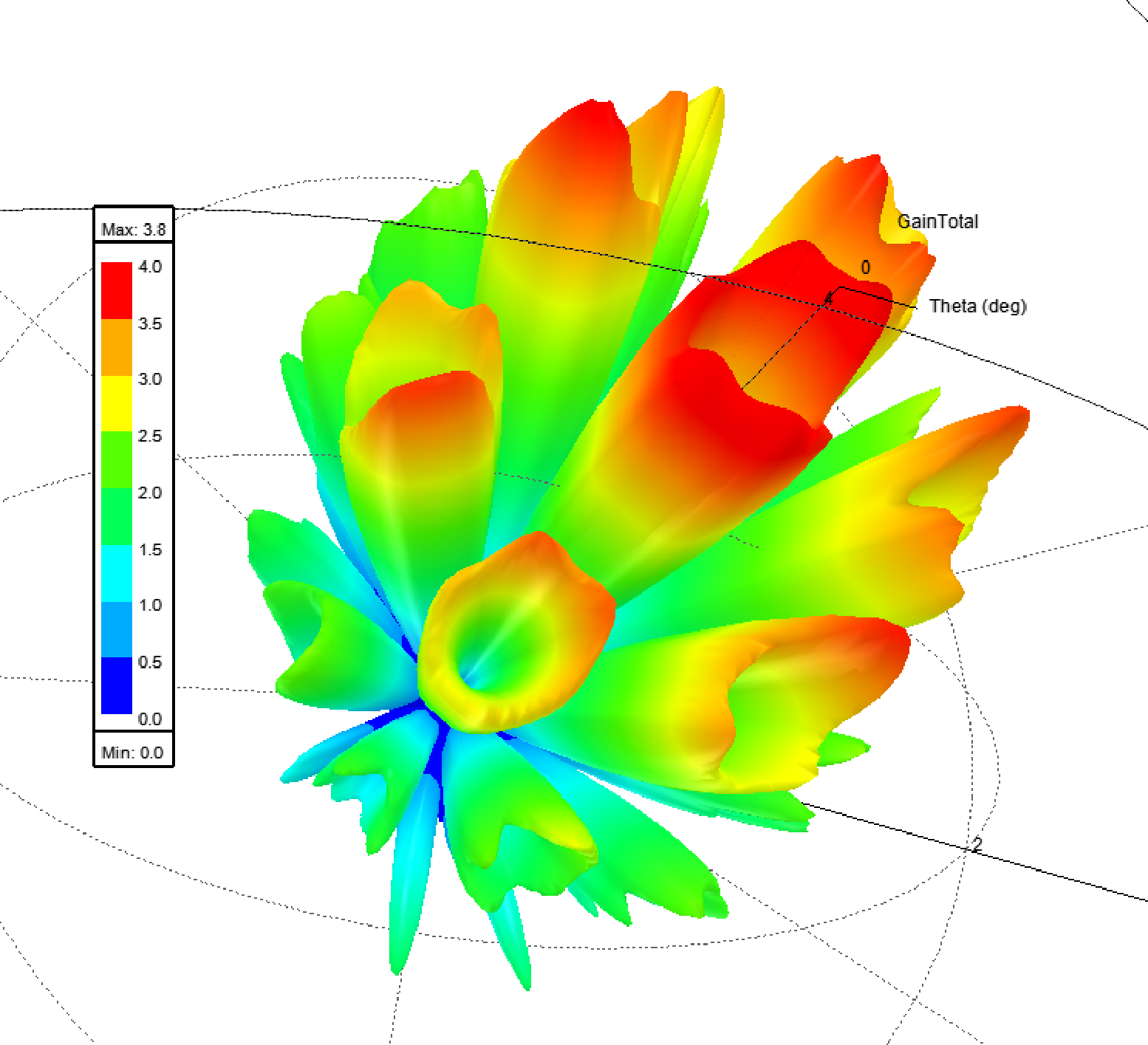}\label{fig:hfssMode1Gain}
%\vspace{1pt}
\end{minipage}
}
\vspace{-5pt}
\subfigure[{Mode 1 gain [dB]}.]{
\begin{minipage}{0.4\linewidth}
\centering
\includegraphics[scale=0.157]{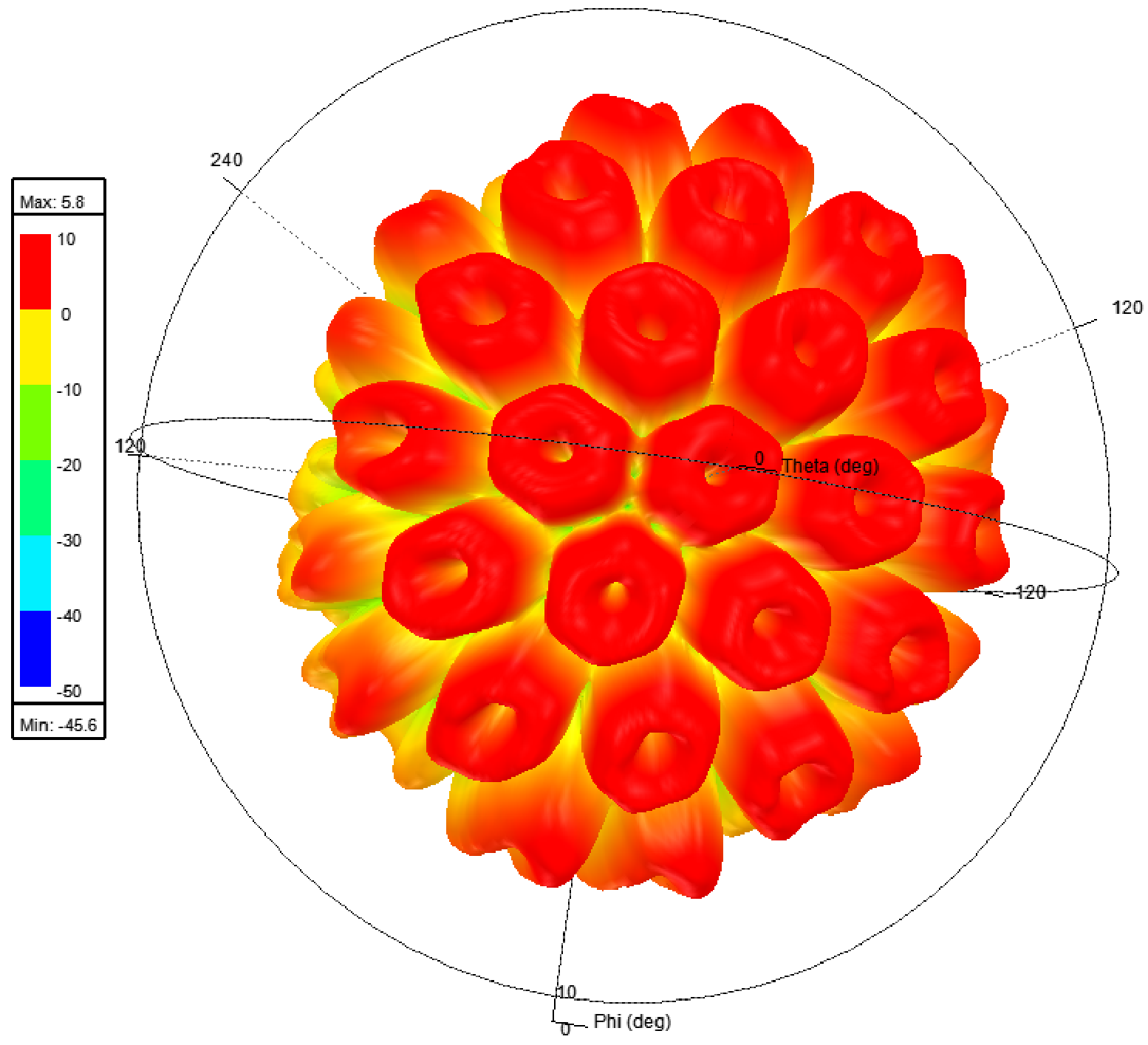}\label{fig:hfssMode1GaindB}
%\vspace{1pt}
\end{minipage}
}
\centering
%\vspace{-5pt}
\caption{HFSS simulated power, phase, and antenna gain distributions for $R_t = 30$ mm.}\label{fig:hfssAmpPhaGain}
%\vspace{-15pt}
\end{figure}
Figure~\ref{fig:hfssAmpPhaGain} shows the amplitude, phase, and antenna gain distributions for the fractal OAM beams with OAM-mode set to $1$ and $R_t = 30$ mm.
%Figure~\ref{fig:hfssAmpPhaGain} shows the amplitude and phase structures for the fractal OAM beams with OAM-mode set to $1$.
The results are plotted based on the model given in Fig.~\ref{fig:hfssModel}. The distance from the transmit UCA to the receive plane is set as $75$ mm. In Fig.~\ref{fig:hfssMode1Power}, we can clearly find a honeycomb amplitude structure of fractal OAM beams, validating that fractal OAM beams can be generated by the UCA. Also, the coordinates of fractal OAM centers are around $(\pm25,\pm14.43,75)$ mm, $(\pm25,\mp14.43,75)$ mm, and $(0,\pm28.87,75)$ mm, which are consistent with Eq.~\eqref{eq:OAM_grid_CarCorrdinates}. In Fig.~\ref{fig:hfssMode1Phase}, the variation of color from blue to red, yellow, green, and back to blue corresponds to the change in phase of $2\pi$. The rotational phase structure, which is a typical feature of OAM beams, can be clearly found in Fig.~\ref{fig:hfssMode1Phase}, also validating the feasibility. Figures~\ref{fig:hfssMode1Gain} and \ref{fig:hfssMode1GaindB} show the antenna gain for the transmit UCA, where Fig.~\ref{fig:hfssMode1GaindB} is in dB. We can also find fractal OAM structures in Figs.~\ref{fig:hfssMode1Gain} and \ref{fig:hfssMode1GaindB}. Furthermore, the closer the fractal OAM beam is to the transmission axis, the higher the gain is. This is because, as the degree of deviation from the transmission axis increases, the average distance from each UCA element increases, leading to a decrease in the total received power.

\begin{figure}[htbp]
\centering
\vspace{-5pt}
\subfigure[Mode 1 Power.]{
\begin{minipage}{0.4\linewidth}
\centering
\includegraphics[scale=0.157]{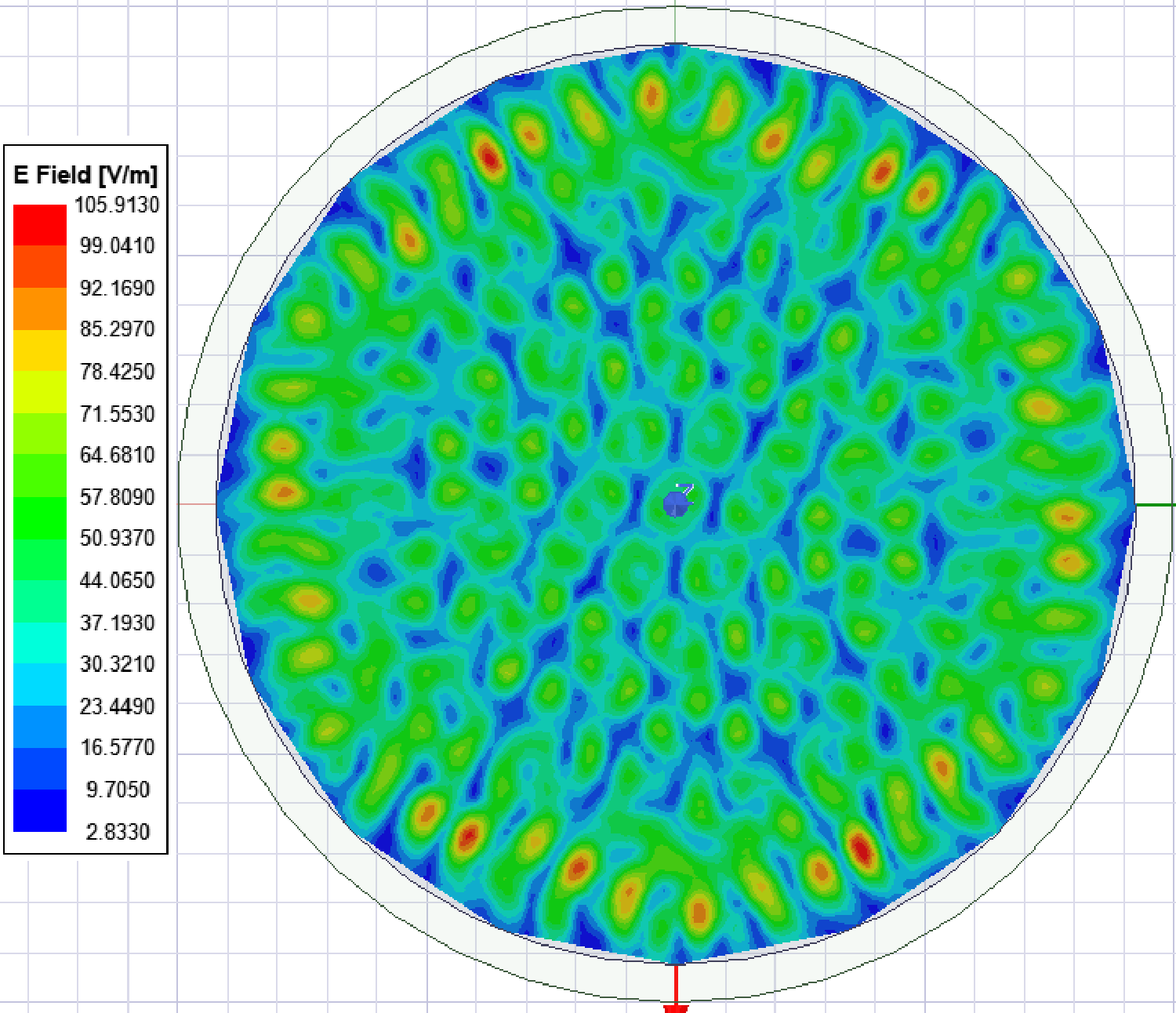}\label{fig:hfssMode1Power_Rt80mm}
%\vspace{1pt}
\end{minipage}
}
%\vspace{-5pt}
\subfigure[Mode 1 Phase.]{
\begin{minipage}{0.4\linewidth}
\centering
\includegraphics[scale=0.157]{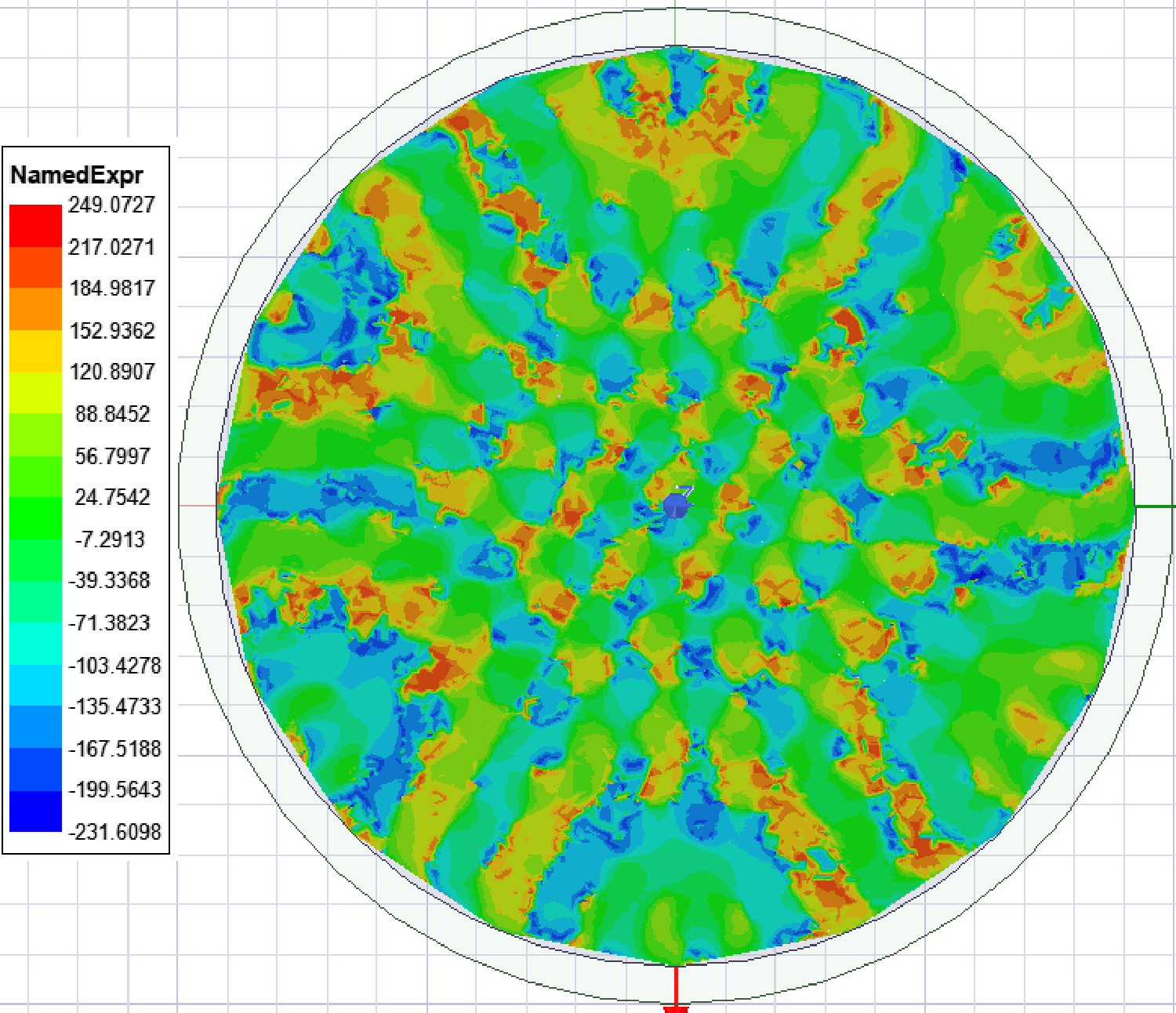}\label{fig:hfssMode1Phase_Rt80mm}
%\vspace{1pt}
\end{minipage}
}\\
%\vspace{-5pt}
\subfigure[Mode 1 gain.]{
\begin{minipage}{0.4\linewidth}
\centering
\includegraphics[scale=0.157]{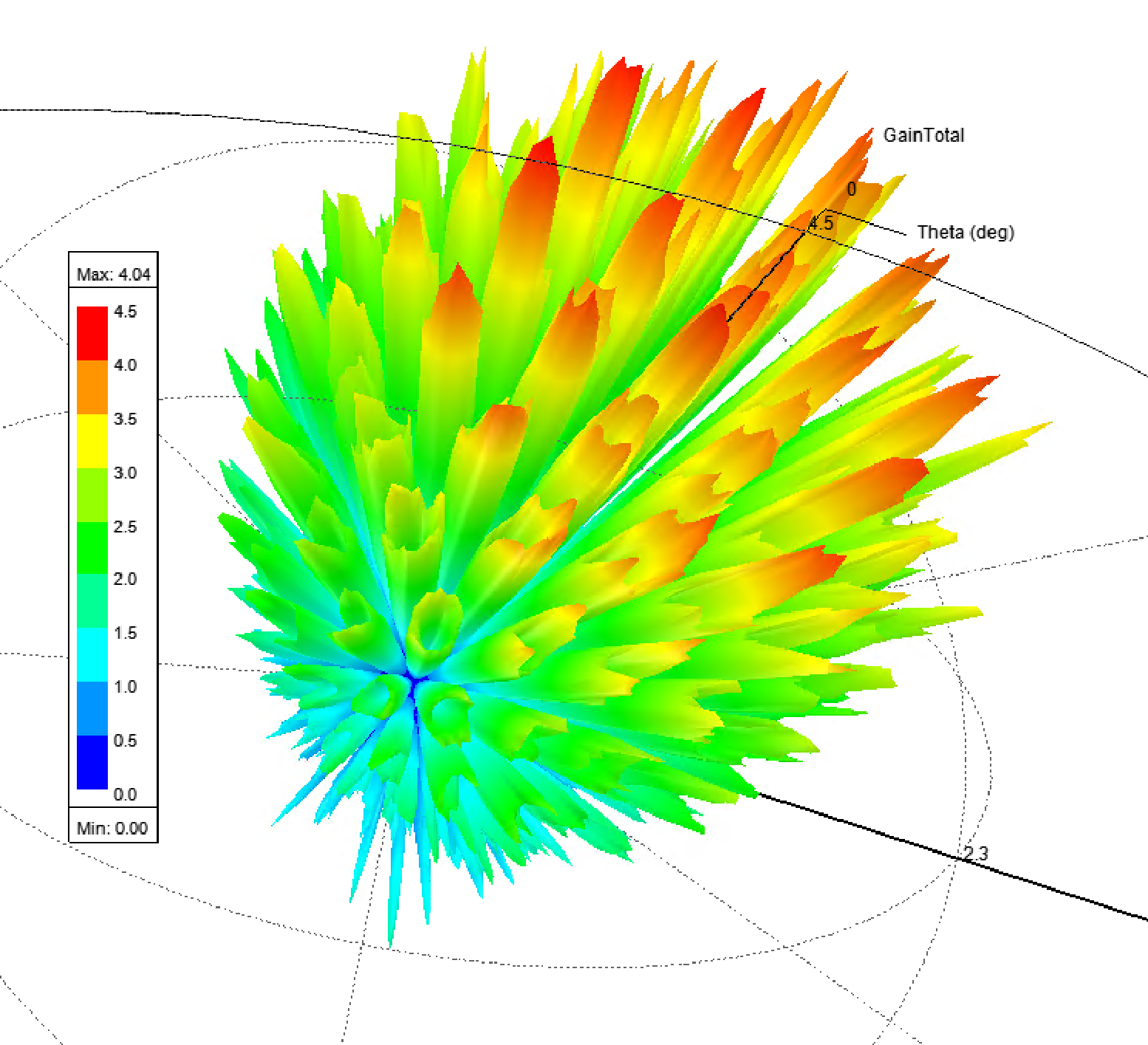}\label{fig:hfssMode1Gain_Rt80mm}
%\vspace{1pt}
\end{minipage}
}
%\vspace{-5pt}
\subfigure[{Mode 1 gain [dB]}.]{
\begin{minipage}{0.4\linewidth}
\centering
\includegraphics[scale=0.157]{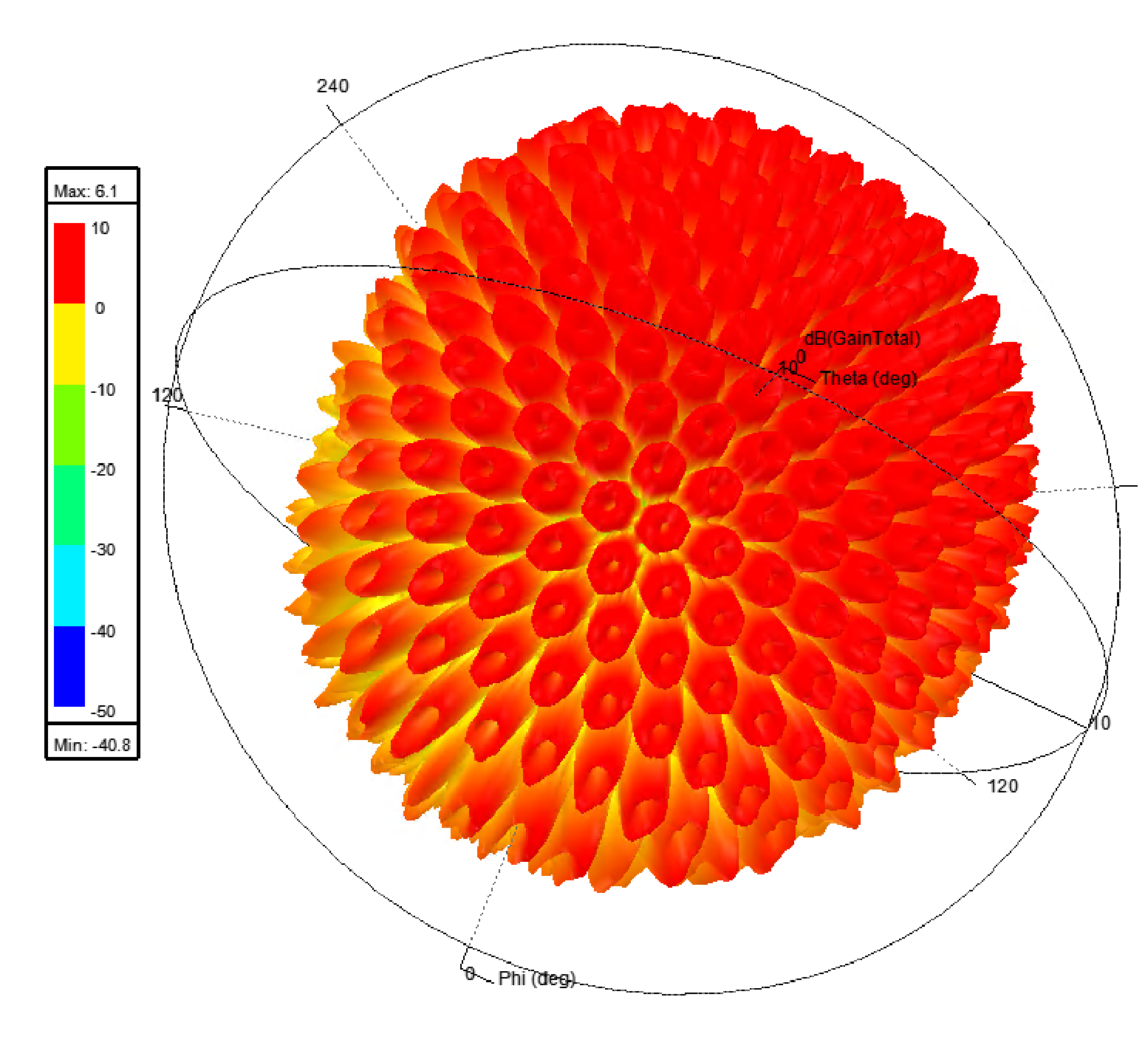}\label{fig:hfssMode1GaindB_Rt80mm}
%\vspace{1pt}
\end{minipage}
}
\centering
\vspace{-5pt}
%\caption{HFSS simulated power, phase, and antenna gain distributions and calculated OAM power and phase for $R_t = 10$ mm.}\label{fig:hfssAmpPhaGain_Rt80mm}
\caption{HFSS simulated power, phase, and antenna gain distributions for $R_t = 80$ mm.}\label{fig:hfssAmpPhaGain_Rt80mm}
%\vspace{-15pt}
\end{figure}
Figure~\ref{fig:hfssAmpPhaGain_Rt80mm} shows the amplitude, phase, and antenna gain distributions for the fractal OAM beams with OAM-mode set to $1$ and $R_t = 80$ mm based on the model given in Fig.~\ref{fig:hfssModel}. The distance from the transmit UCA to the receive plane is also set as $75$ mm. Other variables remain the same as in Fig.~\ref{fig:hfssAmpPhaGain}. Compared Fig.~\ref{fig:hfssAmpPhaGain_Rt80mm} with Fig.~\ref{fig:hfssAmpPhaGain}, we can find that a UCA with a larger transmit UCA radius can generate more fractal OAM beams at the same distance from the transmit UCA. Also, the non-centered OAM phase structures with larger transmit UCA radius exhibit a higher degree of completeness than the phase structures generated by a smaller transmit UCA. However, a larger transmit UCA radius leads to a lower received power at the same distance. Moreover, a larger transmit UCA radius means a larger footprint and higher synchronization requirements. Also, a larger transmit UCA radius results in a larger simulation volume, greatly slowing down the simulation speed. Therefore, we set $R_t = 30$ mm in the following simulations and analyses.

\subsection{Capacity and BER Performances}
\begin{figure}[htbp]
\centering
%\vspace{-15pt}
\subfigure[Aligned UCA model for fractal OAM.]{
\begin{minipage}{0.45\linewidth}
\centering
\includegraphics[scale=0.17]{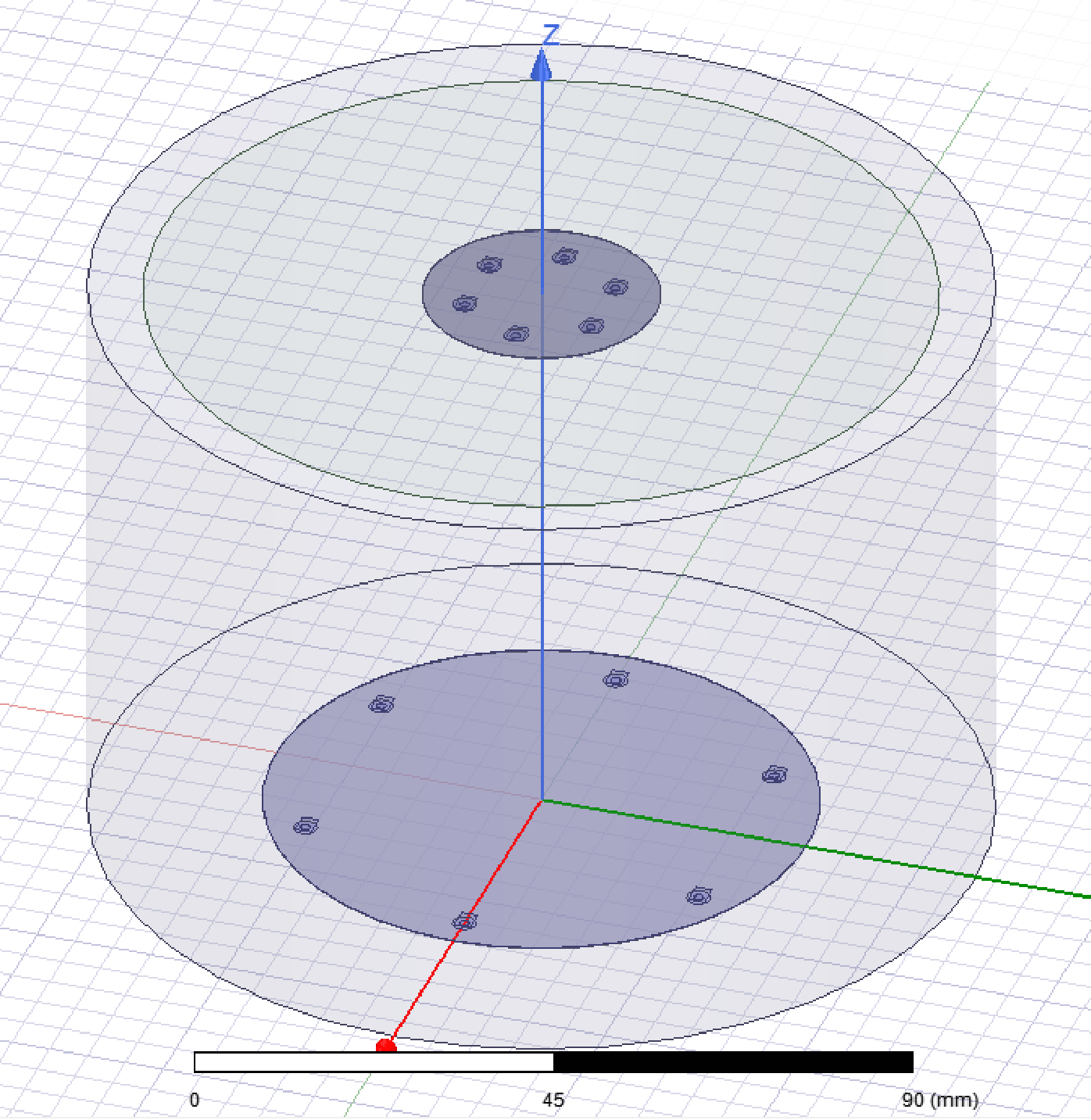}\label{fig:hfssUCAModel_dual_Rt30mm}
\vspace{3pt}
\end{minipage}
}
\subfigure[Unaligned UCA model for fractal OAM.]{
\begin{minipage}{0.45\linewidth}
\centering
\includegraphics[scale=0.17]{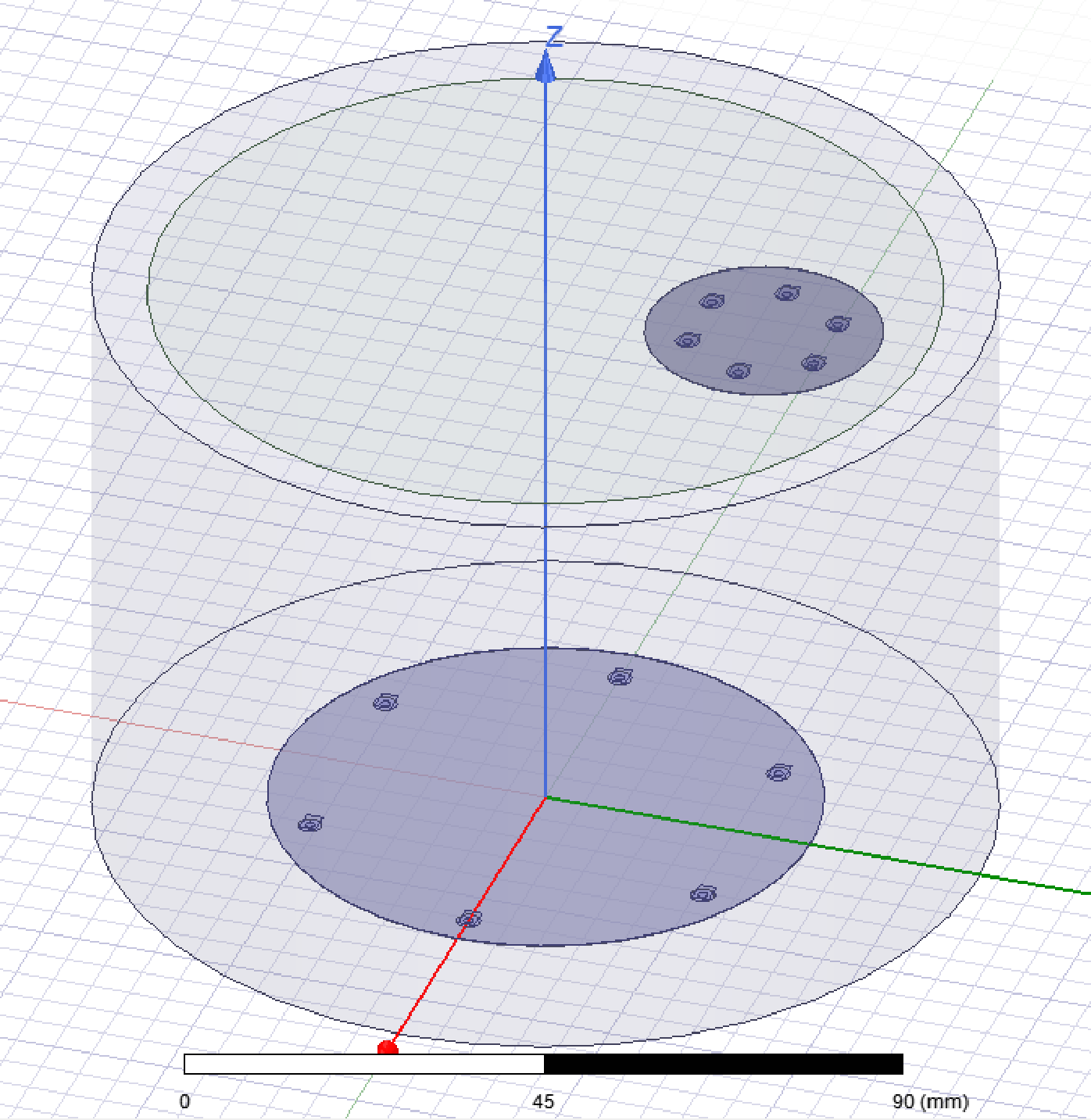}\label{fig:hfssUCAModel_dualUnaligned_Rt30mm}
\vspace{3pt}
\end{minipage}
}\\
\subfigure[Aligned UCA model for normal OAM.]{
\begin{minipage}{0.45\linewidth}
\centering
\includegraphics[scale=0.17]{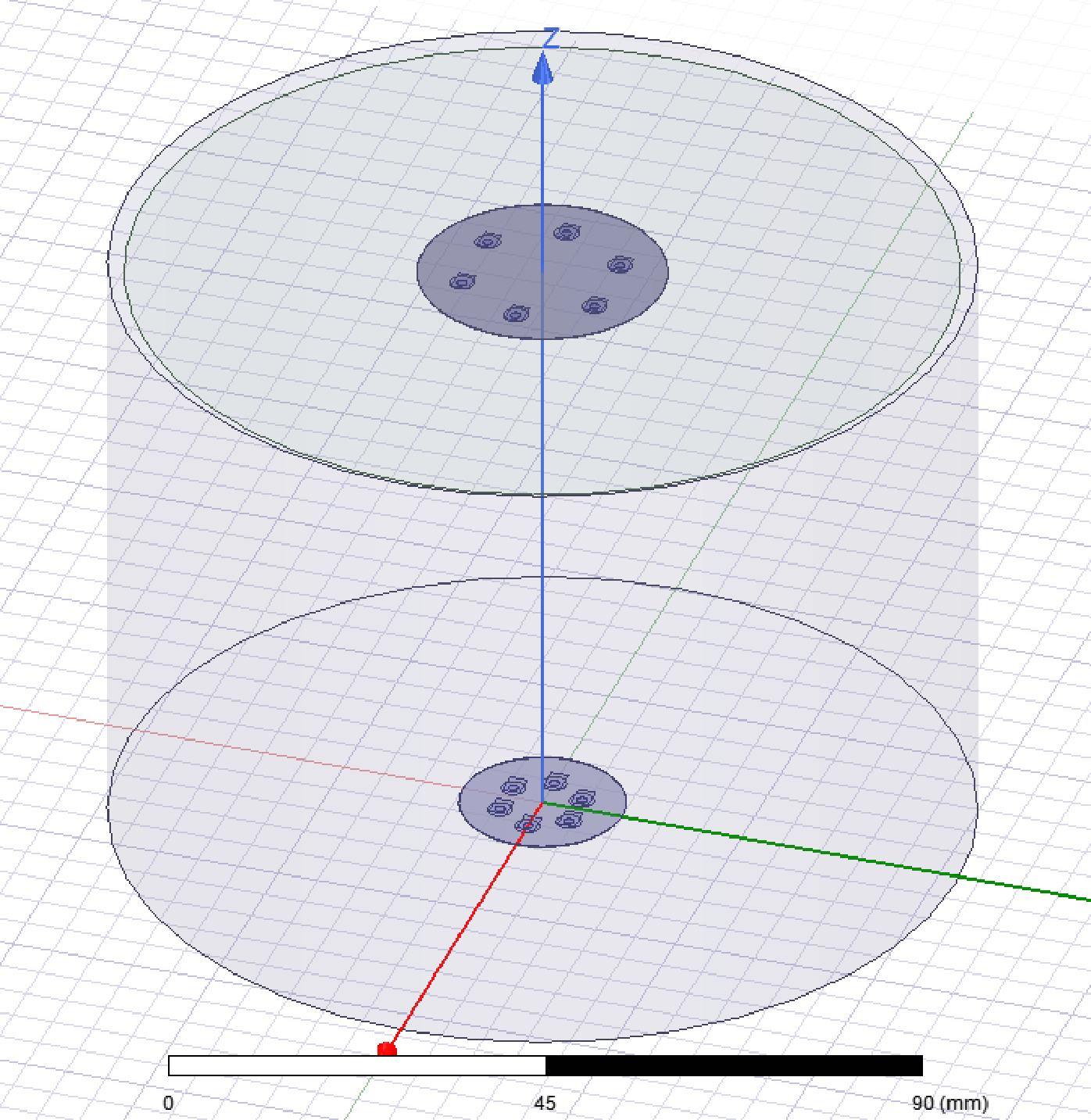}\label{fig:hfssUCAModel_dualNormal_Rt30mm}
\vspace{3pt}
\end{minipage}
}
\subfigure[Unaligned UCA model for normal OAM.]{
\begin{minipage}{0.45\linewidth}
\centering
\includegraphics[scale=0.17]{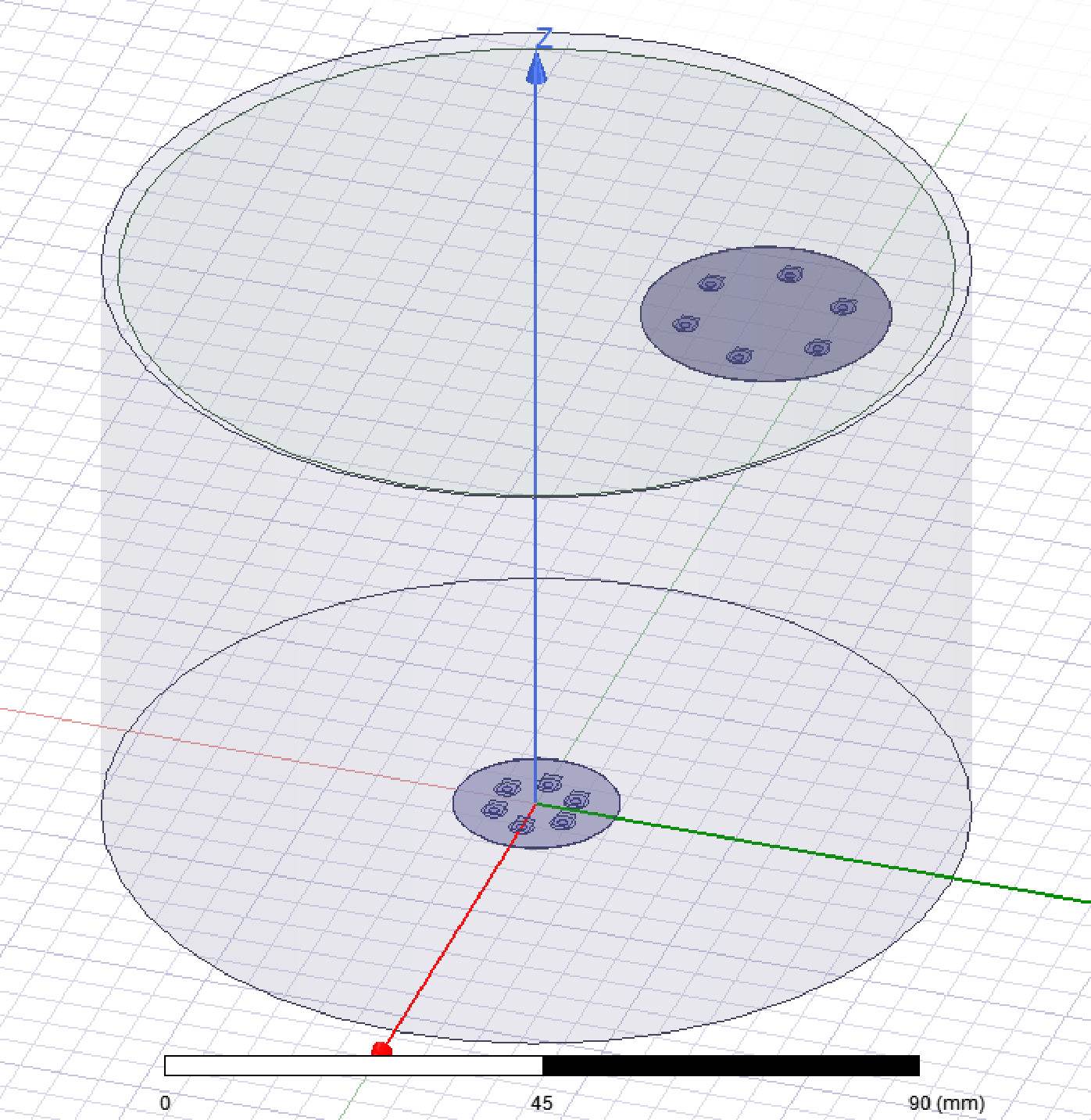}\label{fig:hfssUCAModel_dualUnalignedNormal_Rt30mm}
\vspace{3pt}
\end{minipage}
}
\centering
%\vspace{-10pt}
\caption{HFSS simulation UCA models for fractal OAM and normal OAM.}\label{fig:hfssModel_dual_Rt30mm}
%\vspace{-10pt}
\end{figure}
Figure~\ref{fig:hfssModel_dual_Rt30mm} shows the simulation UCA models for the fractal OAM and normal OAM transmissions. We set the radius of the transmit UCA as $30$ mm for fractal OAM and set the radius of the transmit UCA as $5$ mm for normal OAM. The receive UCAs are set on the plane $75$ mm away from the transmit UCA. We set the radii of the receive UCAs as $9.62$ mm for both the fractal OAM and normal OAM transmissions, which is the upper bound of $R_r$ according to Eq.~\eqref{eq:RrRadius}. Since $R_t$ is set as $30$ mm for the fractal transmission, the non-central fractal OAM beam has only one circle, so there are only binary cases, aligned and unaligned. For aligned transmissions, the receive UCA centers are positioned at $(0,0,75)$ mm. For unaligned transmissions, the receive UCA centers are positioned at $(0,28.87,75)$ mm according to Eq.~\eqref{eq:OAM_grid_CarCorrdinates} by setting $m=0$ and $n=2$. Other variables remain the same as in Fig.~\ref{fig:hfssModel}.

\begin{figure}[htbp]
\centering
%\vspace{-15pt}
\subfigure[Capacity performance.]{
\begin{minipage}{1\linewidth}
\centering
\includegraphics[scale=0.5]{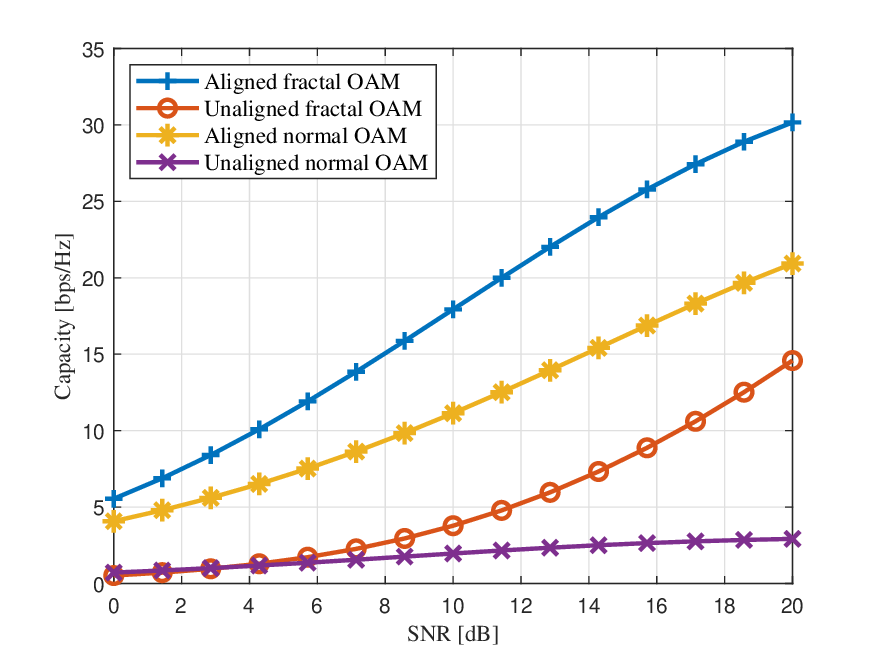}\label{fig:HFSS_Fractal_OAM_capacity}
%\vspace{-20pt}
\end{minipage}
}
%\vspace{-10pt}
\subfigure[BER performance.]{
\begin{minipage}{1\linewidth}
\centering
\includegraphics[scale=0.5]{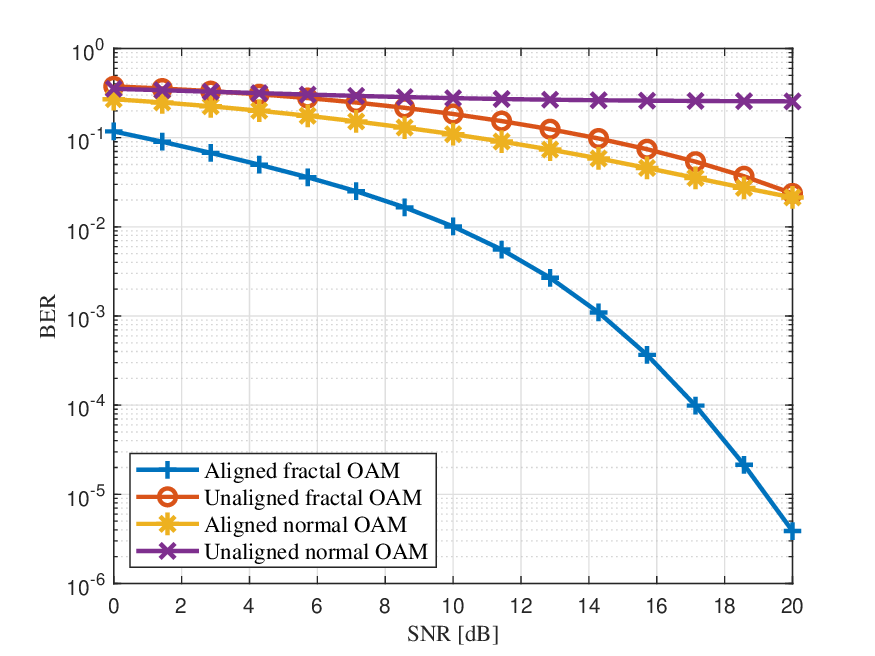}\label{fig:HFSS_Fractal_OAM_BER}
%\vspace{-20pt}
\end{minipage}
}
\centering
%\vspace{-10pt}
\caption{HFSS simulated capacity and BER performances for normal OAM and fractal OAM transmissions with aligned and unaligned UCAs.}\label{fig:HFSS_Fractal_OAM_capacityBER}
%\vspace{-5pt}
\end{figure}
Based on the S-parameter matrices of UCA models in Fig.~\ref{fig:hfssModel_dual_Rt30mm} given by HFSS, which describe the relationship between different ports, we can get the free-space-channel matrices by taking the bottom left quarter of the S-parameter matrices\cite{OAM_NFC}. We then plot the capacity and BER performances for normal OAM and fractal OAM transmissions with aligned and unaligned UCAs in Fig.~\ref{fig:HFSS_Fractal_OAM_capacityBER} by substituting the free-space-channel matrices into Eqs.~\eqref{eq:fractal_OAM_capacity} and \eqref{eq:fractal_OAM_BER}. Figure~\ref{fig:HFSS_Fractal_OAM_capacityBER} shows that, in the transceiver-aligned scenario, fractal OAM exhibits superior capacity and BER performances compared to normal OAM. This observation indicates that fractal OAM effectively alleviates the hollow divergence of OAM beams, leading to better energy concentration and improved communication performance. In the transceiver-unaligned scenario, fractal OAM also outperforms normal OAM by a significant margin in terms of capacity and BER performance. This means that the receive UCA no longer needs to be perfectly aligned with the transmit UCA, allowing for more flexibility in the arrangement of the receive UCA. The simulation results given in Fig.~\ref{fig:HFSS_Fractal_OAM_capacityBER} are consistent with the numerical results given in Fig.~\ref{fig:Fractal_OAM_capacityBER}. Moreover, the simulation BER of transceiver-aligned normal OAM in Fig.~\ref{fig:HFSS_Fractal_OAM_BER} are much lower than the numerical results in Fig.~\ref{fig:Fractal_OAM_BER}. This is because the transceiver distance in simulations is much smaller than in numerical results, resulting in less divergence and greater received power.

\subsection{Impact of the Receive UCA Radius}
\begin{figure}[htbp]
\centering
%\vspace{-15pt}
\subfigure[Aligned UCA model with $R_r = 4.81$ mm.]{
\begin{minipage}{0.45\linewidth}
\centering
\includegraphics[scale=0.17]{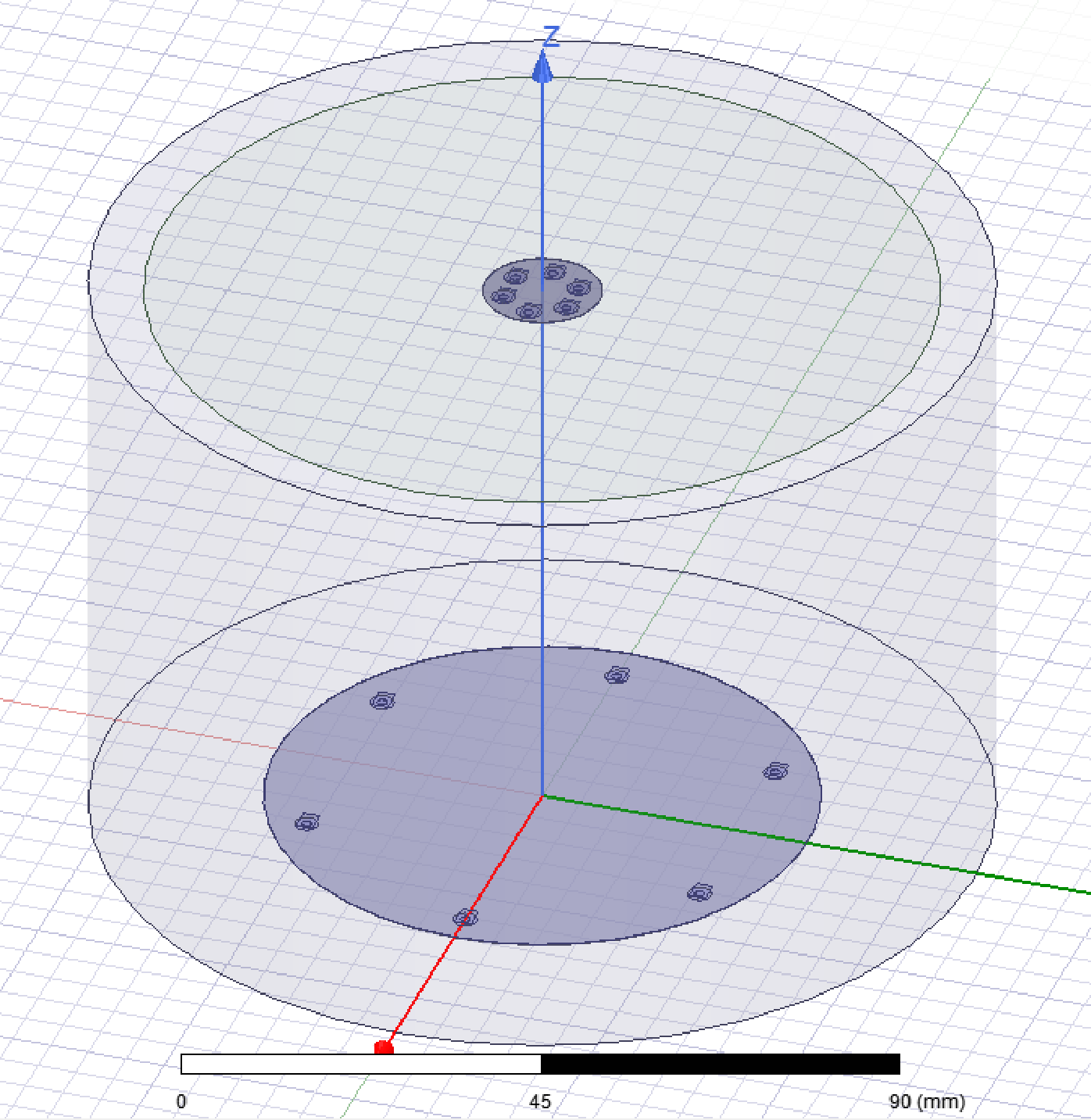}\label{fig:hfssUCAModel_dual_RrDiv2}
\vspace{3pt}
\end{minipage}
}
\subfigure[Unaligned UCA model with $R_r = 4.81$ mm.]{
\begin{minipage}{0.45\linewidth}
\centering
\includegraphics[scale=0.17]{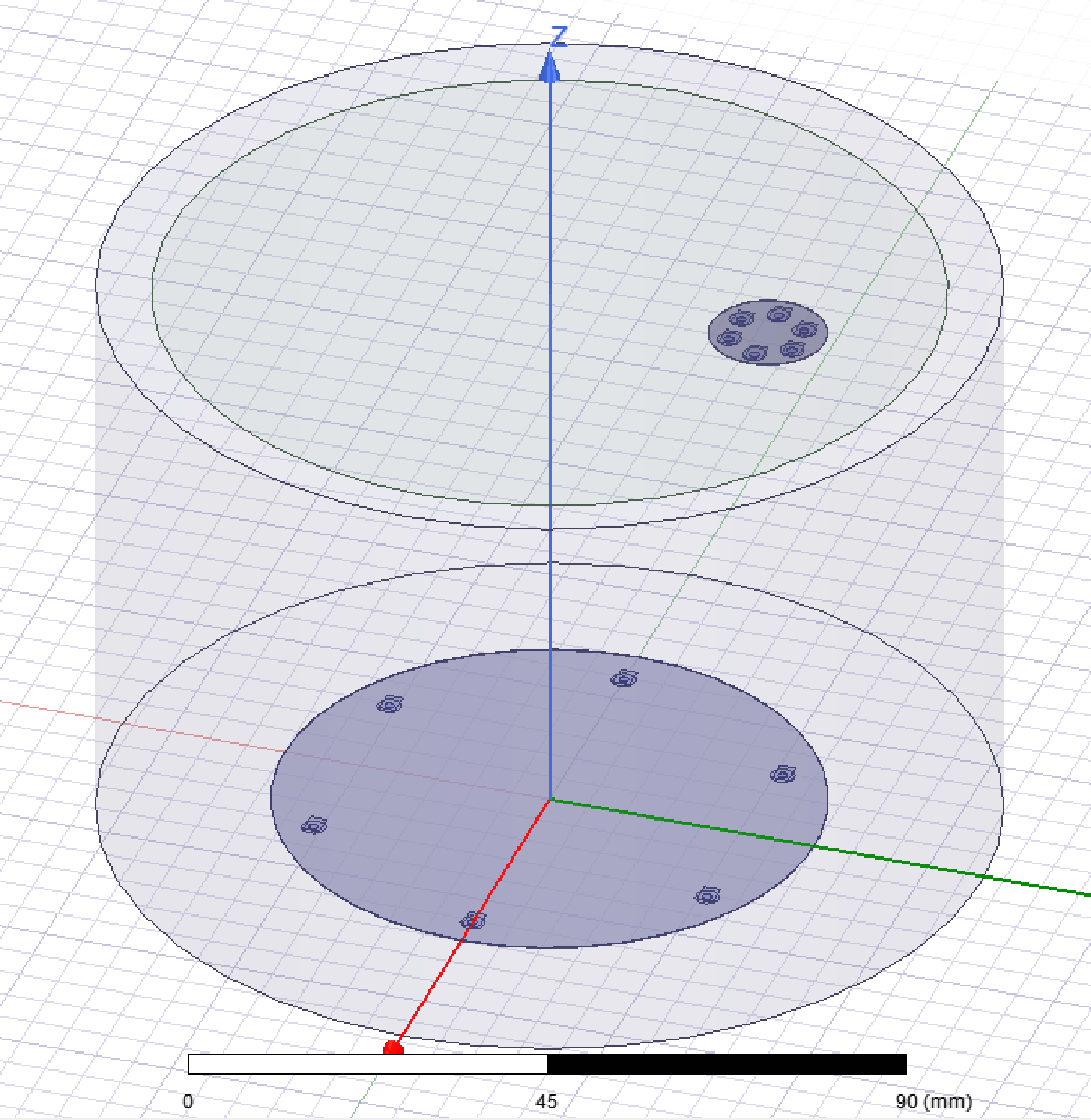}\label{fig:hfssUCAModel_dualUnaligned_RrDiv2}
\vspace{3pt}
\end{minipage}
}\\
\subfigure[Aligned UCA model with $R_r = 3.21$ mm.]{
\begin{minipage}{0.45\linewidth}
\centering
\includegraphics[scale=0.17]{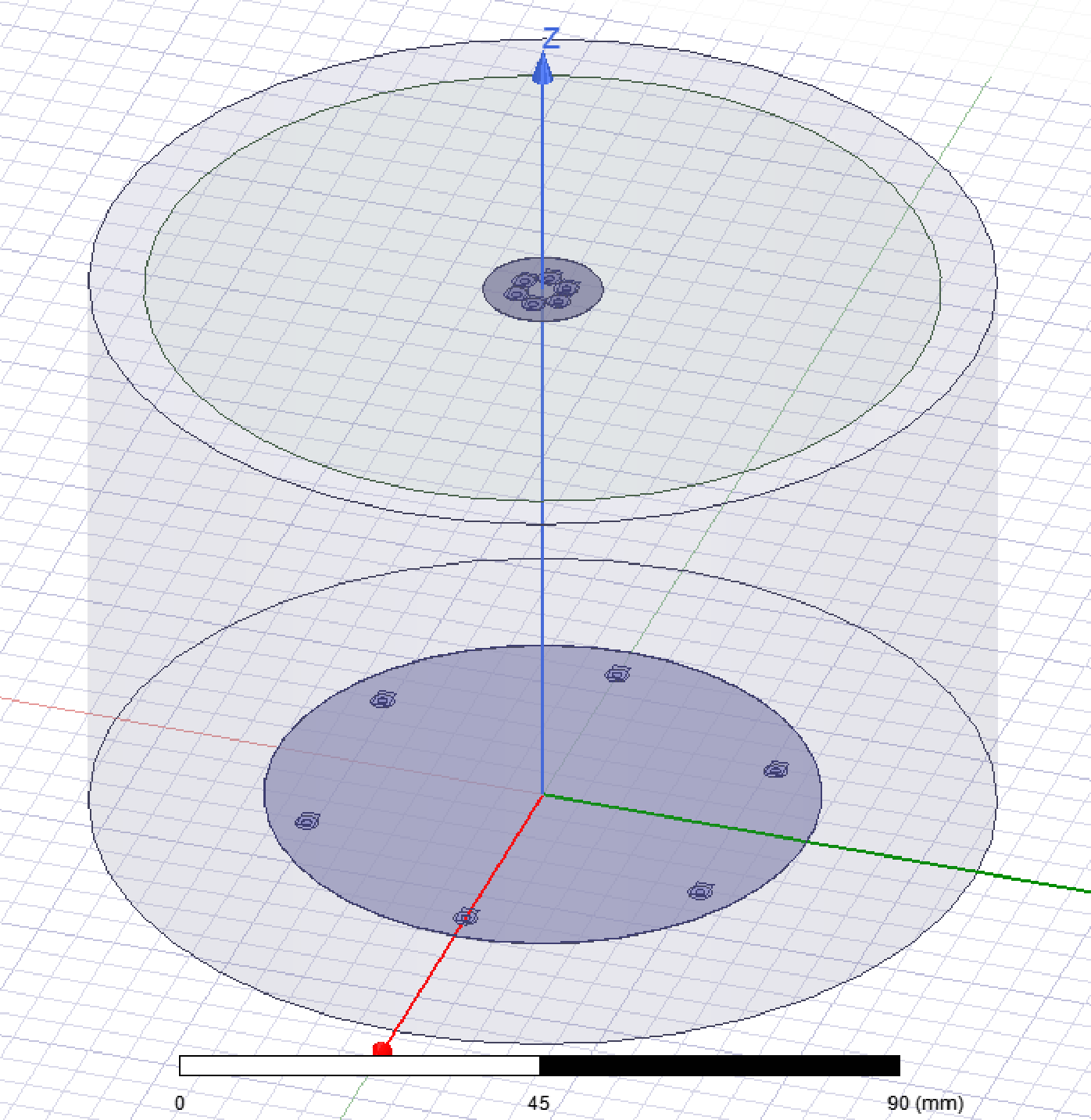}\label{fig:hfssUCAModel_dualNormal_RrDiv3}
\vspace{3pt}
\end{minipage}
}
\subfigure[Unaligned UCA model with $R_r = 3.21$ mm.]{
\begin{minipage}{0.45\linewidth}
\centering
\includegraphics[scale=0.17]{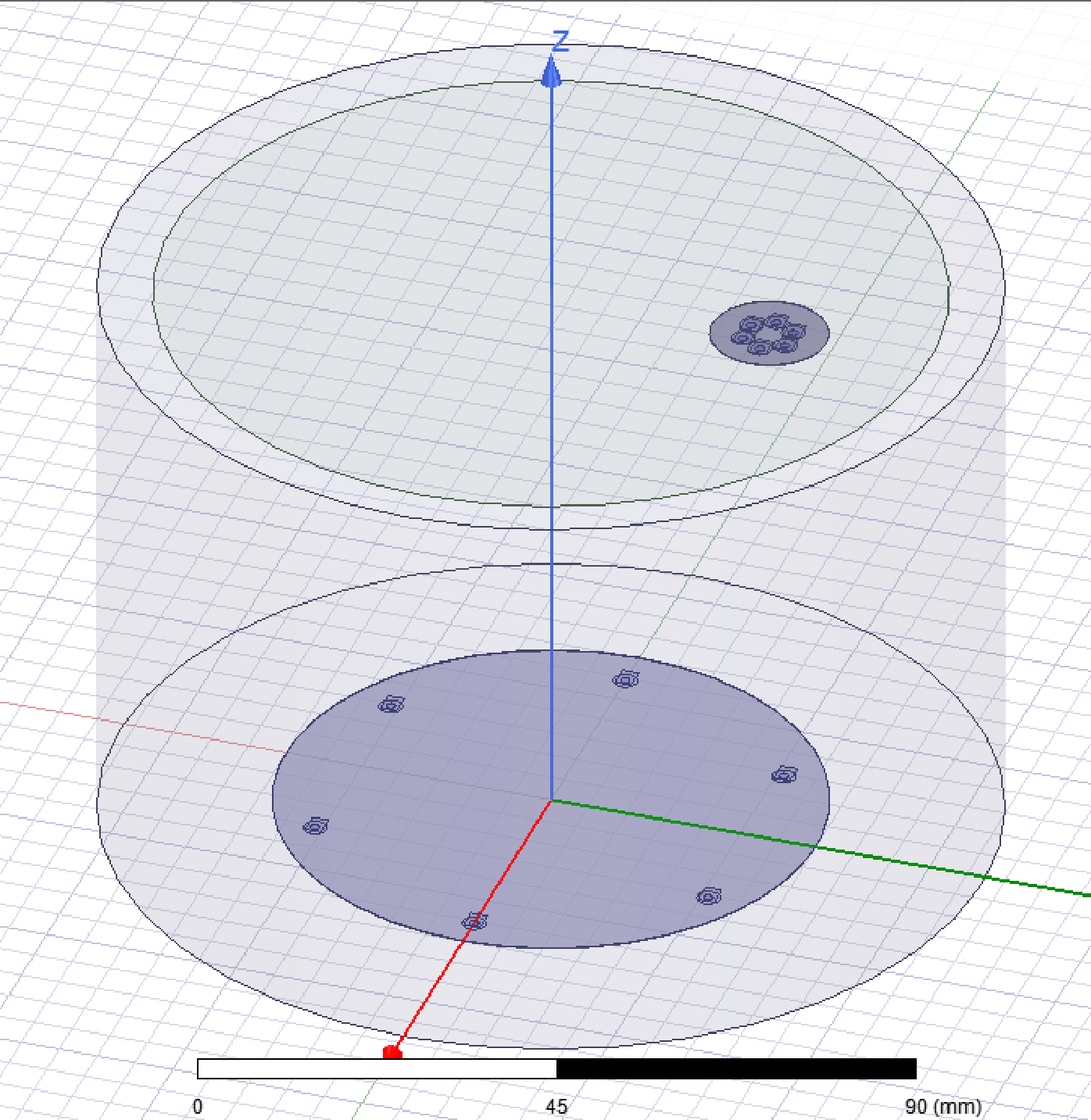}\label{fig:hfssUCAModel_dualUnalignedNormal_RrDiv3}
\vspace{3pt}
\end{minipage}
}
\centering
%\vspace{-10pt}
\caption{HFSS simulation UCA models for fractal OAM with different receive UCA radii.}\label{fig:hfssModel_dual_Rr}
%\vspace{-10pt}
\end{figure}
Figure~\ref{fig:hfssModel_dual_Rr} shows the simulation UCA models for fractal OAM with different receive UCA radii. In addition to the simulation results with a receive UCA radius of $9.62$ mm given in the previous subsection, we also simulated the cases with receive UCA radii of $4.81$ and $3.21$ mm, which are half and one-third as large as $9.62$ mm in Fig.~\ref{fig:hfssModel_dual_Rt30mm}. Other variables remain the same as in Fig.~\ref{fig:hfssModel_dual_Rt30mm}.

\begin{figure}[htbp]
\centering
%\vspace{-15pt}
\subfigure[Capacity performance.]{
\begin{minipage}{1\linewidth}
\centering
\includegraphics[scale=0.5]{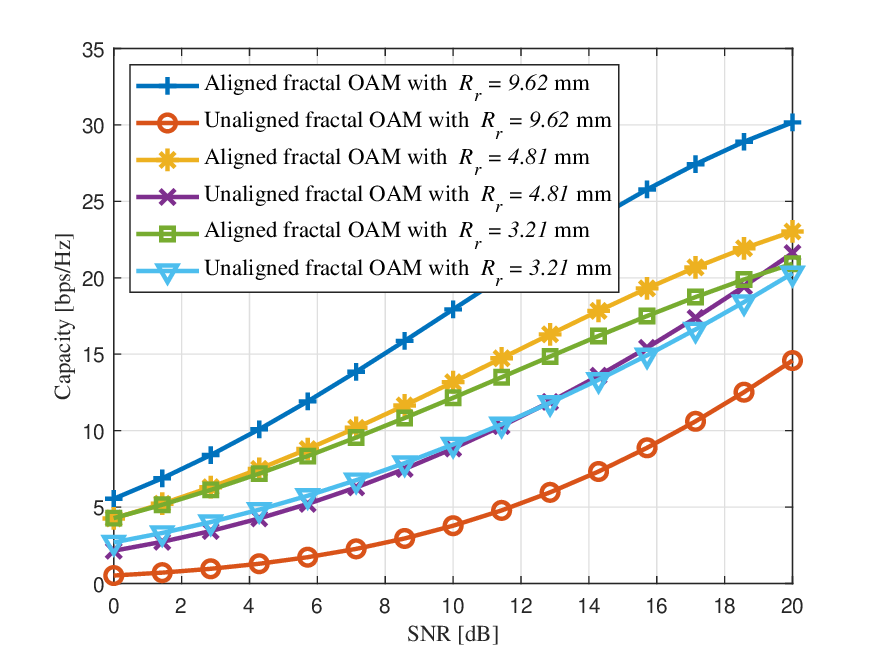}\label{fig:HFSS_Fractal_OAM_capacity_Rr}
%\vspace{-20pt}
\end{minipage}
}
%\vspace{-10pt}
\subfigure[BER performance.]{
\begin{minipage}{1\linewidth}
\centering
\includegraphics[scale=0.5]{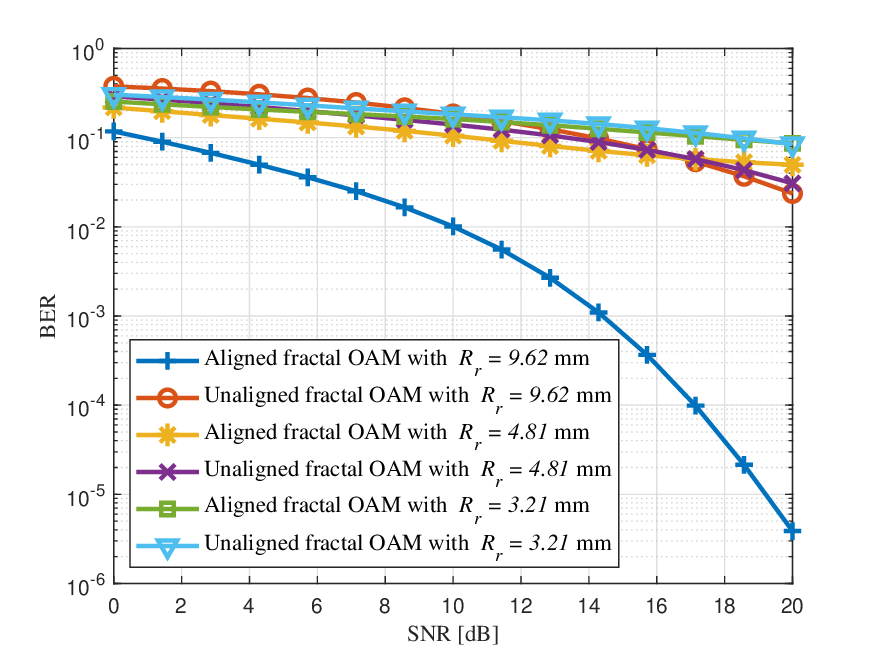}\label{fig:HFSS_Fractal_OAM_BER_Rr}
%\vspace{-20pt}
\end{minipage}
}
\centering
%\vspace{-10pt}
\caption{HFSS simulated impact of the receive UCA radius for fractal OAM transmissions with aligned and unaligned UCAs.}\label{fig:HFSS_Fractal_OAM_capacityBER_Rr}
%\vspace{-5pt}
\end{figure}
Based on the S-parameter matrices of UCA models in Fig.~\ref{fig:hfssModel_dual_Rr} given by HFSS, we plot the capacity and BER performances for fractal OAM with different receive UCA radii in Fig.~\ref{fig:HFSS_Fractal_OAM_capacityBER_Rr}. For aligned transceivers, fractal OAM with $R_r=9.62$ mm have the best capacity and BER performances compared with the other two cases as shown in Fig,~\ref{fig:HFSS_Fractal_OAM_capacityBER_Rr}. Aligned fractal OAM with $R_r=3.21$ mm have the lowest capacity and highest BER among three transceiver-aligned cases. In general, the capacity decreases and BER increases as the receive UCA radius increases for aligned transceivers. However, for transceiver-unaligned cases, fractal OAM with $R_r=9.62$ mm have the lowest capacity and highest BER. Unaligned fractal OAM transmissions with $R_r=4.81$ mm and $R_r=3.21$ mm have similar capacity and BER performances.

\subsection{Impact of the Distance Between the Transceiver UCAs}
\begin{figure}[htbp]
\centering
%\vspace{-15pt}
\subfigure[Aligned UCA model with $z = 50$ mm.]{
\begin{minipage}{0.45\linewidth}
\centering
\includegraphics[scale=0.17]{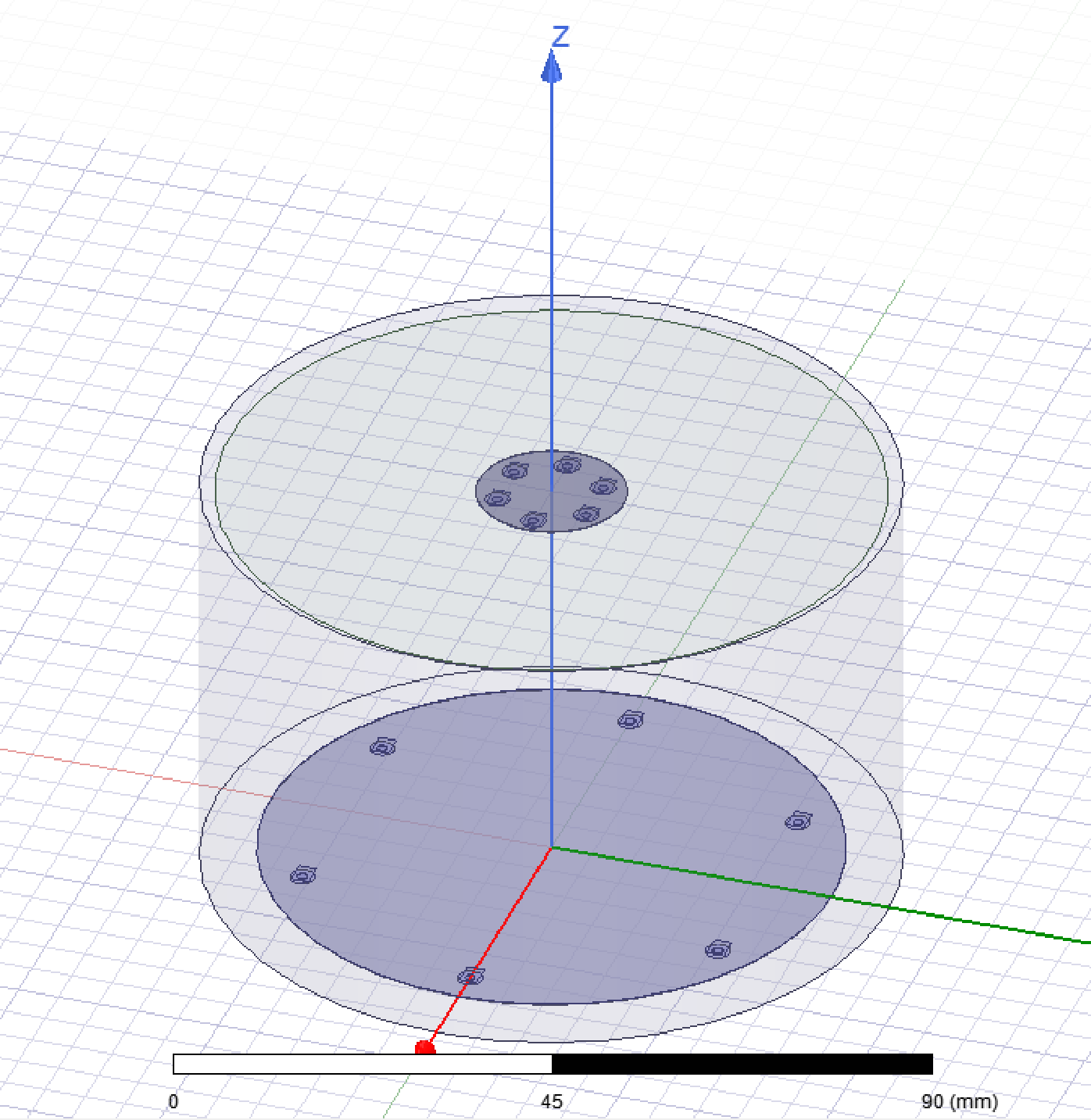}\label{fig:hfssUCAModel_dual_z50mm}
\vspace{3pt}
\end{minipage}
}
\subfigure[Unaligned UCA model with $z = 50$ mm.]{
\begin{minipage}{0.45\linewidth}
\centering
\includegraphics[scale=0.17]{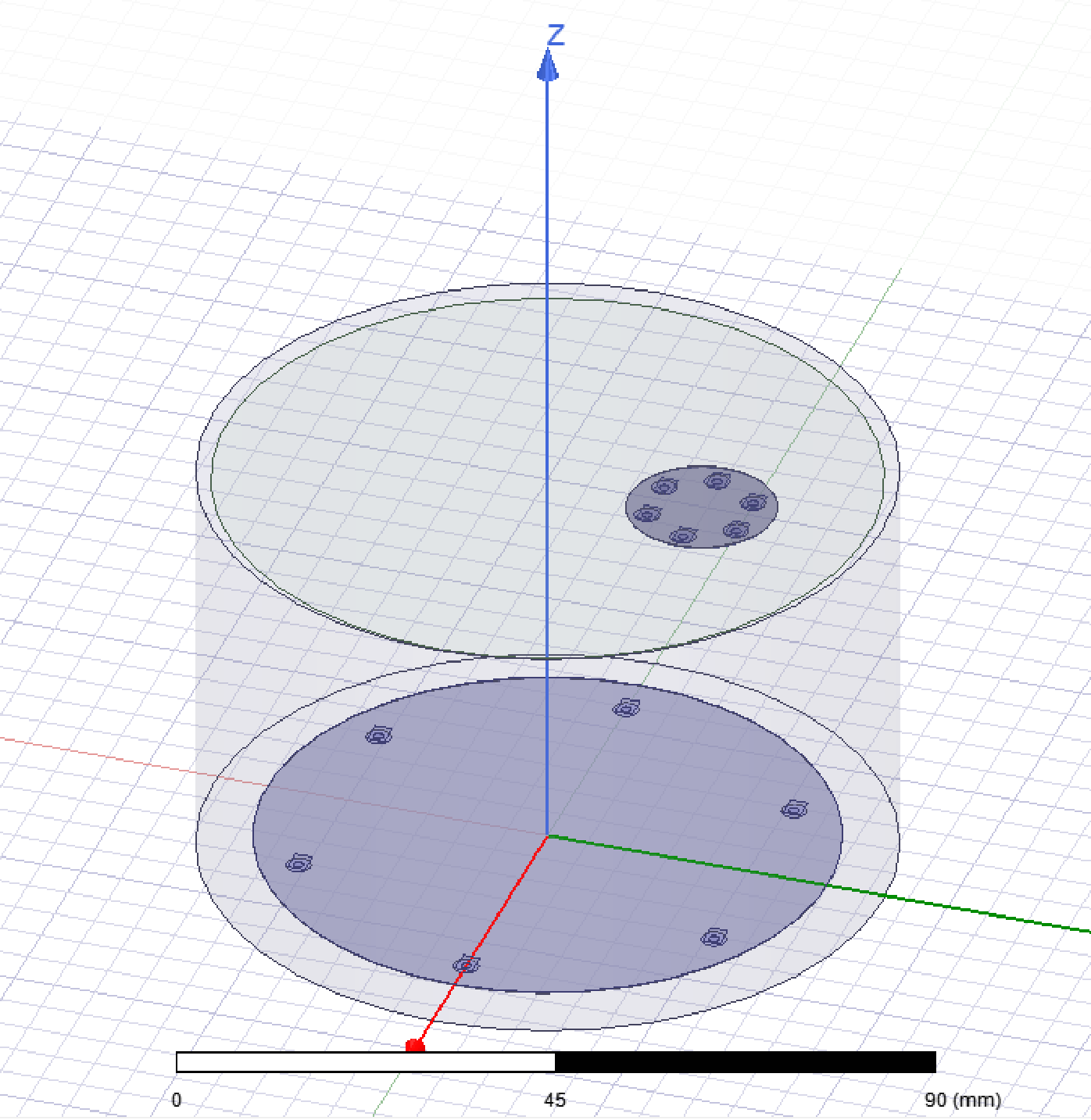}\label{fig:hfssUCAModel_dualUnaligned_z50mm}
\vspace{3pt}
\end{minipage}
}\\
\subfigure[Aligned UCA model with $z = 100$ mm.]{
\begin{minipage}{0.45\linewidth}
\centering
\includegraphics[scale=0.17]{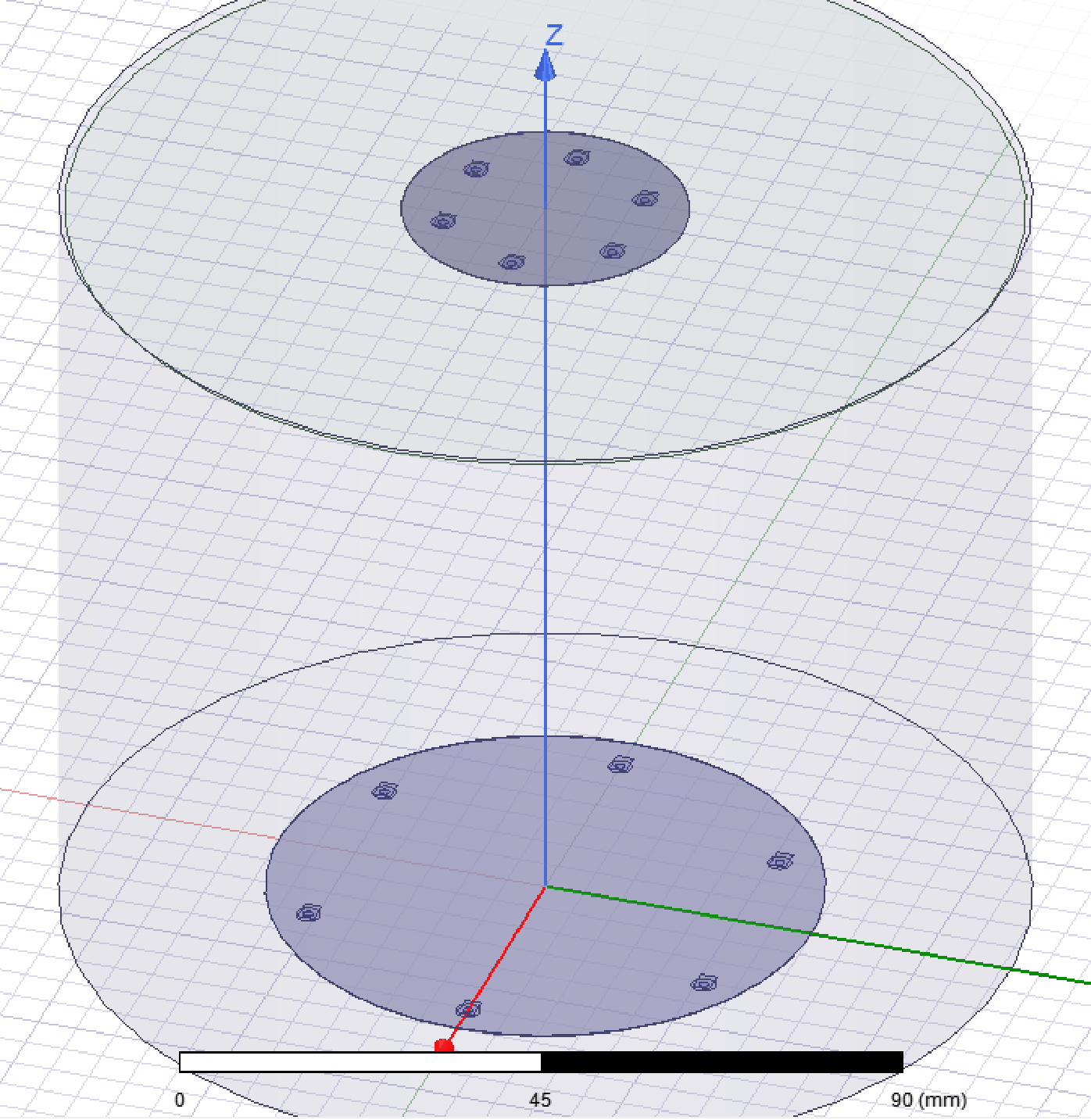}\label{fig:hfssUCAModel_dualNormal_z100mm}
\vspace{3pt}
\end{minipage}
}
\subfigure[Unaligned UCA model with $z = 100$ mm.]{
\begin{minipage}{0.45\linewidth}
\centering
\includegraphics[scale=0.17]{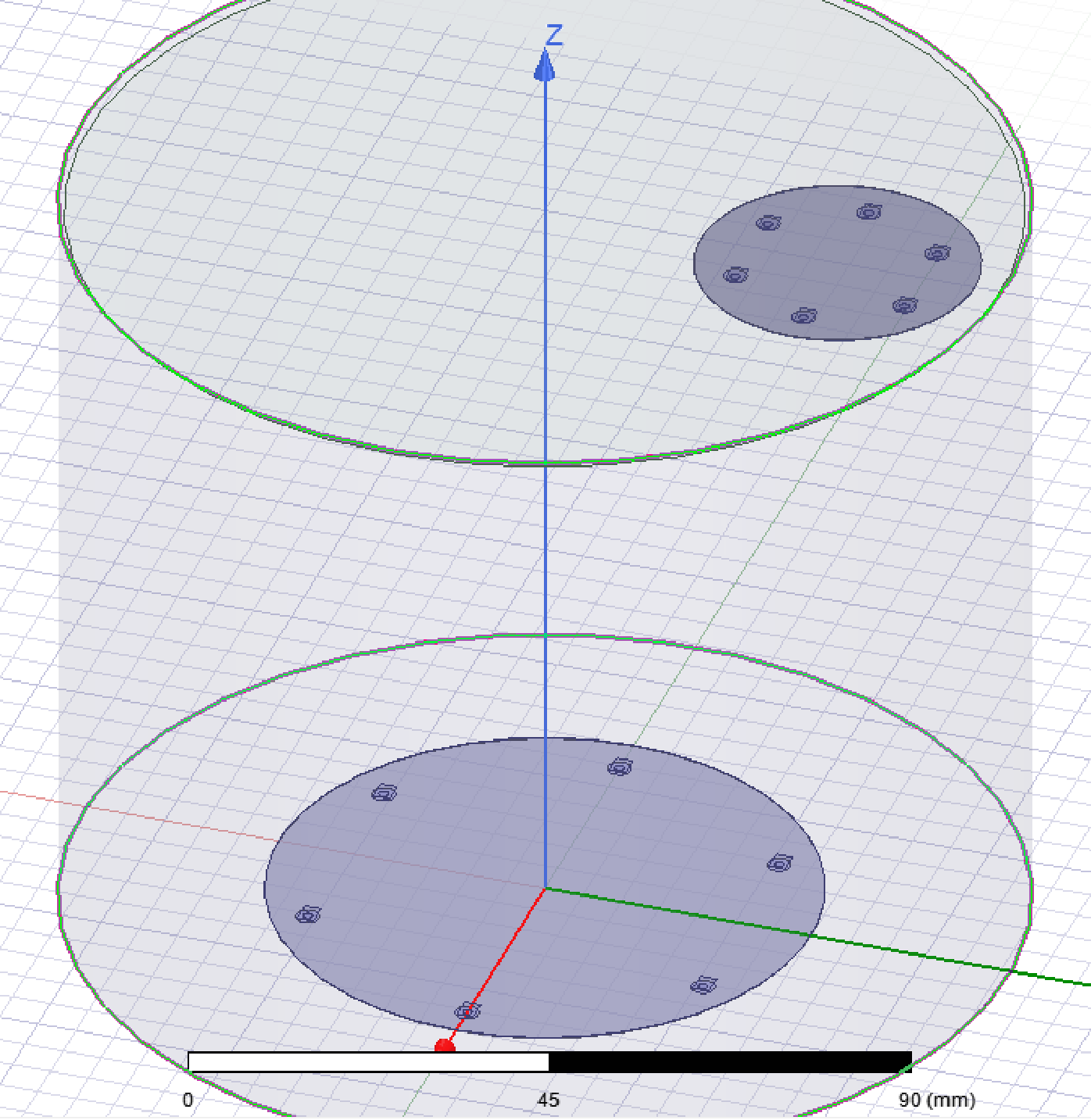}\label{fig:hfssUCAModel_dualUnalignedNormal_z100mm}
\vspace{3pt}
\end{minipage}
}
\centering
%\vspace{-10pt}
\caption{HFSS simulation UCA models for fractal OAM with different transceiver distances.}\label{fig:hfssModel_dual_z}
%\vspace{-10pt}
\end{figure}
Figure~\ref{fig:hfssModel_dual_z} shows the simulation UCA models for fractal OAM with different transceiver distances. In addition to the simulation results with receive UCAs set on the plane $75$ mm away from the transmit UCA given in the previous subsection, we also simulated the cases with receive UCAs set on the plane $50$ and $100$ mm away from the transmit UCA. We set the radii of the receive UCAs as $6.42$ mm and $12.83$ mm for $z=50$ mm and $z=100$ mm, respectively, according to the upper bound of $R_r$ in Eq.~\eqref{eq:RrRadius}. For unaligned transceivers, the receive UCA centers are positioned at $(0,19.25,50)$ mm and $(0,38.49,100)$ mm for $z=50$ mm and $z=100$ mm, respectively, according to Eq.~\eqref{eq:OAM_grid_CarCorrdinates}. Other variables remain the same as in Fig.~\ref{fig:hfssModel_dual_Rt30mm}.

\begin{figure}[htbp]
\centering
%\vspace{-15pt}
\subfigure[Capacity performance.]{
\begin{minipage}{1\linewidth}
\centering
\includegraphics[scale=0.5]{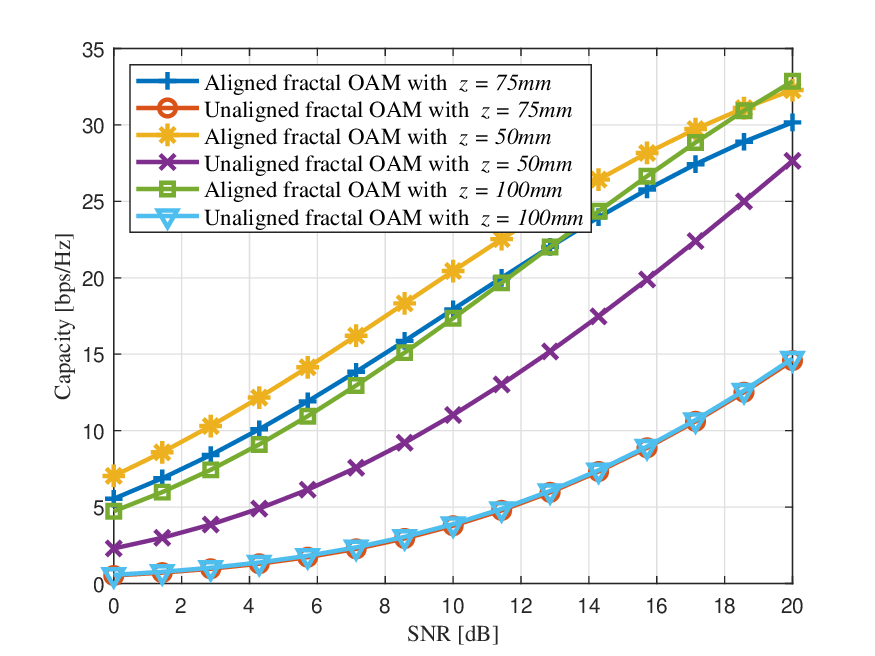}\label{fig:HFSS_Fractal_OAM_capacity_z}
%\vspace{-20pt}
\end{minipage}
}
%\vspace{-10pt}
\subfigure[BER performance.]{
\begin{minipage}{1\linewidth}
\centering
\includegraphics[scale=0.5]{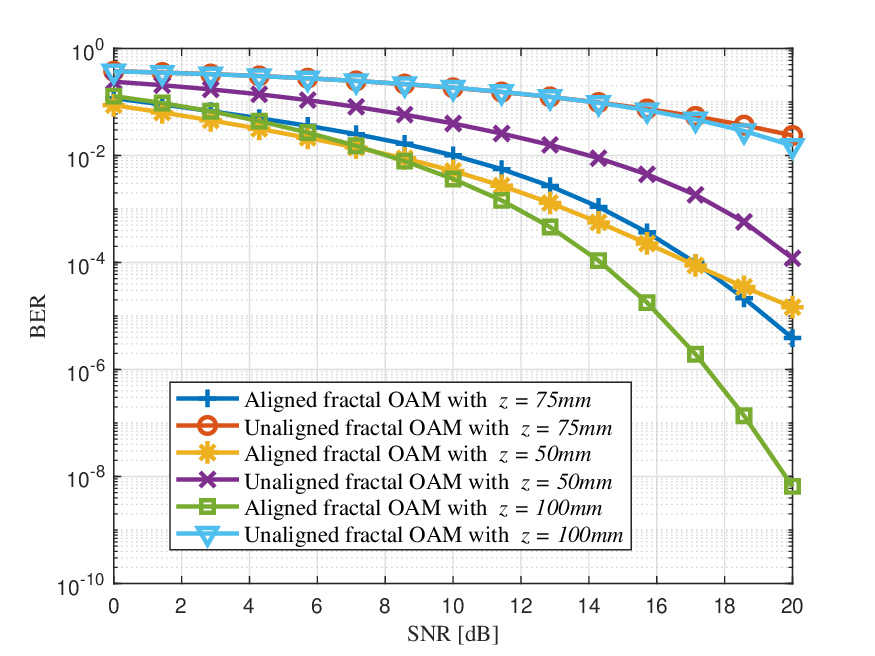}\label{fig:HFSS_Fractal_OAM_BER_z}
%\vspace{-20pt}
\end{minipage}
}
\centering
%\vspace{-10pt}
\caption{HFSS simulated impact of the transceiver distance for fractal OAM transmissions with aligned and unaligned UCAs.}\label{fig:HFSS_Fractal_OAM_capacityBER_z}
%\vspace{-5pt}
\end{figure}
Based on the S-parameter matrices of UCA models in Fig.~\ref{fig:hfssModel_dual_z} given by HFSS, we plot the capacity and BER performances for fractal OAM with different transceiver distances in Fig.~\ref{fig:HFSS_Fractal_OAM_capacityBER_z}. For aligned transceivers, fractal OAM with $z=50$ mm has better capacity and BER performances than fractal OAM with $z=75$ mm. Aligned fractal OAM with $z=100$ mm have the lowest capacity among the transceiver-aligned cases for SNR $<13$ dB and have the highest BER for SNR $<3$ dB as shown in Fig.~\ref{fig:HFSS_Fractal_OAM_capacityBER_z}. For SNR $>8.5$ dB, fractal OAM with $z=100$ mm have the lowest BER among the transceiver-aligned cases. For SNR $>18.5$ dB, aligned fractal OAM with $z=100$ mm have the highest capacity. For unaligned transceivers, fractal OAM transmissions with $z=75$ mm and $z=100$ mm have similar capacity and BER performances. While fractal OAM with $z=50$ mm have the best capacity and BER performances among the transceiver-unaligned cases.

\section{Conclusions}\label{sec:Conclusion}
In this paper, we proposed the fractal OAM generation and detection schemes, which could alleviate the hollow divergence of OAM beams and significantly improve the capacity and BER performances for unaligned OAM transmissions. We first briefly introduced the UCA-based fractal OAM phenomenon. Then, we proposed the fractal OAM beam generation and detection schemes by analyzing the minimum transmit UCA radius as well as the relationship between the transmit UCA radius, the transmission distance, the wavelength, the fractal OAM center coordinates, and the fractal OAM radius. After that, we derived the channel capacity and BER of our fractal OAM beam generation and detection schemes. Numerical results and simulations showed that the capacity and BER performances of our proposed fractal OAM outperform normal OAM transmissions. Simulations also validated the feasibility of our proposed fractal OAM generation scheme.

%\begin{appendices}
%\section{Proof for Theorem~\ref{the:channel_matrix}}\label{pro:multi-coil channel}
%
%\end{appendices}

\bibliographystyle{IEEEtran}
\bibliography{References}

\ifCLASSOPTIONcaptionsoff
  \newpage
\fi

\end{document}